\documentclass[10pt,reqno]{amsart}
\usepackage{latexsym,amsmath}
\usepackage{times}
\usepackage{amsfonts}
\usepackage{amssymb}
\usepackage{latexsym}

\usepackage{bbm}

\newcommand{\beq}{\begin{equation}} \newcommand{\eeq}{\end{equation}}

\numberwithin{equation}{section}

\newcommand\br[1]{\left(#1\right)}

\def\vec#1{\mathchoice{\mbox{\boldmath$\displaystyle#1$}}
{\mbox{\boldmath$\textstyle#1$}}
{\mbox{\boldmath$\scriptstyle#1$}}
{\mbox{\boldmath$\scriptscriptstyle#1$}}}

\DeclareMathOperator{\pr}{\mathbb P}

\newcommand\sh[1]{\textcolor{cyan}{#1}}
\newtheorem{definition}{Definition}[section]

\newtheorem{theorem}[definition]{Theorem}
\newtheorem{lemma}[definition]{Lemma}
\newtheorem{proposition}[definition]{Proposition}
\newtheorem{corollary}[definition]{Corollary}

\newtheorem{algorithm}[definition]{Algorithm}
\newtheorem{fact}[definition]{Fact}

\usepackage[dvips]{epsfig}
\graphicspath{{./Fig_eps/}{./Eps/}}
\DeclareGraphicsExtensions{.ps,.eps}
\usepackage{color}
\usepackage{fullpage}

\newcommand\rk{r_{k\mathrm{-SAT}}}
\newcommand\sign{\mathrm{sign}}

\newcommand\suc{\mathrm{success}}

\newcommand\cA{\mathcal{A}}
\newcommand\cB{\mathcal{B}}
\newcommand\cC{\mathcal{C}}

\newcommand\cE{\mathcal{E}}
\newcommand\cU{\mathcal{U}}
\newcommand\cN{\mathcal{N}}
\newcommand\cQ{\mathcal{Q}}

\newcommand\cS{\mathcal{S}}
\newcommand\cT{\mathcal{T}}

\newcommand\cL{\mathcal{L}}
\newcommand\cM{\mathcal{M}}

\newcommand\cP{\mathcal{P}}

\newcommand\cV{\mathcal{V}}

\newcommand\cZ{\mathcal{Z}}

\def\cC{{\mathcal C}}
\def\cE{{\mathcal E}}

\newcommand\eul{\mathrm{e}}
\newcommand\eps{\varepsilon}

\newcommand\Erw{\mathbb{E}}

\newcommand{\vecone}{\vec{1}}

\newcommand{\Bin}{{\rm Bin}}

\newcommand\ra{\rightarrow}

\newcommand\brk[1]{\left\lbrack{#1}\right\rbrack}

\newcommand\RR{\mathbb{R}}

\newcommand{\whp}{w.h.p.}

\newcommand\el{{[\ell]}}
\newcommand\elp{{[\ell+1]}}
\newcommand\elm{{[\ell-1]}}
\newcommand\xa{{x\ra a}}
\newcommand\ya{{y\ra a}}
\newcommand\ax{{a\ra x}}

\newcommand\xb{{x\ra b}}
\newcommand\yb{{y\ra b}}
\newcommand\bx{{b\ra x}}

\newcommand\zn{{2/\tau\el}}
\newcommand\nb{{|N(b)|}}
\newcommand\na{{|N(a)|}}

\newcommand\de{{\varDelta}}
\newcommand\sq{{E}}
\newcommand\les{{\leq1}}
\newcommand\vphi{{\vec{\varPhi}}}

\newcommand\Lem{Lemma}
\newcommand\Prop{Proposition}
\newcommand\Thm{Theorem}

\usepackage{courier}
\newcommand\algstyle{\small\sffamily}
\newcommand{\algn}[1]{\textnormal{\texttt{#1}}}
\newcommand{\spd}{\algn{SPdec}}
\newcommand{\pspd}{\algn{PermSPdec}}

\begin{document}
\title{Analysing Survey Propagation Guided Decimation on Random Formulas}

\author[]{Samuel Hetterich}
\thanks{$^\star$ The research leading to these results has received funding from the European Research Council under the European Union's Seventh Framework
			Programme (FP/2007-2013) / ERC Grant Agreement n.\ 278857--PTCC}

\address{Samuel Hetterich, {\tt hetteric@math.uni-frankfurt.de}, Goethe University, Mathematics Institute, 10 Robert Mayer St, Frankfurt 60325, Germany.}
\maketitle

\begin{abstract}
Let $\vphi$ be a uniformly distributed random $k$-SAT formula with $n$ variables and $m$ clauses. For clauses/variables ratio $m/n \leq r_{k\text{-SAT}} \sim 2^k\ln2$ the formula $\vphi$ is satisfiable with high probability. However, no efficient algorithm is known to provably find a satisfying assignment beyond $m/n \sim 2k \ln(k)/k$  with a non-vanishing probability.
Non-rigorous statistical mechanics work on $k$-CNF led to the development of a new efficient ``message passing algorithm'' called \emph{Survey Propagation Guided Decimation} [M\'ezard et al., Science 2002]. Experiments conducted for $k=3,4,5$ suggest that the algorithm finds satisfying assignments close to $r_{k\text{-SAT}}$. However, in the present paper we prove that the basic version of Survey Propagation Guided Decimation fails to solve random $k$-SAT formulas efficiently already for $m/n=2^k(1+\varepsilon_k)\ln(k)/k$ with $\lim_{k\to\infty}\varepsilon_k= 0$ almost a factor $k$ below $r_{k\text{-SAT}}$.  
 \end{abstract}

\section{Introduction}


Random $k$-SAT instances have been known as challenging benchmarks for decades~\cite{Cheeseman, MitchellSelmanLevesque, Pearl}.
The simplest and most intensely studied model goes as follows.
Let $k\geq3$ be an integer, fix a density parameter $r>0$, let $n$ be a (large) integer and let $m=\lceil rn\rceil$.
Then $\varPhi = \varPhi_k(n,m)$ signifies a $k$-CNF chosen uniformly at random among all $(2n)^{km}$ possible formulas.
With $k,r$ fixed the random formula is said to enjoy a property {\em with high probability} if the probability that the property holds tends to $1$ as $n\to\infty$.

The conventional wisdom about random $k$-SAT has been that the problem of finding a satisfying assignment is computationally most challenging for $r$ below but close to the {\em satisfiability threshold} $\rk$ where the random formula ceases to be satisfiable \whp~\cite{MitchellSelmanLevesque}.
Whilst the case $k=3$ may be the most accessible from a practical (or experimental) viewpoint, the picture becomes both clearer and more dramatic for larger values of $k$.
Asymptotically the $k$-SAT threshold reads $\rk=2^k\ln2-(1+\ln 2)/2+\eps_k$, where $\eps_k\to0$ in the limit of large $k$ \cite{DSS}.
However, the best current algorithms are known to find satisfying assignments in polynomial time merely up to $r\sim 2^k\ln k/k$~\cite{BetterAlg}.
In fact, standard heuristics such as Unit Clause Propagation bite the dust for even smaller densities, namely $r=c2^k/k$ for a certain
absolute constant $c>0$~\cite{FrSu}.
The same goes (provably) for various DPLL-based solvers.
Hence, there is a factor of about $k/\ln k$ between the algorithmic threshold and the actual satisfiability threshold.

In the early 2000s physicists put forward a sophisticated but non-rigorous approach called the
{\em cavity method} to tackle problems such as random $k$-SAT both analytically and algorithmically.
In particular, the cavity method yields a {\em precise} prediction as to the value of $\rk$ for any $k\geq3$~\cite{Mertens, MM}, which was recently
verified rigorously for sufficiently large values of $k$~\cite{DSS}.
Additionally, the cavity method provided a heuristic explanation for the demise of simple combinatorial or DPLL-based algorithms well below $\rk$.
Specifically, the density $2^k\ln k/k$ marks the point where the geometry of the set of satisfying assignments changes from (essentially)
a single connected component to a collection of tiny well-separated clusters~\cite{pnas}.
In fact, a typical satisfying assignment belongs to a ``frozen'' cluster, i.e., there are extensive long-range correlations between the variables. The cluster decomposition as well as the freezing prediction have largely been verified rigorously~\cite{Molloy} and
we begin to understand the impact of this picture on the performance of algorithms~\cite{Barriers}.

But perhaps most remarkably, the physics work has led to the development of a new efficient ``message passing algorithm'' called {\em Survey Propagation Guided Decimation}
to overcome this barrier~\cite{BMZ,Kroc,MPZ,Allerton}. More precisely, the algorithm is based on a heuristic that is designed to find whole frozen clusters not only single satisfying assignments by identifying each cluster by the variables determined by long-range correlations and locally ``free'' variables. Thus, by its very design {\em Survey Propagation Guided Decimation} is build to work at densities where frozen clusters exist. Although the experimental performance for small $k$ is outstanding this yields no evidence of a relation between the occurrence of frozen clusters and the success of the algorithm. 
Yet not even the physics methods lead to a precise explanation of these empirical results or to a prediction as to the density up to which
we might expect SP to succeed for general values of $k$.
In effect, analysing SP has become one of the most important challenges in the context of random constraint satisfaction problems.

The present paper furnishes the first rigorous analysis of $\spd$ (the basic version of) Survey Propagation Guided Decimation for random $k$-SAT. We give a precise definition and detailed explanation below. Before we state the result let us point out that two levels of randomness are involved:~the choice of the random formula $\vec\varPhi$, and the ``coin tosses'' of the randomized algorithm \algn{SPdec}. For a (fixed, non-random) $k$-CNF $\varPhi$ let $\suc(\varPhi)$ denote the probability that \algn{SPdec}$(\varPhi)$ outputs a satisfying assignment. Here, of course, ``probability'' refers to the coin tosses of the algorithm only. Then, if we apply \algn{SPdec} to the {\it random} $k$-CNF $\vec\varPhi$, the success probability $\suc(\vec{\varPhi})$ becomes a random variable. Recall that $\vec\varPhi$ is unsatisfiable for $r >2^k\ln 2$ \whp.

\begin{theorem} \label{theo_1}
 There is a sequence $(\varepsilon_k)_{k\geq 3}$ with $\lim_{k\ra \infty} \varepsilon_k = 0$ such that for any $k,r$ satisfying $	2^k(1+\varepsilon_k)\ln(k) /k \leq r \leq 2^k\ln 2$
we have $\suc(\vec\varPhi) \leq \exp(-\Omega(n))$ \whp
\end{theorem}

If the success probability is exponential small in $n$ sequentially running $\spd$ a sub-exponential number of times will not find a satisfying assignment \whp\ rejecting the hypotheses that $\spd$ solves random $k$-SAT formulas efficiently for considered clauses/variables ratio. 
Thus, \Thm~\ref{theo_1} shows that \spd~does not outclass far simpler combinatorial algorithms for general values of $k$.
Even worse, in spite of being designed for this very purpose,
the SP algorithm does {\em not} overcome the barrier where the set of satisfying assignments decomposes into tiny clusters asymptotically. This is even more astonishing since it is possible to {\em prove} the existence of satisfying assignments up to the satisfiability threshold rigorously based on the cavity method but algorithms designed by insights of this approach fail far below that threshold.  

We are going to describe the SP algorithm in the following section.
Let us stress that \Thm~\ref{theo_1} pertains to the ``vanilla'' version of the algorithm.
Unsurprisingly, more sophisticated variants with better empirical performance have been suggested, even ones that involve backtracking~\cite{Marino}. However, the basic version of the SP algorithm analysed in the present paper arguably encompasses all the conceptually important features of the SP algorithm. 

The only prior rigorous result on the Survey Propagation algorithm is the work of Gamarnik and Sudan~\cite{Gamarnik2}
on the $k$-NAESAT problem (where the goal is to find a satisfying assignment whose binary inverse is satisfying as well). 
However, Gamarnik and Sudan study a ``truncated'' variant of the algorithm where only a bounded number of message passing iterations is performed. The main result of~\cite{Gamarnik2} shows that this version of Survey Propagation fails for densities about a factor of $k/\ln^2k$ below the NAE-satisfiability threshold
and about a factor of $\ln k$ above the density where the set of NAE-satisfying assignments shatters into tiny clusters.
Though, experimental data and the conceptional design of the SP algorithm suggest that it exploits its strength in particular by iterating the message passing iterations a unbounded number of times that depends on $n$. In particular, to gather information from the set of messages they have to converge to a fixed point which turns out to  happen only after a number of iterations of order $\ln(n)$. 

An in-depth introduction to the cavity method and its impact on combinatorics, information theory and computer science can be found in~\cite{MM}.

\section{The \spd \ algorithm}

The proof of \Thm~\ref{theo_1} is by extension of the prior analysis~\cite{BP} of the much simpler {\em Belief Propagation Guided Decimation} algorithm. To outline the proof strategy and to explain the key differences, we need to discuss the SP algorithm in detail.
For a $k$-CNF $\varPhi$ on the variables $V= \{x_1,\ldots,x_n\}$ we generally represent truth assignments as maps $\sigma:V\rightarrow \{-1,1\}$, with $-1$ representing ``false'' and $1$ representing ``true''. Survey Propagation is an efficient message passing heuristic on the factor graph $G(\varPhi)$. Before explaining the Survey Propagation heuristic, we explain the simpler Belief Propagation heuristic and emphasize the main extensions later on. To define the messages involved we denote the ordered pair $(x,a)$ with $x\ra a$ and similarly $(a,x)$ with $a\ra x$ for each $x\in V$ and $a\in N(x)$. The messages are iteratively sent probability distributions $\br{\mu_{x\ra a}(\zeta)}_{x\in V_t, a\in N(x), \zeta \in \{-1,1\}}$ over $\{-1,1\}$. In each iteration messages are sent from variables to adjacent clauses and back. After setting initial messages due to some initialization rule the messages send are obtained by applying a function to the set of incoming messages at each vertex. Both, the initialization and the particular update rules at the vertices are specifying the message passing algorithm. The messages are updated $\omega(n)$ times which may or may not depend on $n$. A detailed explanation of the Belief Propagation heuristic can be found in~\cite[p.~519]{BraZec}.

It is well known that the Belief Propagation messages on a tree converge after updating the messages two times the depth of the tree to a fixed point. Moreover, in this case for each variable the marginal distribution of the uniform distribution on the set of all satisfying assignments can be computed by the set of the fixed point messages. Since $G(\varPhi)$ for constant clauses/variables ratio contains only a small number of short cycles one may expect that on the base of the Belief Propagation messages a good estimate of the marginal distribution of the uniform distribution on the set of all satisfying assignments of $\varPhi$ could be obtained. Besides the fact that it is not even clear that the messages converge to a fixed point on arbitrary graphs this is of course only a weak heuristic explanation which is refuted by \cite{BP}. However, at each decimation step using the Belief Propagation heuristic the Belief Propagation guided decimation algorithm assigns one variable due to the estimated marginal distribution to $-1$ or $1$. Simplifying the formula and running Belief Propagation on the simplified formula and repeating this procedure would lead to a satisfying assignment chosen uniformly at random for sure if the marginals were correct at each decimation step. 

Let us now introduce the Survey Propagation heuristic. As mentioned above the geometry of the set of satisfying assignments comes as a collection of tiny well-separated clusters above density $2^k\ln(k)/k$. In that regime a typical solution belongs to a ``frozen'' cluster. That is all satisfying assignments in such a frozen cluster agree on a linear number of frozen variables. Flipping one of these variables leads to a set of unsatisfied clauses only containing additional frozen variables. Satisfying one of these clauses leads to further unsatisfied clauses of this kind ending up in an avalanche of necessary flippings to obtain a satisfying assignment. This ends only after a linear number of flippings. Thus, identifying these frozen variables gives a characterization of the whole cluster. Given a satisfying assignment with identified frozen variables each satisfying assignment that disagrees on one of these frozen variables has linear distance therefore belonging to a different cluster. 

\begin{figure}
For real numbers $0\leq x, y\leq 1$ such that $\max\{x,y\}> 0$ we define 
\begin{align*}
	\psi_\zeta(x,y) =  \begin{cases} 
			xy\cdot\Psi(x,y) & \text{if } \zeta = 0\\		  
			(1-x)y\cdot\Psi(x, y) & \text{if } \zeta =1\\
			(1-y)x\cdot\Psi(x, y) & \text{if } \zeta =-1\\
	\end{cases}	, \qquad\Psi(x,y) = (x+y-xy)^{-1}
\end{align*}
If $x=y=0$ set $\psi_0(0)=0$ and $\psi_{\pm1}(0)=\frac12$.
Define for all $x \in V_t, a,b\in N(x), \zeta\in \{-1,0,1\}$ and $\ell\geq 0$
\begin{align}
	\mu_{x\ra a}^{[0]}(\pm1)&=\frac12 ,\qquad \mu_{x\ra a}^{[0]}(0)=0,\qquad
	\mu_{b\ra x}^{[\ell]}(0)= 1-\prod_{y\in N(b)\setminus \{x\}}\mu_{y \ra b}^{[\ell]}(-\sign(y,b))\label{update_b_to_x}\\
	\pi^\elp_\xa(\pm1)&= \prod_{b\in N(x,\pm1)\setminus\{a\}} \mu_\bx^\el(0)\label{update_pi_x_to_a}\\
	\mu_{x\ra a}^{[\ell+1]}(\zeta) &= (SP(\mu^{[\ell]}))_{x\ra a}(\zeta)=\psi_\zeta(\pi_\xa^\el(1),\pi_\xa^\el(-1)).\label{update_x_to_a}
\end{align}
Let $\omega = \omega(k,r,n) \geq 0$ be any integer-valued function. Define
\begin{align}
	\pi_x^{[\omega+1]}(\varPhi_t,\pm1) &=\prod_{b\in N(x,\pm1)}\mu_{\bx}^{[\omega]}(0)\\
	\mu_x^{[\omega]}(\varPhi_t,\zeta) &= \psi_\zeta(\pi_x^{[\omega+1]}(\varPhi_t,1)\cdot\pi_x^{[\omega+1]}(\varPhi_t,-1))\label{def_pi_omegap}\\
	\mu_x^{[\omega]}(\varPhi_t)  &= \frac{\mu_x^{[\omega]}(\varPhi_t,1)}{\mu_x^{[\omega]}(\varPhi_t,1) + \mu_x^{[\omega]}(\varPhi_t,-1)} = \mu_x^{[\omega]}(\varPhi_t,1) + \frac12 \mu_x^{[\omega]}(\varPhi_t,0).\label{def_marg_4}
\end{align} 
\caption{The Survey Propagation equations that are the Belief Propagation equations on covers.}\label{Fig_sp}
\end{figure}

This picture inspires the definition of \emph{covers} as generalized assignments $\sigma \in \{-1,0,1\}^n$ such that 
\begin{itemize}
\item each clause either contains a true literal or two $0$ literals and
\item for each variable $x\in V$ that is assigned $-1$ or $1$ exists a clause $a\in N(x)$ such that for all $y\in N(a)\setminus \{x \}$ we have $\sign(y,a)\cdot\sigma(y) = -1$.
\end{itemize}  
These two properties mirrors the situation in frozen clusters where assigning a variable to the value $0$ indicates that these variable supposes to be free in the corresponding cluster which is obtained by only flipping $0$ variables to one of the values $-1$ or $1$. However, Implementing the concept of covers, Survey Propagation is a heuristic of computing the marginals over the set of covers by using the Belief Propagation update rules on covers. This leads to the equations given by Figure \ref{Fig_sp}. For a more detailed explanation of the freezing phenomenon we point the reader to \cite{Molloy}. For a deeper discussion on covers we refer to \cite{Covers}.

We are now ready to state the \spd algorithm. 


\begin{algorithm}
\spd$(\varPhi)$ \\
Input: \emph{A $k$-CNF $\varPhi$ on $V=\{x_1,\ldots,x_n\}$.} Output: \emph{An assignment $\sigma: V \ra \{-1,1\}$.}\\
\emph{\algstyle{
0.\qquad Let $\varPhi_0 = \varPhi$. \\
1.\qquad For $t=0,\ldots,n-1$ do \\
2.\qquad\qquad Use SP to compute $\mu_{x_{t+1}}^{[\omega]}(\varPhi_t)$. \\
3.\qquad\qquad Assign 
\begin{eqnarray}
	\sigma(x_{t+1}) = \begin{cases}
	1 &\text{ with probability } \mu_{x_{t+1}}^{[\omega]}(\varPhi_t) \\
	-1&\text{ with probability } 1-\mu_{x_{t+1}}^{[\omega]}(\varPhi_t).
	\end{cases}
\end{eqnarray} 
4.\qquad\qquad Obtain a formula $\varPhi_{t+1}$ from $\varPhi_t$ by substituting the value $\sigma(x_{t+1})$ for $x_{t+1}$ and simplifying.  \\
5.\qquad Return the assignment $\sigma$. 
}}
\end{algorithm}

Let us emphasize that the value $\mu_{x_{t+1}}^{[\omega]}(\varPhi_t)$ in Step 2 of \spd~is the estimated marginal probability over the set of covers of variable $x_{t+1}$ in the simplified formula to take the value $1$ plus one half the estimated marginal probability over the set of covers in the simplified formula to take the value $0$. This makes sense since by the heuristic explanation a variable assigned to the value $0$ is free to take either value $1$ or $-1$. 
Thus, our task is to study the $SP$ operator on the decimated formula $\varPhi_t$.

\section{Proof of Theorem~\ref{theo_1}}
The probabilistic framework used in our analysis of $\spd$ was introduced in \cite{BP} for analysing the \emph{Belief Propagation Guided Decimation} algorithm. The most important technique in analysing algorithms on the random formula $\vphi$ is the ''method of deferred decisions'', which traces the dynamics of an algorithm by differential equations, martingales, or Markov chains. It actually applies to algorithms that decide upon the value of a variable $x$ on the basis of the clauses or variables at small bounded distance from $x$ in the factor graph \cite{AchSor}. Unfortunately, the \spd~algorithm at step $t$ explores clauses at distance $2\omega$ from $x_t$ where $\omega=\omega(n)$ may tend to infinity with $n$. Therefore, the ``defered decisions'' approach does not apply and to prove Proposition~\ref{theo_1} a fundamentally different approach is needed. 

We will basically reduce the analysis of \spd~to the problem of analysing the SP operator on the random formula $\vphi^t$ that is obtained from $\vphi$ by substituting ``true'' for the first $t$ variables $x_1,\ldots,x_t$ and simplifying (see Theorem \ref{theo_balanced} below). In the following sections we will prove that this decimated formula has a number of simple to verify quasirandomness properties with very high probability. Finally, we will show that it is possible to trace the Survey Propagation algorithm on a formula $\varPhi$ enjoying this properties.

Applied to a fix, non-random formula $\varPhi$ on $V=\{x_1,\ldots,x_n\}$, \spd \ yields an assignment $\sigma: V\ra\{-1,1\}$ that may or may not be satisfying. This assignment is random, because \spd \ itself is randomized. Hence, for any fixed $\varPhi$ running \spd$(\varPhi)$ induces a probability distribution $\beta_\varPhi$ on $\{-1,1\}^V$. With $\cS(\varPhi)$ the set of all satisfying assignments of $\varPhi$, the ``success probability'' of \spd \ on $\varPhi$ is just 
\begin{equation}
\suc(\varPhi) = \beta_\varPhi(\cS(\varPhi)).
\end{equation}
Thus, to establish Theorem \ref{theo_1} we need to show that in the \emph{random} formula, 
\begin{equation}
	\suc(\vphi) = \beta_\vphi(s(\vphi)) = \exp\br{-\Omega(n)}
\end{equation}
is exponentially small \whp \ To this end, we are going to prove that the measure $\beta_\vphi$ is ``rather close'' to the uniform distribution on $\{-1,1\}^V$ \whp, of which $\cS(\vphi)$ constitutes only an exponentially small fraction. 
However, to prove Theorem \ref{theo_1} we prove that the entropy of the distribution $\beta_\vphi$ is large. Let us stress that this is not by Mosers entropy compression argument which works up to far smaller clauses/variables ratios.

\subsection{Lower bounding the entropy}

{\it Throughout the paper we let $\rho_k = (1+\varepsilon_k)\ln(k)$ where  $(\varepsilon_k)_{k\geq 3}$ is the sequence promised by Theorem \ref{theo_1} and let $r$ be such that $\rho_k\leq\rho= kr/2^k$.}

For a number $\delta > 0$ and an index $l > t$ we say that $x_{l}$ is \emph{$(\delta,t)$-biased} if 
\begin{eqnarray}
\left|\mu_{x_{l}}^{[\omega]}(\varPhi^t,1) - \frac12 \br{1-\mu_{x_{l}}^{[\omega]}(\varPhi^t,0)}\right| >\delta.
\end{eqnarray}
Moreover $\varPhi$ is \emph{$(\delta, t)$-balanced} if no more than $\delta(n-t)$ variables are $(\delta,t)$-biased.

If $\vphi$ is $(\delta,t)$-balanced, then by the basic symmetry properties of $\vphi$ the probability that $x_{t+1}$ is $(\delta, t)$-biased is bounded by $\delta$. Furthermore, given that $x_{t+1}$ is not $(\delta,t)$-biased, the probability that \spd \ will set it to ``true'' lies in the interval $[\frac12-\delta,\frac12+\delta]$. Consequently,
\beq \label{equ_12}
\left|\frac12-\pr\brk{\sigma(x_{t+1})=1 | \vphi\text{ is } (\delta,t)\text{-balanced}}\right| \leq 2\delta.
\eeq
Thus, the smaller $\delta$ the closer $\sigma(x_{t+1})$ comes to being uniformly distributed. Hence, if $(\delta,t)$-balancedness holds for all $t$ with a ``small'' $\delta$, then $\beta_\varPhi$ will be close to the uniform distribution on $\{-1,1\}^V$.

To put this observation to work, let $\theta = 1-t/n$ be the fraction of unassigned variables and define
	\begin{eqnarray} \label{equ_def_delta}
	\delta_t=\exp(-c\theta k), \qquad\Delta_t=\sum_{s=1}^t \delta_t\qquad\text{and}\qquad\hat{t} = \br{1- \frac{\ln(\rho)}{c^2 k}}n,
	\end{eqnarray}
where $c>0$ is a small enough absolute constant.

The following result provides the key estimate by providing that at any time $t$ up to $\hat t$ with sufficiently high probability $\vphi$ is $(\delta_t,t)$-balanced  with a sufficiently small $\delta_t$ to finally prove Theorem \ref{theo_1}.

\begin{proposition}\label{theo_balanced}
For any $k,r$ satisfying $2^k\rho_k/k < r\leq 2k\ln 2$ there is $\xi = \xi(k,r) \in [0,\frac1k]$ so that for $n$ large enough the following holds. 
For any $0 \leq t\leq \hat{t}$ we have
\begin{eqnarray}
\Pr\brk{\vphi \text{ is } (\delta_t,t)\text{-balanced}}\geq 1-\exp\brk{-3\xi n-10 \varDelta_t}.
\end{eqnarray}
\end{proposition}

\subsection{Tracing the Survey Propagation Operator}
\label{sec_tracing_sp}

To establish Proposition \ref{theo_balanced} we have to prove that $\vphi$ is $(\delta_t,t)$-balanced with probability very close to one. Thus, our task is to study the SP operator defined in (\ref{update_b_to_x}) to (\ref{update_x_to_a}) on $\vphi^t$.
Roughly speaking, Proposition \ref{theo_balanced} asserts that with probability very close to one, most of the messages $\mu_\xa^\el(\pm1)$ are close to $\frac12(1-\mu_\xa^\el(0))$. To obtain this bound, we are going to proceed in two steps: we will exhibit a small number \emph{quasirandomness properties} and show that these hold in $\vphi^t$ with the required probability. Then, we are prove that \emph{deterministically} any formula that has these properties is $(\delta_t,t)$-balanced.

\subsubsection{The ``typical'' value of $\pi_\xa^\el(\zeta)$}
\label{sec_typical_pi}

First of all recall that the messages send from a variable $x$ to a clause $a\in N(x)$ are obtained by 
\begin{align}
\psi_\zeta(\pi_\xa^\el(1),\pi_\xa^\el(-1)) \qquad\text{for } \zeta \in \{-1,0,1\}.
\end{align}
This in mind, we claim a strong statement that both $\pi_\xa^\el(1)$ and $\pi_\xa^\el(-1)$ are very close to a ``typical'' value $\pi\el$ for most of the variables $x\in V_t$ and clauses $a\in N(x)$ at any iteration step $\ell$ under the assumption that the set of biased variables is small at time $\ell-1$. Assuming that
$$\pi_\xa^\el(1)=\pi_\xa^\el(-1) = \pi\el$$ 
we of course obtain unbiased messages by 
$$\mu_\xa^\el(\pm1)= \psi_{1}(\pi\el) = \psi_{-1}(\pi\el) = \frac12(1-\mu_\xa^\el(0)).$$

The products $\pi_\xa^\el(\zeta)$ are nothing else but the product of the messages $$\mu_\bx^\elm(0)=1-\prod_{y\in N(b)\setminus\{x\}} \mu_\yb^{\elm}(-\sign(y,b))$$ send from all clauses $b\in N(x,\zeta)\setminus \{a\}$ to $x$. Therefore, we define inductively $0\leq\pi\el\leq 1$ to be the product of this kind over a ``typical'' neighborhood. The term ``typical'' refers to the expected number of clauses of all lengths that contain at most one additional biased variable. Focusing on those clauses will suffice to get the tightness result of the biases. Moreover, we assume that all of the messages $\mu_\yb^\elm(-\sign(y,b))$ send from variables to clauses in such a typical neighborhood are $\psi_{\sign(y,b)}(\pi\elm,\pi\elm)$ which is claimed to be a good estimation of most of the messages send at time $\ell-1$. Additionally, define $\tau\el = (1-\psi_0(\pi\el))$ as the estimate of the sum $\mu_\xa^\el(1)+ \mu_\xa^\el(-1)$. Let us emphasize that there is no ``unique'' $\pi\el$ and the way it is obtained in the following is in some sense the canonical and convenient choice to sufficiently bound the biases for most of the messages.

Generally, let $T\subset V_t$ and $x\in V_t$. Then the expected number of clauses of length $j$ that contain $x$ and at most one other variable from the set $T$ is asymptotically  
\begin{eqnarray}\label{equ_exp_number_clauses_length_j}
	\mu_{j,\les}(T) = 2^j\rho\cdot \Pr\brk{\Bin( k-1,\theta)=j-1}\cdot \Pr\brk{\Bin\br{j-1,\frac{|T|}{\theta n}}<2}.
\end{eqnarray}
Indeed, the expected number of clauses of $\vphi$ that $x$ appears in equals $km/n=kr=2^k\rho$. Furthermore, each of these gives rise to a clause of length $j$ in $\vphi^t$ iff exactly $j-1$ among the other $k-1$ variables in the clauses are from $V_t$ while the $k-j$ remaining variables are in $V\setminus V_t$ and occur with negative signs. (If one of them had a positive sign, the clause would have been satisfied by setting the corresponding variable to true. It would thus not be present in $\vphi^t$ anymore.) 
Moreover, at most one of the $j-1$ remaining variables is allowed to be from the set $T$. The fraction of variables in $T$ in $V_t$ equals $\frac{|T|}{\theta n}$.
Finally, since $x$ appears with a random sign in each of these clauses the expected number of clauses of length $j$ that contain $x$ and at most one other variable from the set $T$ is asymptotically $\mu_{j,\les}(t)/2$.  

Additionally let $0\leq p \leq 1$ and define 
\begin{eqnarray}
	\tau(p)=1-\psi_0\br{p} \label{def_nu}\qquad \text{and} \qquad
	\pi(T,p) = \prod_{j=0.1\theta k}^{10\theta k}\br{1-\br{2/\tau(p)}^{-j+1}}^{\mu_{j,\les}(T)/2} \label{def_mu}.
\end{eqnarray}

Moreover, let
\begin{eqnarray*}
	\Pi(T,p) = \sum_{j= 0.1\theta k}^{10\theta k} \frac{\mu_{j,\les}(T)}{2}\cdot\br{2/\tau(p)}^{-j+1}
\end{eqnarray*}
be the approximated absolute value of the logarithm of $\pi(T,p)$.

For a fixed variable $x\in V_t$ the expected number of clauses that contain more than one additional variable from a ``small'' set $T$ for a ``typical'' clause length $0.1\theta k\leq j\leq 10\theta k$ is very close to the expected number of all clauses of that given length. Thus, the actual size of $T$ will influence $\pi(T,p)$ but this impact is small if $T$ is small and the following bounds on $\pi(T,p)$ can be achieved.

\begin{lemma}\label{lem_bound_mu_l}
Let $T\subset V_t$ of size $|T|\leq \delta\theta n$ and $0\leq p\leq2\exp(-\rho)$. Then $\exp\br{-2\rho}\leq\pi(T,p)\leq 2\exp\br{-\rho}$.
\end{lemma}

\subsubsection{Bias}

First of all let us define the bias not only for the $1$ and $-1$ messages but also for the $0$ messages. Hence, for $\ell\geq 0, x\in V_t$ and $a\in N(x)$ let
\begin{eqnarray}
	\varDelta_{x \ra a}^{[\ell]} &=& \mu_{x \ra a}^{[\ell]}(1) - \frac12 \br{1-\mu_{x\ra a}^{[\ell]}(0)} \qquad\qquad\text{and} \label{equ_def_delta_xa}\\
	E_{x \ra a}^{[\ell]} &=& \frac12 \br{ \mu_{x \ra a}^{[\ell]}(0) - \psi_0(\pi\el)}.\label{equ_def_square_xa}
\end{eqnarray}
We say that $x \in V_t$ is \emph{$\ell$-biased} if 
\begin{align}
	\max_{a\in N(x)}|\varDelta_{x\ra a}^{[\ell]}| > 0.1\delta  \qquad\text{or}\qquad\max_{a\in N(x)}|E_{x\ra a}^{[\ell]}| > 0.1\delta \pi\el
\end{align}
and \emph{$\ell$-weighted} if 
\begin{align}
	\max_{a\in N(x)}|E_{x\ra a}^{[\ell]}| > 10 \pi\el.
\end{align}
Let $B[\ell]$ be the set of all $\ell$-biased variables and $B'[\ell]$ be the set of all $\ell$-weighted variables. 
Obviously, by definition, we have $B'\el \subset B'\el$.

Writing $\mu_\xa^\el(\sign(x,a))$ in terms of the biases we obtain 
\begin{eqnarray}
\mu_{\xa}^\el(\sign(x,a)) &=& \frac12 (1-\psi_0(\pi\el)) - \br{E_\xa^\el +\sign(x,a)\varDelta_\xa^\el} \nonumber\\
& =&\tau\el/2 - \br{E_\xa^\el +\sign(x,a)\varDelta_\xa^\el}\label{equ_def_mu_ya}
\end{eqnarray}
We are going to prove that $|\varDelta_{x\ra a}^{[\ell]}|$ and $|E_{x\ra a}^{[\ell]}|$ are small for most $x$ and $a\in N(x)$. That is, given the $\varDelta_{x\ra a}^{[\ell]}$ and $E_{x\ra a}^{[\ell]}$ we need to prove that the biases $\varDelta_{x\ra a}^\elp$ and $E_{x\ra a}^\elp$ do not 'blow up'. The proof is by induction where the hypothesis is that at most $\delta_t\theta n$ variables are $\ell$-biased and at most $\delta^2\theta n$ variables are $\ell$-weighted and our goal is to show that the same holds true for $\ell+1$. 

\subsubsection{The quasirandomness property}
\label{sec_tracing_sp_3}
We will now exhibit a few simple quasirandomness properties that $\vphi^t$ is very likely to possess. Based only on these graph properties we identify potentially $\ell$-biased or $\ell$-weighted variables. In turn, we prove that variables in the complement of these sets are surely not $\ell$-biased resp. $\ell$-weighted. Moreover, we show that these sets are small enough with sufficiently high probability. 

To state the quasirandomness properties, fix a $k$-CNF $\varPhi$. Let $\varPhi^t$ denote the CNF obtained from $\varPhi$ by substituting ``true'' for $x_1,\ldots,x_t$ and simplifying $(1\leq t\leq n)$. Let $V_t=\{x_{t+1},\ldots,x_n\}$ be the set of variables of $\varPhi^t$.
Let $\delta=\delta_t$. With $c>0$ we let $k_1 = \sqrt{c}\theta k$. 
For a variable $x \in V_t, \zeta \in \{1,-1\}$ and a set $T \subset V_t$ let 
 	\begin{eqnarray*} 
 	\cN(x,\zeta) &=& \left\{b\in N(x,\zeta): 0.1\theta k \leq |N(b)| \leq 10\theta k\right\}, \\
	\cN_{\leq 1}(x,T,\zeta) &=& \{b \in \cN(x,\zeta): |N(b) \cap T\setminus \{x\}| \leq 1 \}, \\
	\cN_{i}(x,T,\zeta) &=& \left\{b\in \cN(x,\zeta):|N(b) \cap T\setminus\{x\}|= i  \right\} \text{ for  } i\in\{0,1\},\\
	N_{1}(x,T,\zeta) &=& \{b \in N(x,\zeta): |N(b)\setminus T| \geq k_1 \wedge |N(b) \cap T\setminus \{x\}| = 1 \}, \\
	N_{>1}(x,T,\zeta) &=& \{b \in N(x,\zeta): |N(b)\setminus T| \geq k_1 \wedge |N(b) \cap T\setminus \{x\}| > 1 \}. 
	\end{eqnarray*}
Thus, $\cN_{\leq 1}(x,T,\zeta)$ is the set of all clauses $a$ that contain $x$ with $\sign(x,a)=\zeta$ (which may or may not be in $T$) and at most one other variable from $T$. In addition, there is a condition on the length $\nb$ of the clauses $b$ in the decimated formula $\varPhi^t$. Having assigned the first $t$ variables, we should ``expect'' the average clause length to be $\theta k$. The sets $\cN_i(x,T,\zeta)$ are a partition of $\cN_{\leq1}(x,T,\zeta)$ separating clauses that contain exactly one additional variable from $T\setminus\{x\}$ and clauses that contain none.  

\begin{itemize}
 \item[\textbf{Q1}] No more than $10\delta\theta n$ variables occur in clauses of length less than $\theta k/10$ or greater than $10\theta k$ in $\varPhi_t$. Moreover, there are at most $10^{-4}\delta \theta n$ variables $x \in V_T$ such that 
 	\begin{eqnarray*} 
	(\theta k)^3 \delta \cdot \sum_{b\in N(x,\zeta)} 2^{-|N(b)|}& >& 1. 
	\end{eqnarray*}
\item[\textbf{Q2}] For any set $T \subset V_t$ of size $|T| \leq s\theta n$ such that $\delta^5\leq s\leq 10\delta$ and any $p \in (0,1]$ there are at most $10^{-3}\delta^2 \theta n$ variables $x$ such that for one $\zeta \in \{-1,1\}$ either
	 \begin{eqnarray*}
	 \left|\Pi(T,p) - \sum_{b \in \cN_{\leq 1}(x, T,\zeta)} \br{2/\tau(p)}^{1-|N(b)|}\right| &> 2\delta/1000 &\qquad\text{or} \\
	 \sum_{b \in \cN_{1}(x, T,\zeta)} 2^{-|N(b)|} &> 10^4\rho \theta k s  &\qquad \text{or} \\
	 \sum_{b \in \cN_{\les}(x, T,\zeta)} 2^{-|N(b)|} &> 10^4\rho.
	 \end{eqnarray*}
 \item[\textbf{Q3}] If $T\subset V_t$ has size $|T| \leq \delta\theta n$, then there are no more than $10^{-4}\delta\theta n$ variables $x$ such that at least for one $\zeta \in \{-1,1\}$ 
 	\begin{eqnarray*} 
 	\sum_{b\in N_{>1}(x,T,\zeta)} 2^{|N(b)\cap T\setminus \{x\}|-|N(b)|} > \delta/(\theta k).
	\end{eqnarray*}
 \item[\textbf{Q4}] For any $0.01 \leq z \leq 1$ and any set $T\subset V_t$ of size $|T| \leq 100 \delta\theta n$ we have 
 	\begin{eqnarray*} 
	\sum_{b:|N(b)\cap T|\geq z|N(b)|}|N(b)|\leq \frac{1.01}{z}|T| + 10^{-4}\delta\theta n.
	\end{eqnarray*}
 \item[\textbf{Q5}] For any set $T \subset V_t$ of size $|T| \leq 10 \delta \theta n$, any $p \in (0,1]$ and any $\zeta\in\{-1,1\}$ the linear operator $\varLambda(T,\mu,\zeta): \mathbb{R}^{V_t} \ra  \mathbb{R}^{V_t}$,
 	\begin{eqnarray*} 
	\varGamma = (\varGamma_y)_{y\in V_t} \mapsto \left\{\sum_{b\in \cN_{\leq 1}(x,T,\zeta)} \sum_{y \in N(b)\setminus \{x\}} \br{2/\tau(p)}^{-|N(b)|}\sign(y,b)\varGamma_y\right\}
	\end{eqnarray*}
has norm $\parallel \varLambda(T,\mu,\zeta)\parallel_\square \leq \delta^4\theta n$.
\end{itemize}

\begin{definition}\label{def_quasi}
 Let $\delta >0$. We say that $\varPhi$ is \emph{$(\delta,t)$-quasirandom} if \emph{\textbf{Q0}-\textbf{Q5}} are satisfied. 
\end{definition}

Condition \textbf{Q0} simply bounds the number of redundant clauses and the number of variables of very high degree; it is well-known to hold for random $k$-CNFs \whp \ Apart from a bound on the number of very short/very long clauses, \textbf{Q1} provides a bound on the ``weight'' of clauses in which variables $x\in V_t$ typically occur, where the weight of a clause $b$ is $2^{-\nb}$. Moreover, \textbf{Q2} and \textbf{Q3} provide that there is no small set $T$ for which the total weight of the clauses touching that set is very big. In addition, \textbf{Q2} (essentially) requires that for most variables $x$ the weights of the clauses where $x$ occurs positively/negatively should approximately cancel. Further, \textbf{Q4} provides a bound on the lengths of clauses that contain many variables from a small set $T$. Finally, the most important condition is \textbf{Q5}, providing a bound on the cut norm of a signed, weighted matrix, representation of $\varPhi^t$.

\begin{proposition} \label{prop_quasirandom}
There is a sequence $(\varepsilon_k)_{k\geq 3}$ with $\lim_{k\ra \infty} \varepsilon_k = 0$ such that for any $k,r$ satisfying $2^k(1+\varepsilon_k)\ln(k) /k \leq r \leq 2^k\ln 2$ there is $\xi= \xi(k,r)\in [0,\frac1k]$ so that for $n$ large and $\delta_t, \hat{t}$ as in (\ref{equ_def_delta}) for any $1\leq t\leq \hat{t}$ we have 
	\begin{eqnarray*}
	P\brk{\varPhi \text{ is } (\delta_t,t)\text{-quasirandom}}\geq 1- \exp\br{-10(\xi n+ \varDelta_t)}
	\end{eqnarray*}
\end{proposition}

\begin{theorem}\label{theo_25}
There is a sequence $(\varepsilon_k)_{k\geq 3}$ with $\lim_{k\ra \infty} \varepsilon_k = 0$ such that for any $k,r$ satisfying $2^k(1+\varepsilon_k)\ln(k) /k \leq r \leq 2^k\ln 2$ and $n$ sufficiently large the following is true. 
 \begin{equation*}\label{prop_s_less_Lower}
 \parbox{13cm}{
  Let $\varPhi$ be a $k$-CNF with $n$ variables and $m$ clauses that is $(\delta_t,t)$-quasirandom for some $1\leq t\leq \hat{t}$. Then $\varPhi$ is $(\delta_t,t)$- balanced.}
 \end{equation*}
\end{theorem}

The proof of Proposition \ref{prop_quasirandom} is a necessary evil: it is long, complicated and based on standard arguments. Theorem \ref{theo_25} together with Proposition \ref{prop_quasirandom} yields Proposition \ref{theo_balanced}

\subsubsection{Setting up the induction}

To prove Theorem \ref{theo_25} we succeed by induction over $\ell$. In particular we define sets $T\el$ and $T'\el$ that contain variables that are potentially $\ell$-biased or $\ell$-weighted only depending on the graph structure and the size of the sets $T[\ell-1]$ and $T'[\ell-1]$. The exact definition of the sets $T\el$ and $T'\el$ can be found in the Appendix~\ref{sec_def_of_T} and actually it will turn out that $T\el\subset B\ell$ and $T'\el\subset B'\ell$. Since we are going to trace the SP operator on $\varPhi^t$ iterated from the initial set of messages $\mu_{x\ra a}^{[0]}(\pm1)=\frac12$ and $\mu_{x\ra a}^{[0]}(0)=0$ for all $x \in V_t$ and $a\in N(x)$ we set $T\el=T'\el=\emptyset$ and $\pi[0]=0$ such that $\tau[0]=1$. Now we 
define inductively 
\begin{eqnarray*}
	\pi{[\ell +1]} = \pi\br{T\el,\pi\el},\quad \Pi\elp = \Pi\br{T[\ell],\pi\el}\quad \text{and}\quad \tau{\elp} = \tau\br{\pi\elp}.
\end{eqnarray*}
\begin{proposition}\label{prop_B_subset_T}
Assume that $\pi\el \leq 2\exp\br{-\rho}$. We have $B[\ell]\subset T[\ell]$ and $B'[\ell]\subset T'[\ell]$ for all $\ell \geq 0$. 
\end{proposition} 

Furthermore, we establish the following bounds on the size of $T[\ell]$ and $T'[\ell]$. Since the sets are defined by graph properties independent from the actual state of the algorithm the quasirandomness properties suffice to obtain 
\begin{proposition}\label{prop_bound_T}
If $\varPhi \text{ is } (\delta_t,t)\text{-quasirandom}$, we have $T[\ell]<\delta \theta n, T'[\ell]<\delta^2 \theta n$ and $\pi\el \leq 2\exp\br{-\rho}$ for all $\ell \geq 0$.
\end{proposition} 

Finally, let us give an idea how this is actually proved. We aim to prove that for most variables $x\in V_t$ for all $a\in N(x)$ simultaneously for both $\zeta \in \{-1,1\}$ the values $\pi^{[\ell]}_\xa(\zeta)$ are close to a typical value which is estimated by $\pi\el$ for each iteration. Let us define for $x\in V_t, a\in N(x)$ and $\zeta \in \{1,-1\}$ 
\begin{eqnarray*}
	P_{\leq 1}^\elp(x\ra a, \zeta) &=& \prod_{b \in \cN_{\leq 1}(x,T[\ell],\zeta)\setminus \{a\}} \mu_{b \ra x}^{[\ell]}(0) \\ 
	P_{> 1}^\elp(x\ra a, \zeta) &=& \prod_{b \in N(x,\zeta) \setminus (\{a\} \cup \cN_{\leq 1}(x,T[\ell],\zeta))} \mu_{b \ra x}^{[\ell]}(0).
\end{eqnarray*}
We obtain
\begin{eqnarray}\label{def_mu_el_xa_zeta}
	\pi^{[\ell]}_\xa (\zeta) &=& P_{\leq 1}^{[\ell]}(x\ra a, \zeta) \cdot P_{> 1}^{[\ell]}(x\ra a, \zeta).
\end{eqnarray}
We show that the first factor representing the product over messages send by clauses of typical length (regarding the decimation time $t$) and exposed to at most one additional variable from $T\el$ is close to $\pi{[\ell+1]}$ simultaneously for $\zeta\in \{-1,1\}$ for all variables $x\in V\setminus T'\elp$ and all $a\in N(x)$. Additionally, we prove that the second factor representing the product over messages send by clauses of atypical length or exposed to at least two additional variables from $T\el$ is close to one simultaneously for $\zeta\in \{-1,1\}$ for all variables $x\in V\setminus T\elp$ and all $a\in N(x)$.

\subparagraph*{Acknowledgements}

I thank my supervisor Amin Coja-Oghlan for supportive conversation and helpful comments on the final version of this paper.

\appendix

\section{Proofdetails}

\subsection{Preliminaries and notation}

In this section we collect a few well-known results and introduce a bit of notation. First of all, we note for later reference a well-known estimate of the expected number of satisfying assignments (see e.g \cite{AchPer} for a derivation).

\begin{lemma}\label{lem_8}
We have $\Erw\brk{\cS(\vphi)}=\Theta(2^n(1-2^{-k})^m)\leq 2^n\exp\br{-rn/2^k}$.
\end{lemma}

Furthermore we are going to need the following Chernoff bound on the tails of a binomially distributed random variable or, more generally, a sum of independent Bernoulli trials \cite[p.~21]{JLR}.

\begin{lemma}\label{Lemma_Chernoff}
Let $\varphi(x)=(1+x)\ln(1+x)-x$.
Let $X$ be a binomial random variable with mean $\mu>0$.
Then for any $t>0$ we have
	\begin{eqnarray*}
	\pr\brk{X>\mu+t}&\leq&\exp(-\mu\cdot\varphi(t/\mu)),\quad
	\pr\brk{X<\mu-t}\leq\exp(-\mu\cdot\varphi(-t/\mu)).
	\end{eqnarray*}
In particular, for any $t>1$ we have
	$\pr\brk{X>t\mu}\leq\exp\brk{-t\mu\ln(t/\eul)}.$
\end{lemma}

For a real $b\times a$ matrix $\Lambda$ let
\begin{eqnarray*}
	\|\Lambda\|_{\square} = \max_{\zeta \in \RR^{a}\setminus \{0\}} \frac{\|\Lambda\zeta\|_1}{\|\zeta\|_\infty}.
\end{eqnarray*}
Thus, $\|\Lambda\|_\square$ is the norm of $\Lambda$ viewed as an operator from $\RR^a$ equipped with the $L^\infty$-norm to $\RR^b$ endowed with the $L^1$-norm. For a set $A\subset [a]=\{1,\ldots ,a \}$ we let $\vecone_A \in \{0,1\}$ denote the indicator vector of $A$. the following well-known fact about the norm $\|\cdot\|_\square$ of matrices with diagonal entries equal to zero is going to come in handy.
\begin{fact}\label{fac_10}
For a real $b\times a$ matrix $\Lambda$ with zeros on the diagonal we have
\begin{eqnarray*}
	\|\Lambda\|_\square \leq 24 \max_{A\subset [a],B\subset [b]: A\cap B=\emptyset} |\langle\Lambda\vecone_A,\vecone_B\rangle|.
\end{eqnarray*} 
\end{fact} 
By definition we have
\begin{eqnarray}
	1 &=& 2\psi_{1}(x_1) + \psi_{0}(x_1) = 2\psi_{-1}(x_1) + \psi_{0}(x_1).\label{pro_psi_3}
\end{eqnarray}

\begin{lemma}\label{lem_bound_psi}
Let  $0<x_1, x_2,p_1,p_2,\eps_1,\eps_2 \leq 1$. Assume that $\left|x_1-p_1\right| \leq \eps_1$ and $ \left|x_2-p_2\right| \leq \eps_2$. Then
\begin{eqnarray}
	\left|\psi_0(x_1,x_2) - \psi_0(p_1, p_2)\right| \leq \eps_1 + \eps_2.
\end{eqnarray}
Suppose $\eps_1 \leq p_1/2$ and $\eps_2\leq p_2/2$. Then for $\zeta \in \{-1,1\}$ we have
\begin{eqnarray}
	\left|\psi_\zeta(x_1,x_2) - \psi_\zeta(p_1, p_2)\right| \leq 2\cdot\br{\frac{\eps_1}{p_1} + \frac{\eps_2}{p_2}}.
\end{eqnarray}
\end{lemma}
\begin{proof}
By the mean value theorem there exist $0<\xi_i^\zeta\leq 1$ such that for $i=1,2$ we have
\begin{eqnarray}
\left|p - \xi_i^\zeta\right|& \leq& \eps_i\qquad\text{and}\label{equ_lem_psi_1}\\
\psi_\zeta(x_1,x_2)& =& \psi_\zeta(p_1,p_2) + \sum_{i=1}^2(p_i-\xi_i^\zeta)\cdot \frac{\partial \psi_\zeta}{\partial x_i}(\xi_1^\zeta,\xi_2^\zeta).\label{equ_lem_psi_2}
\end{eqnarray}
Thus, we have to bound the first derivatives of the functions $\psi_\zeta$ which are given by
\begin{align*}
\frac{\partial \psi_0}{\partial x_1}
 &= x_2^2\cdot\Psi(x_1,x_2)^{-2}  &
\frac{\partial \psi_0}{\partial x_2}
 &= x_1^2\cdot\Psi(x_1,x_2)^{-2} \\ 
\frac{\partial \psi_1}{\partial x_1}
 &= -x_2\cdot\Psi(x_1,x_2)^{-2} & 
\frac{\partial \psi_1}{\partial x_2} 
&= x_1(1-x_1)\cdot\Psi(x_1,x_2)^{-2} \\
\frac{\partial \psi_{-1}}{\partial x_1} 
&= x_2(1-x_2)\cdot\Psi(x_1,x_2)^{-2}  & 
\frac{\partial \psi_{-1}}{\partial x_2}
 &= -x_1\cdot\Psi(x_1,x_2)^{-2}.
\end{align*}
For all $0<\xi_1, \xi_2\leq 1$ we have $\Psi(\xi_1,\xi_2)= \xi_1 + \xi_2 -\xi_1 \xi_2 \geq \xi_1, \xi_2$ and thus $\frac{\partial \psi_0}{\partial x_i}(\xi_1,\xi_2) \leq 1$. Together with (\ref{equ_lem_psi_1}) and (\ref{equ_lem_psi_2}) the first assertion follows.

For all $0<\xi_1,\xi_2\leq 1$ such that $|\xi_1-p_1| \leq \eps_1\leq p_1/2$ and $|\xi_2-p_2| \leq \eps_2\leq p_2/2$ we have 
\begin{eqnarray}
	\frac{\xi_1}{\br{\xi_1 + \xi_2 -\xi_1 \xi_2}^2} &\leq& \frac{\xi_1}{\br{\max\{\xi_1,\xi_2\}}^2} \leq \frac{1}{\max\{\xi_1,\xi_2\}}\leq \xi_2^{-1} \leq 2/p_2  \qquad\text{and}\\
	\frac{\xi_2}{\br{\xi_1 + \xi_2 -\xi_1 \xi_2}^2} &\leq& \frac{\xi_2}{\br{\max\{\xi_1,\xi_2\}}^2} \leq \frac{1}{\max\{\xi_1,\xi_2\}}\leq \xi_1^{-1} \leq 2/p_1 .
\end{eqnarray}
Thus, $\left|\frac{\partial \psi_\zeta}{\partial x_1}(\xi_1,\xi_2)\right| \leq 2/p_1 $ and $\left|\frac{\partial \psi_\zeta}{\partial x_2}(\xi_1,\xi_2)\right| \leq 2/p_2 $. Together with (\ref{equ_lem_psi_1}) and (\ref{equ_lem_psi_2}) the second assertion follows.
\end{proof}

\begin{lemma}\label{lem_mu_el_w_el}
Let $T \subset V_t$ and $0\leq p\leq 1$. We have
\begin{eqnarray*}
	\left|\Pi(T,p) + \ln \pi(T,p) \right| \leq \delta^4.
\end{eqnarray*}
\end{lemma}
\begin{proof}
Using the approximation $|\ln(1-z) + z| \leq z^2$ for $|z|\leq \frac12$ we obtain
\begin{eqnarray*}
	\left|\Pi(T,p) + \ln \pi(T,p)\right| &=& \left| \sum_{j= 0.1\theta k}^{10\theta k} \frac{\mu_{j,\les}(T)}{2}\cdot\br{2/\tau(p)}^{-j+1}\ \right.\\
	&&\left.\qquad\qquad\qquad+ \ln\br{\prod_{j=0.1\theta k}^{10\theta k}\br{1-\br{2/\tau(p)}^{-j+1}}^{\mu_{j,\les}(T)/2}} \right| \\ 
	&\leq& \sum_{j= 0.1\theta k}^{10\theta k} \frac{\mu_{j,\les}(T)}{2}\cdot \left| \br{2/\tau(p)}^{-j+1} + \ln\br{1-\br{2/\tau(p)}^{-j+1}} \right| \\
	&\leq& \sum_{j= 0.1\theta k}^{10\theta k} \frac{\mu_{j,\les}(T)}{2}\cdot \br{2/\tau(p)}^{-2j+2} \\
	&\leq&  \sum_{j= 0.1\theta k}^{10\theta k} 2^{-j+1}\rho \qquad\text{[by (\ref{equ_exp_number_clauses_length_j}) and as $0\leq\tau(p)\leq1$]} \\
	&\leq& 20\theta k  \rho2^{-0.1\theta k} \leq \delta^{-4}\qquad \text{[as $\theta k \geq \ln(\rho)/c^2$ and $c\ll 1$]}
\end{eqnarray*}
as claimed.
\end{proof}

\begin{proof}[Proof of Lemma~\ref{lem_bound_mu_l}] 
We start by establishing bounds on $\tau(p)$ as
\begin{eqnarray}\label{equ_bound_mu_4}
	1\geq\tau(p)= 1- \psi_0(p) = 1- \frac{p}{2 - p}\geq 1-p.
\end{eqnarray}
To get the lower bound we use the elementary inequality $\ln (1-z) \geq -2z$ for $z\in [0,0.5]$ and find
\begin{eqnarray*}
	\ln \pi(T,p) &=&  \sum_{j=0.1\theta k}^{10\theta k}\frac{\mu_{j,\leq 1}(T)}{2}\cdot \ln\br{1-\br{2/\tau(p)}^{1-j}}\geq -2\sum_{j=0.1\theta k}^{10\theta k}\frac{\mu_{j,\les}(T)}{2}\cdot\br{2/\tau(p)}^{-j+1}\\
	&=&-2\rho \sum_{j=0.1\theta k}^{10\theta k}\tau(p)^{j-1}\Pr\brk{\Bin\br{k-1,\theta}=j-1}\Pr\brk{\Bin\br{j-1,|T|/\theta n}<2} \\
	&&\qquad\qquad\qquad\qquad\qquad\qquad\qquad\qquad\qquad\qquad \qquad \brk{ \text{by (\ref{equ_exp_number_clauses_length_j})}} \\
	&\geq&-2\rho \qquad\qquad   \brk{ \text{by (\ref{equ_bound_mu_4})}} .
\end{eqnarray*}
To obtain the upper bound we apply Lemma \ref{Lemma_Chernoff} (the Chernoff bound) and get
\begin{eqnarray}
	\Pr\brk{0.1\theta k <\Bin(k-1,\theta)< 10\theta k} \geq 1- \exp\br{-\theta k/2}\label{equ_bound_mu_3}
\end{eqnarray}
and since $|T|/\theta n \leq \delta$ we have
\begin{eqnarray}
	\Pr\brk{\Bin\br{j-1,|T|/\theta n}<2}\geq \Pr\brk{\Bin\br{j-1,|T|/\theta n}=0}\geq (1-\delta)^{j-1}. \label{equ_bound_mu_2}
\end{eqnarray}
Therefore,
\begin{eqnarray}
	\ln \pi(T,p) &=&  \sum_{j=0.1\theta k}^{10\theta k}\frac{\mu_{j,\leq 1}(T)}{2}\cdot \ln\br{1-\br{2/\tau(p)}^{1-j}}\geq -\sum_{j=0.1\theta k}^{10\theta k}\frac{\mu_{j,\les}(T)}{2}\cdot\br{2/\tau(p)}^{-j+1}\nonumber\\
	&=& -\rho \sum_{j=0.1\theta k}^{10\theta k}\tau(p)^{j-1}\Pr\brk{\Bin\br{k-1,\theta}=j-1}\Pr\brk{\Bin\br{j-1,|T|/\theta n}<2} \nonumber\\
	&&\qquad\qquad\qquad\qquad\qquad\qquad\qquad\qquad\qquad\qquad\qquad \qquad \brk{ \text{by (\ref{def_mu})}} \nonumber \\
	&\leq& -\rho (1-\delta)^{10\theta k}(1-p)^{10\theta k}\sum_{j=0.1\theta k}^{10\theta k}\Pr\brk{\Bin\br{k-1,\theta}=j-1} \nonumber\\
	&&\qquad\qquad\qquad\qquad\qquad\qquad\qquad\qquad\qquad\qquad\qquad\qquad \brk{ \text{by (\ref{equ_bound_mu_4}) and (\ref{equ_bound_mu_2})}} \nonumber \\
	&\leq& -\rho (1-\delta)^{10\theta k}(1-p)^{10\theta k}(1-\exp\br{-\theta k/2})\qquad \brk{ \text{by (\ref{equ_bound_mu_3})}} \label{equ_bound_mu_1}.
\end{eqnarray}	
As $\delta,p, \exp\br{-\theta k/2} < 0.2$ due to the elementary inequality $1-z \geq \exp\br{-2z}$ for $z \in [0,0.2]$ and by (\ref{equ_bound_mu_1}) we obtain
\begin{eqnarray*}	
	\ln\pi(T,p)&\leq& -\rho \cdot\br{1-\br{20\delta\theta k+20p\theta k+2\exp\br{-\theta k/2}}} \\	
	&\leq&-\rho \cdot\br{1-\br{20\rho^{-1/c}\ln(\rho)/c^2+40\exp(-\rho)\ln(\rho)/c^2+2\rho^{-1/(2c^2)}}}\qquad\\
	&&\qquad\qquad\qquad\qquad\qquad\qquad\qquad\qquad\qquad\qquad \qquad\text{[as $\theta k\geq \ln(\rho)/c^2$]} \\
	&=&  -\rho + o_k\br{1} \leq -\rho + \ln2 \qquad\text{[as $c\ll 1$]},
\end{eqnarray*}
as desired.
\end{proof}

Finally, throughout the paper we let $S_n$ denote te set of permutations of $[n]$.

\subsection{Proof of Theorem \ref{theo_1} - details and computations}
\label{sec_proof_theo_1}

To facilitate the analysis, we are going to work with a slightly modified version of \spd. While the original \spd \ assigns the variables in the natural order $x_1,\ldots,x_n$, the modified version \pspd \ chooses a permutation $\pi$ of $[n]$ uniformly at random and assigns the variables in the order $x_{\pi(1)},\ldots, x_{\pi(n)}$. 

Let $\bar{\beta}_\varPhi$ denote the probability distribution induced on $\{-1,1\}^V$ by \pspd$(\varPhi)$. Because the uniform distribution over $k$-CNFs is invariant under permutations of the variables, we obtain
\begin{fact}\label{fact_11}
If $\bar{\beta}_\vphi(\cS(\vphi))\leq \exp\br{-\Omega(n)}$ \whp, then $\suc(\vphi)= \beta_\vphi(\cS(\vphi))\leq\exp\br{-\Omega(n)}$ \whp
\end{fact} 

Let $\varPhi$ be a $k$-CNF. Given a permutation $\pi$ and a partial assignment $\sigma: \{x_{\pi(s)}:s\leq t\} \ra \{-1,1\}$ we let $\varPhi_{t,\pi,\sigma}$ denote the formula obtained from $\varPhi$ by substituting the values $\sigma(x_{\pi(s)})$ for the variables $x_{\pi(s)}$ for $1\leq s\leq t$ and simplifying. Formally, $\varPhi_{t,\pi,\sigma}$ is obtained from $\varPhi$ as follows:
\begin{itemize}
\item remove all clauses $a$ of $\varPhi$ that contain a variable $x_{\pi(s)}$ with $1\leq s\leq t$ such that $\sigma(x_{\pi(s)}) = \sign(x_{\pi(s)},a)$.
\item for all clauses $a$ that contain a $x_{\pi(s)}$ with $1\leq s\leq t$ such that $\sigma(x_{\pi(s)})= \sign(x_{\pi(s)},a)$, remove $x_{\pi(s)}$ from $a$.
\item remove any empty clauses (resulting from clauses of $\varPhi$ that become unsatisfied if we set $x_{\pi(s)}$ to $\sigma(x_{\pi(s)})$ for $1\leq s\leq t$) from the formula.
\end{itemize}

For a number $\delta > 0$ and an index $l > t$ we say that $x_{\pi(l)}$ is \emph{$(\delta,t)$-biased} if 
\begin{eqnarray}
\left|\mu_{x_{\pi(l)}}^{[\omega]}(\varPhi_{t,\pi,\sigma},1) - \frac12 \br{1-\mu_{x_{\pi(l)}}^{[\omega]}(\varPhi_{t,\pi,\sigma},0)}\right| >\delta.
\end{eqnarray}
Moreover the tripel $(\varPhi,\pi,\sigma)$ is \emph{$(\delta, t)$-balanced} if no more than $\delta(n-t)$ variables are $(\delta,t)$-biased.

\begin{lemma}[\cite{BP}]\label{lem_12}
 For any $0\leq t\leq \hat{t}$ we have
	\begin{eqnarray*} 
	\varDelta_t = (1+o(1))\delta_t n/(c k).
	\end{eqnarray*}
Furthermore,  $\varDelta_{\hat t} \sim \frac{n}{ck} \brk{(\rho)^{-\frac1c} - \exp(-ck)}$.
\end{lemma}
For $\xi >0$ we say that $\varPhi$ is \emph{$(t,\xi)$-uniform} if 
	\begin{eqnarray*} 
	|\left\{(\pi,\sigma) \in S_n \times \{-1,1\}^{V}: (\varPhi,\pi,\sigma) \text{ is not } (\delta_t,t)\text{-balanced}\right\}|\leq 2^nn!\cdot \exp\brk{-10(\xi n+\varDelta_t)}.
	\end{eqnarray*}

Now it is possible to relate the distribution $\bar{\beta}_\varPhi$ to the uniform distribution on $\{-1,1\}^V$ for $(t,\xi)$-uniform formulas.
\begin{proposition}[\cite{BP}]\label{prop_13}
Suppose that $\varPhi$ is $(t,\xi)$-uniform for all $0 \leq t\leq \hat{t}$. Then
 	\begin{eqnarray*} 
	\bar{\beta}_{\varPhi}(\mathcal{E}) \leq \frac{|\mathcal{E}|} {2^{\hat{t}}} \cdot \exp\brk{6(\varDelta_{\hat{t}}+\xi n)} + \exp(-\xi n/2) \quad \text{for any } \mathcal{E} \subset \{-1,1\}^V.
	\end{eqnarray*}
\end{proposition}
Proposition \ref{prop_13} reduces the proof of Theorem \ref{theo_1} to showing that $\vphi$ is $(t,\xi)$-uniform with some appropriate probability. 

We call a clause $a$ of a formula $\varPhi$ \emph{redundant} if $\varPhi$ has another clause $b$ such that $a$ and $b$ have at least two variables in common. Furthermore, we call the formula $\varPhi$ \emph{tame} if 
\begin{enumerate}
\item[i.] $\varPhi$ has no more than $\ln n$ redundant clauses, and
\item[ii.] no more than $\ln n$ variables occur in more than $\ln n$ clauses of $\varPhi$. 
\end{enumerate} 
The following is a well-known fact.
\begin{lemma} \label{lem_14}
The random formula $\vec\varPhi$ is tame \whp
\end{lemma}

The following Corollary is the formulation of Proposition \ref{theo_balanced} for $(\vphi,\pi,\sigma)$ implied by Proposition \ref{theo_balanced} by the basic symmetry properties of $\vphi$. 

\begin{corollary}\label{cor_balanced}
For any $k,r$ satisfying $2^k\rho_k/k < r\leq 2k\ln 2$ there is $\xi = \xi(k,r) \in [0,\frac1k]$ so that for $n$ large enough the following holds. 
Fix any permutation $\pi$ of $[n]$ and any assignment $\sigma \in \{-1,1\}^V$. Then for any $0 \leq t\leq \hat{t}$ we have
\begin{eqnarray}
\Pr\brk{(\vec\varPhi,\pi,\sigma) \text{ is } (\delta_t,t)\text{-balanced}|\vec\varPhi \text{ is tame}}\geq 1-\exp\brk{-3\xi n-10 \varDelta_t}.
\end{eqnarray}
\end{corollary}

\begin{corollary}[\cite{BP}]\label{cor_16}
In the notation of Corollary~\ref{cor_balanced} 
\begin{eqnarray}
\Pr\brk{\forall t\leq \hat{t}: \vec\varPhi \text{ is } (t,\xi)\text{-uniform}|\vec\varPhi \text{ is tame}}\geq 1-\exp\brk{-3\xi n}.
\end{eqnarray}
\end{corollary}

\begin{proof}[Proof of Theorem~\ref{theo_1}] 
Let us keep the notation of Theorem \ref{theo_balanced}. By Lemma \ref{lem_14} we may condition on $\vphi$ being tame. Let $\cU$ be the event that $\vphi$ is $(t,\xi)$-uniform for all $1\leq t\leq \hat t$. Let $\cS$ be the event that $|\cS(\vphi)|\leq n\cdot\Erw\brk{|S(\vphi)|}$. By Corollary \ref{cor_16} and Markov's inequality, we have $\vphi\in \cU\cap \cS$ \whp, then by Proposition \ref{prop_13}
\begin{eqnarray}
	\bar{\beta}_\vphi(\cS(\vphi))&\leq& \frac{\cS(\vphi)}{2^{\hat t}} \cdot\exp\br{6(\varDelta_{\hat t}+\xi n)} + \exp\br{-\xi n/2}\nonumber\\
	&\leq& n\cdot \Erw\brk{|\cS(\vphi)|} \cdot2^{\hat t} \exp\br{6(\varDelta_{\hat t}+\xi n)} + \exp\br{-\xi n/2}.\label{equ_1_1}
\end{eqnarray}
By Lemma \ref{lem_8} and \ref{lem_12} we have $\Erw\brk{|\cS(\vphi)|}\leq 2^n\exp\br{-rn/2^k}$ and $\varDelta_{\hat t} \leq \frac{n}{ck}(kr/2^k)^{-\frac1c}$. Plugging these estimates and the definition (\ref{equ_def_delta}) of $\hat t$ into (\ref{equ_1_1}), we find that given $\vphi \in \cU\cap \cS$,
\begin{eqnarray*}
	\bar{\beta}_\vphi(\cS(\vphi)) \leq n\exp\br{n\br{-\frac{r}{2^k} + \frac{\ln(kr/2^k)\ln(2)}{c^2k} + \frac{6}{ck}(kr/2^k)^{-\frac1c}+6\xi}} + \exp\br{-\xi n/2}.
\end{eqnarray*}
Recalling that $\rho=kr/2^k$ and $\xi\leq 1/k$, we thus obtain
\begin{eqnarray}
	\bar{\beta}_\vphi(\cS(\vphi))\leq n\exp\br{-\frac{n}{k} \br{\rho - \frac{\ln \rho\ln2}{c^2} - \frac{6}{c\rho^{\frac1c}}+6 }} + \exp\br{-\xi n/2}.\label{equ_1_2}
\end{eqnarray} 
Hence, since $\rho\geq \ln k$, (\ref{equ_1_2}) yields $\bar{\beta}_\vphi(\cS(\vphi)) = \exp\br{-\Omega(n)}$. Finally, Theorem  \ref{theo_1} follows from Fact \ref{fact_11}.
\end{proof}

\subsection{Sketch of proof}
Before we dive into the proofs of the rather technical statements let us give a sketch of the proof in order to develop an intuition of the underlying idea of the proof.

Writing $\mu_\xa^\el(\sign(x,a))$ in terms of the biases we obtain 
\begin{eqnarray}
\mu_{\xa}^\el(\sign(x,a)) &=& \frac12 (1-\psi_0(\pi\el)) - \br{E_\xa^\el +\sign(x,a)\varDelta_\xa^\el} \nonumber\\
& =&\tau\el/2 - \br{E_\xa^\el +\sign(x,a)\varDelta_\xa^\el}\label{equ_def_mu_ya}
\end{eqnarray}
We are going to prove that $|\varDelta_{x\ra a}^{[\ell]}|$ and $|E_{x\ra a}^{[\ell]}|$ are small for most $x$ and $a\in N(x)$. That is, given the $\varDelta_{x\ra a}^{[\ell]}$ and $E_{x\ra a}^{[\ell]}$ we need to prove that the biases $\varDelta_{x\ra a}^\elp$ and $E_{x\ra a}^\elp$ do not 'blow up'. The proof is by induction where the hypothesis is that at most $\delta_t\theta n$ variables are $\ell$-biased and at most $\delta^2\theta n$ variables are $\ell$-weighted and our goal is to show that the same holds true for $\ell+1$. To establish this, we need to investigate one iteration of the update rules (\ref{update_b_to_x}) and (\ref{update_x_to_a}). 

Now, to estimate how far $\pi^\elp_\xa(\zeta)$ actually strays from $\pi\elp$ we start by rewriting (\ref{update_b_to_x}) in terms of the biases $\varDelta_{x\ra a}^{[\ell]}$ and $E_{x\ra a}^{[\ell]}$, we obtain
\begin{eqnarray}
	\mu_{a \ra x}^{[\ell]}(0) &=& 1 - \prod_{y\in N(a)\setminus\{x\}} {\tau\el}/{2}  - \br{E_\ya^\el + \sign(y,a)\varDelta_{y \ra a}^{[\ell]}}\nonumber \\
	&=& 1 - \br{2/\tau\el}^{1-|N(a)|}\hspace{-0.23cm}\prod_{y\in N(a)\setminus\{x\}}\hspace{-0.23cm} 1-2/\tau\el\br{E_{y\ra a}^{[\ell]}(0)+ \sign(y,a)\varDelta_{y \ra a}^{[\ell]}}.\label{equ_def_mu_ax}
\end{eqnarray}
Under the assumption that $0.1\theta \leq |N(a)| \leq 10\theta k$, and $|\varDelta_{y \ra a}^{[\ell]}| \leq 0.1\delta_t = \exp(-c\theta k)$ as well as $|E_{y \ra a}^{[\ell]}| \leq 0.1\pi\el\delta_t \leq \exp(-c\theta k)$ for \emph{all} $y \in N(a) \setminus \{x\}$, and since by induction and \Lem~\ref{lem_bound_mu_l} $\tau\el$ is close to $1$ we can approximate (\ref{equ_def_mu_ax}) by
\begin{align*}
	\mu_{a \ra x}^{[\ell]}(0) 	&= 1 - \br{2/\tau\el}^{1-|N(a)|}\prod_{y\in N(a)\setminus\{x\}} 1-2/\tau\el\br{E_{y\ra a}^{[\ell]}(0)+ \sign(y,a)\varDelta_{y \ra a}^{[\ell]}}\nonumber\\
				&\sim \exp\br{- \br{2/\tau\el}^{1-|N(a)|}\br{1-2/\tau\el\hspace{-0.23cm}\sum_{y\in N(a)\setminus\{x\}}\hspace{-0.23cm}\br{E_{y\ra a}^{[\ell]}(0)+ \sign(y,a)\varDelta_{y \ra a}^{[\ell]}}}}
\end{align*}

Finally, we approximate 
\begin{eqnarray}
	\hspace{-1cm}\ln\pi^\elp_\xa (\zeta) &=& \ln\prod_{b\in N(x,\zeta)}\mu_\bx^\el(0)\nonumber \\
			  &\sim& - \hspace{-0.6cm}\sum_{b\in N(x,\zeta))\setminus\{a\}}\hspace{-0.6cm} \br{2/\tau\el}^{1-|N(b)|}\br{1-2/\tau\el\hspace{-0.35cm}\sum_{y\in N(b)\setminus\{x\}}\hspace{-0.5cm}\br{E_{y\ra b}^{[\ell]}(0)+ \sign(y,b)\varDelta_{y \ra b}^{[\ell]}}}\label{equ_nonrig_3}
\end{eqnarray}
which we claim to be very close to $\pi\el$. To prove that, we show that $\Pi\elp-\ln \pi^\elp_\xa (\zeta)$ is close to zero which by induction, \Lem~\ref{lem_bound_mu_l} and (\ref{equ_nonrig_3}) is the case if
\begin{eqnarray*}
	\Pi\elp - \sum_{b\in N(x,\zeta)\setminus\{a\}} \br{2/\tau\el}^{1-|N(b)|}\br{1-2/\tau\el\sum_{y\in N(b)\setminus\{x\}}\br{E_{y\ra b}^{[\ell]}(0)+ \sign(y,b)\varDelta_{y \ra b}^{[\ell]}}}
\end{eqnarray*}
is close to zero.

The first contribution to that sum is just the weight of clauses in which $x$ appears in with sign $\zeta$. By definition this should be close to $\pi\elp$ for many variables. 

The second contribution comes from the biases of the 'zero-messages'. This influence is small since the bound on $E_{y\ra b}^{[\ell]}$ is so tight and the set of $\ell$-weighted variables is so small that only a little number of variables are influenced by $\ell$-weighted variables. 

The third contribution 
\begin{eqnarray}
 \sum_{b\in N(x,\zeta)\setminus\{a\}} \sum_{y\in N(b)\setminus \{x\}} \br{2/\tau\el}^{2-\nb}\sign(y,b)\varDelta_{y \ra a}^{[\ell]}
\end{eqnarray}
is a \emph{linear} function of the bias vector $\varDelta^\el$ from the previous round. Indeed, this operator can be represented by a matrix 
\begin{align*}
	\hat\Lambda^\zeta &= (\hat\Lambda_{\xa,\yb}^\zeta)_{\xa,\yb}\qquad \text{with entries}\\
	\hat\Lambda_{\xa,\yb}^\zeta &= \begin{cases}
				\br{2/\tau\el}^{2-\nb}\sign(y,b) &\text{if }  a\neq b, x \neq y, \text{ and }b \in N(x,\zeta),\\
				0 &\text{otherwise.} 
	                      \end{cases}
\end{align*}
with $\xa,\yb$ ranging over all edges of the factor graph of $\vphi^t$.

Since $\hat\Lambda^\zeta$ is based on $\vphi^t$, it is a random matrix. One could therefore try to use standard arguments to bound it in some norm (say, $\|\hat\Lambda^\zeta\|_\square$). The problem with this approach is that $\hat\Lambda^\zeta$ is very high-dimensional: it operates on a space whose dimension is equal to the number of \emph{edges} of the factor graph. In effect, standard random matrix arguments do not apply.

To resolve this problem, consider a ``projection'' of $\hat\Lambda^\zeta$ onto a space of dimension merely $|V_t|\theta n$, namely
\begin{eqnarray}
 \Lambda^\zeta: \mathbb{R}^{V_t} \rightarrow \mathbb{R}^{V_t}, \Gamma = (\Gamma_y)_{y\in V_t} \mapsto \left\{\sum_{b\in N(x,\zeta)}\sum_{y\in N(b)\setminus\{x\}} \br{2/\tau\el}^{2-\nb}\sign(y,b)\Gamma_y\right\}_{x\in V_t}
\end{eqnarray}
One can think of $\Lambda^\zeta$ as a signed and weighted adjacency matrix of $\vphi^t$. Standard arguments easily show that $\|\Lambda^\zeta\|_\square\leq \delta_t^4\theta n$ with a very high probability. In effect, we expect that for all but a very small number of variables $x\in V_t$ we have simultaneously for $\zeta \in \{-1,1\}$ that
\begin{eqnarray}
 \max_{a\in N(x)}\left| \sum_{b\in N(x,\zeta)\setminus\{a\}}\sum_{y\in N(b)\setminus\{x\}} \br{2/\tau\el}^{2-\nb}\sign(y,b)\varDelta_\yb^\el \right| \leq \delta_t/4.
\end{eqnarray}

The quasirandomness properties are designed to identify graphs such that the number of variables where the $\sim$ signs in the above discussion is not appropriate is small and the influence of each small potentially set of biased variables is small.

Let us now turn this sketch into an actual proof. In Section \ref{sec_sp_proof_theo_25_1}, we prove \Prop~\ref{prop_B_subset_T}. In Sections \ref{sec_sp_proof_theo_25_2} to \ref{sec_sp_proof_theo_25_4} we prove \Prop~\ref{prop_bound_T}. In Section \ref{sec_sp_proof_theo_25_5} we prove Theorem \ref{theo_25}. Finally, in Section \ref{sec_proof_quasi} we establish that the quasirandomness property holds on $\vphi^t$ with the required probability.

\subsubsection{Definition of $T\el$ and $T'\el$}
\label{sec_def_of_T}
We like to show that for most variables $x\in V_t$ for all $a\in N(x)$ simultaneously for both $\zeta \in \{-1,1\}$ the values $\pi^{[\ell]}_\xa(\zeta)$ are close to a typical value which is estimated by $\pi\el$ for each iteration of SP.

We are going to trace the SP operator on $\varPhi^t$ iterated from the initial set of messages $\mu_{x\ra a}^{[0]}(\pm1)=\frac12$ and $\mu_{x\ra a}^{[0]}(0)=0$ for all $x \in V_t$ and $a\in N(x)$. Therefore, we define sets $T_1\el,\ldots,T_4\el\subset V_t$ and parameters $\pi\el$ and $\tau\el$ inductively that will allow us to identify biased variables. Let $T\el = T_1\el\cup T_2\el \cup T_3\el \cup N(T_4\el)$ and $T'\el=T_1\el \cup T_2\el$. It will turn out that $T\el$ is a superset of the set of biased variables and $T'\el$ a superset of the variables $x\in V_t$ such that for one clause $a\in N(x)$ we find $|\psi_0(\pi\el)-\mu_\xa^\el(0)|$ is large.

Let us define for $x\in V_t, a\in N(x)$ and $\zeta \in \{1,-1\}$ 
\begin{eqnarray*}
	\cN_{\leq 1}^{[\ell +1]}(x\ra a, \zeta) &=&  \cN_{\leq 1}(x,T[\ell],\zeta)\setminus \{a\}\\
	\cN_{1}^{[\ell +1]}(x\ra a, \zeta) &=&  \cN_{1}(x,T[\ell],\zeta)\setminus \{a\}\\
	\cN_{0}^{[\ell +1]}(x\ra a, \zeta) &=&  \cN_{0}(x,T[\ell],\zeta)\setminus \{a\}\\
	N_{> 1}^{[\ell +1]}(x\ra a, \zeta) &=& N(x,\zeta) \setminus (\{a\} \cup \cN_{\leq 1}(x,T[\ell],\zeta)).
\end{eqnarray*}

First of all, for $\ell=0$ we set $T_1[0]= T_2[0] =  T_3[0] =  T_4[0] =\emptyset$. 
Now, we let 
\begin{eqnarray*}
	T_1\elp&=&\left\{x\in V_t: \max_{(a,\zeta)\in N(x)\times \{-1,1\}}\left|P_{\leq 1}^\elp(\xa,\zeta)-\pi\elp\right|>0.01\delta\pi\elp\right\}
\end{eqnarray*}
contain all variables for which $P_{\leq 1}^\elp(\xa,\zeta)$ fails to be close enough to the typical value.

Let $T_2\elp$ be the set of all variables $x$ that have for at least  one $\zeta=\{-1,1\}$ at least one of the following properties.
\begin{enumerate}
\item [\textbf{T2a.}] $\left|\Pi\elp - \sum_{b \in \cN_{\leq 1}(x, T\el,\zeta)} \br{\zn}^{1-|N(b)|}\right| >2\delta/1000$.
\item [\textbf{T2b.}]  Either 
\begin{eqnarray*}
\sum_{b \in \cN_{1}(x, T\el,\zeta)} 2^{-|N(b)|} >10^4 \rho \theta k \delta\quad \text{ or } \quad \sum_{b \in \cN_{1}(x, T'\el,\zeta)} 2^{-|N(b)|} >10^4\rho \theta k \delta^2.
\end{eqnarray*}
\item [\textbf{T2c.}] $\sum_{b \in \cN_{\les}(x, T\el,\zeta)} 2^{-|N(b)|} >10^4\rho$.
\end{enumerate}
A variable $x$ is \emph{$(\ell+1)$-harmless} if it enjoys the following four properties simultaneously for $\zeta\in \{-1,1\}$.
\begin{enumerate}
\item [\textbf{H1.}] We have $\delta(\theta k)^3\sum_{b\in N(x)}2^{-|N(b)|}\leq 1$, and $0.1\theta k \leq |N(b)| \leq 10\theta k$ for all $b\in N(x)$.
\item [\textbf{H2.}] $\sum_{b\in \cN_{1}(x,T\el,\zeta)}2^{-|N(b)|} \leq \rho(\theta k)^5\delta$ and $\sum_{b\in N_{>1}(x,T\el,\zeta)}2^{|N(b)\cap T\el\setminus \{x\}|-|N(b)|}\leq \delta/(\theta k)$.
\item [\textbf{H3.}] There is at most one clause  $b\in N(x)$ such that $|N(b)\setminus T\el|\leq k_1$.
\item [\textbf{H4.}] $\left|\Pi\elp - \sum_{b\in \cN_{\leq 1}(x,T\el,\zeta)} \br{\zn}^{1-|N(b)|}\right|\leq 0.01\delta$.
\end{enumerate}
Let $H\elp$ signify the set of all $(\ell+1)$-harmless variables and $H[0] = \emptyset$. Further, let $T_3\elp$ be the set of all variables $x$ that have at least one of the following properties.
\begin{enumerate}
\item [\textbf{T3a.}] There is a clause $b\in N(x)$ that is either redundant, or $|N(b)| < 0.1\theta k$, or $|N(b)| > 10\theta k$. 
\item [\textbf{T3b.}] $\delta(\theta k)^3 \sum_{b\in N(x)}2^{-|N(b)|}>1$.
\item [\textbf{T3c.}] At least for one $\zeta=\{-1,1\}$ we have $\sum_{b \in N_{>1}(x, T\el,\zeta)} 2^{|N(b)| \cap T\el \setminus \{x\}|-|N(b)|} >\delta/(\theta k)$.
\item [\textbf{T3d.}] $x$ occurs in more than $100$ clauses from $T_3\el$.
\item [\textbf{T3e.}] $x$ occurs in a clause $b$ that contains fewer than $3|N(b)|/4$ variables form $H\el$.
\end{enumerate}

Furthermore, we let 
\begin{eqnarray}\label{equ_T_4}
	T_4\elp=\left\{a\in \phi^t:|N(a)|\geq 100k_1 \wedge |N(a)\setminus T\el|\leq k_1\right\} \setminus T_4\el.
\end{eqnarray}

In Section \ref{sec_sp_proof_theo_25_1}, we prove \Prop~\ref{prop_B_subset_T}. In Sections \ref{sec_sp_proof_theo_25_2} to \ref{sec_sp_proof_theo_25_4} we prove \Prop~\ref{prop_bound_T}. In Section \ref{sec_sp_proof_theo_25_5} we prove Theorem \ref{theo_25}. Finally, in Section \ref{sec_proof_quasi} we establish that the quasirandomness property holds on $\vphi^t$ with the required probability.

\subsection{Proof of Proposition \ref{prop_B_subset_T}}
\label{sec_sp_proof_theo_25_1}
Throughout this section we assume that 
\begin{align}\label{bound_pi_ell}
	\pi\el\leq2\exp\br{-\rho} \qquad \text{for all }  \ell\geq 0
\end{align}
and thus 
\begin{align}\label{bound_tau_ell}
\tau\el = 1-\psi_0(\pi\el) \geq 1-\pi\el \geq 1- 2\exp\br{-\rho}\geq 1- 2k^{-(1+\varepsilon)}.
\end{align}
The proof will be by induction on $\ell$. We start with a tightness result regarding $\pi^\elp_\xa(\zeta)$.

\begin{proposition}\label{prop_mu_el}
Let $x \in V_t$. Suppose $B\el \subset T\el$. Then simultaneously for $\zeta\in \{-1,1\}$ we have
\begin{eqnarray*}
	\max_{a\in N(x,\zeta)}\left|\pi^\elp_\xa(\zeta)-\pi\elp\right| \leq \begin{cases}
	\delta\pi\elp/80 & \text{ if } x\notin T\elp \\
	2 \pi\elp & \text{ if } x \notin T'\elp.
	\end{cases}
\end{eqnarray*}
\end{proposition}

To prove \Prop~\ref{prop_mu_el} we establish an elementary estimate of the messages $\mu_\bx$ from clauses to variables.
\begin{lemma}\label{lem_28}
Let $x$ be a variable and let $b \in N(x)$ be a clause. Let $t_b = |N(b) \cap B[\ell] \setminus \{x\}|$. Then 
\begin{eqnarray*}
	0 \geq 1- \mu_{b\ra x}^{[\ell]}(0) \leq (2/\tau{[\ell]})^{1-|N(b)|+t_b} \exp(\delta |N(b)|). 
\end{eqnarray*}
\end{lemma}
\begin{proof}
For any $y\in N(b)\setminus \{x\}$ by (\ref{equ_def_mu_ya}) we have 
\begin{eqnarray*}
\mu_{\yb}^\el(\sign(y,b)) =  \tau\el/2 - \br{E_\yb^\el +\sign(y,b)\varDelta_\yb^\el}.
\end{eqnarray*}
Therefore, by definition (\ref{update_b_to_x}) we have
\begin{eqnarray*}
0 &\leq& 1- \mu_\bx^\el(0) = \prod_{y\in N(b)\setminus \{x\}} \tau\el/2 - \br{E_\yb^\el + \sign(y,b)\varDelta_\yb^\el}\\
&=& \br\zn^{1-\nb}\prod_{y\in N(b)\setminus \{x\}} 1 - 2/\tau\el\br{E_\yb^\el + \sign(y,b)\varDelta_\yb^\el}\\
&\leq& \br\zn^{1-\nb}\cdot \br{\zn}^{t_b} \cdot \prod_{y\in N(b)\setminus (\{x\} \cup B\el)} 1 + 2/\tau\el\left|E_\yb^\el + \sign(y,b)\varDelta_\yb^\el\right| \\
&& \qquad \qquad\qquad\qquad \text{[as $|E_\yb^\el + \sign(y,b)\varDelta_\yb^\el| \leq \tau\el/2$ for all $y$]}\\
&\leq& \br{\zn}^{1-\nb+t_b}\cdot \exp\br{2\sum_{y\in N(b)\setminus (\{x\} \cup B\el)}\left|E_\yb^\el + \sign(y,b)\varDelta_\yb^\el\right|} \\
&\leq& \br{\zn}^{1-\nb+t_b}\cdot \exp\br{\nb\delta}  \\
&& \qquad \qquad\qquad\qquad\text{ [as $|\varDelta_\yb^\el|\leq 0.1\delta$ and $|E_\yb^\el|\leq 0.1\delta $ for all $y\notin B\el$]}.
\end{eqnarray*}
\end{proof}

\begin{corollary}\label{cor_29}
Let $x$ be a variable and let $\cT \subset N(b) \setminus\{x\}$ be a set of clauses. For each $b \in \cT$ let $t_b = |N(b) \cap B[\ell] \setminus \{x\}|$. Assume that $t_b < |N(b)|-2$ and $|N(b)| \leq 10 \theta k$ for all $b \in \cT$. Then $\mu_\bx^\el(0) > 0$ for all $b\in \cT$ and 
\begin{eqnarray}
	\left| \ln \prod_{b\in \cT} \mu_{b\ra x}^{[\ell]}(0)\right| \leq \sum_{b\in \cT} \br{2/\tau{[\ell]}}^{4-|N(b)|+t_b}\exp\br{\delta |N(b)|}.
\end{eqnarray}
\end{corollary}
\begin{proof}
For each $b\in \cT$ there is $y\in N(b)\setminus \{x\}$ such that $y\notin B\el$, because $t_b < \nb-2$. Therefore, by (\ref{def_message_ax}) $\mu_\bx^\el(0)>0$. Lemma \ref{lem_28} implies that
\begin{eqnarray}\label{equ_cor29_1}
1\geq \mu_\bx^\el(0)\geq 1-\br{\zn}^{1-\nb+t_b} \exp\br{\delta |N(b)|}.
\end{eqnarray}
Our assumptions $t_b < \nb - 2$ and $\nb \leq 10\theta k$ ensure that 
\begin{eqnarray}
\br{\zn}^{1-\nb+t_b}\leq 1/2 \qquad \text{and}\qquad \exp\br{\delta\nb} \leq 1.1,
\end{eqnarray}
whence $\br{\zn}^{1-\nb+t_b}\exp\br{\delta\nb} \leq 0.6$. Due to the elementary inequality $1-z \geq \exp(-2z)$ for $z \in [0,0.6]$, (\ref{equ_cor29_1}) thus yields
\begin{eqnarray}\label{equ_cor29_2}
\mu_\bx^\el(0) \geq \exp \br{-\br{\zn}^{3-\nb+t_b}\exp\br{\delta\nb}}\geq \exp\br{-\br{\zn}^{4-\nb+t_b}}.
\end{eqnarray}
Multiplying (\ref{equ_cor29_2}) up over $b\in\cT$ and taking logarithms yields
\begin{eqnarray}\label{equ_cor29_3}
0\geq\ln \prod_{b\in \cT} \mu_\bx^\el(0) \geq -\sum_{b\in \cT}\br{\zn}^{4-\nb+t_b}\exp\br{\delta\nb}
\end{eqnarray}
as desired.
\end{proof}

\begin{corollary}\label{cor_30}
Suppose that $x \in H[\ell]$ and that $a\in N(x)$ is a clause such that $|N(a)\setminus T[\ell-1]| \leq k_1$. Moreover, assume that $B[\ell-1] \subset T[\ell-1]$. Then $|\varDelta_{x\ra a}^{[\ell]}|\leq 0.01$.
\end{corollary}
\begin{proof}
Since $x\in H\el$ for each $b\in N(x,\zeta)\setminus\{a\}$ we have the following properties.
\begin{enumerate}
\item [\textbf{P1.}] By \textbf{H1} we have $0.1\theta k\leq \nb \leq 10\theta k$.
\item [\textbf{P2.}] By \textbf{H3} we have $|N(b)\setminus T\elm| \geq k_1$.
\item [\textbf{P3.}] Let $t_b = |N(b)\cap B\elm\setminus \{x\}|$. Our assumption that $B\elm \subset T\elm$ and condition \textbf{H3} ensure that 
\begin{eqnarray}
t_b\leq |N(b) \cap T\elm|\leq \nb - k_1 <\nb-2.
\end{eqnarray}
\end{enumerate}

Since $|N(a)\setminus T[\ell-1]| \leq k_1$ by property \textbf{P2} we find
\begin{eqnarray}
\cT=N_{>1}^\el(\xa,\zeta)=N_{>1}(x,T\elm,\zeta).
\end{eqnarray}
By \textbf{P1} and \textbf{P3} Corollary \ref{cor_29} applies to $\cT$ and yields
\begin{eqnarray}\label{equ_cor30_1}
\left|\ln P_{>1}^\el(\xa,\zeta)\right| = \left|\ln \prod_{b\in \cT} \mu_\bx^\elm(0)\right| \leq \sum_{b\in \cT} \br{\zn}^{4-\nb + t_b}
\end{eqnarray} 
and \textbf{H2} ensures that $\sum_{b\in \cT} \br{\zn}^{-\nb+t_b}\leq \delta$, whence (\ref{equ_cor30_1}) entails
\begin{eqnarray}\label{equ_cor30_5}
\left|P_{>1}^\el(\xa,\zeta) - 1\right| \leq 10^{-4}.
\end{eqnarray}

Moreover, $x\in H\el$ and therefore by \textbf{H1} and since $|N(a)\setminus T\elm|\leq k_1$ we have $|N(a) \cap T\elm| > 1$. Thus we get 
\begin{eqnarray}
\cN_{\leq1}^\el(\xa,\zeta) = \cN_{\leq 1}(x,T\elm,\zeta).
\end{eqnarray}
This yields the factorization
\begin{eqnarray}\label{equ_cor30_8}
P_{\leq 1}^\el(\xa,\zeta) = \prod_{b\in \cN_0(x,T\elm,\zeta)} \mu_\bx^\elm (0)\cdot \prod_{b\in \cN_1(x,T\elm,\zeta)}\mu_\bx^\elm (0).
\end{eqnarray}
With respect to the second product, Corollary \ref{cor_29} yields
\begin{eqnarray}
\left|\ln \prod_{b\in \cN_1(x,T\elm,\zeta)}\mu_\bx^\elm (0)\right| &\leq& \sum_{b\in \cN_1(x,T\elm,\zeta)} \br{\zn}^{5-\nb}\exp\br{\delta|N(b)|}\\
&\leq& 32\rho(\theta k)^5 \delta \qquad \text{[by \textbf{H2} and (\ref{bound_tau_ell})]}\\
&\leq& 10^{-6} \qquad \text{[as $\delta = \exp\br{-c\theta k}$ with $\theta k \geq \ln(\rho)/c^2$]}
\end{eqnarray}
and thus
\begin{eqnarray}\label{equ_cor30_6}
\left|1-\prod_{b\in \cN_1(x,T\elm,\zeta)}\mu_\bx^\elm (0)\right| &\leq& 10^{-5}
\end{eqnarray}
Furthermore, for any $b\in \cN_{0}(x,T\elm,\zeta)$ we have 
\begin{eqnarray}
\mu_\bx^\elm(0)&=& 1- \prod_{y\in N(b)\setminus \{x\}} \tau\el/2 - \br{E_\yb^\el + \sign(y,b)\varDelta_\yb^\el}\\
&=& 1- \br\zn^{1-\nb}\prod_{y\in N(b)\setminus \{x\}} 1 - 2/\tau\el\br{E_\yb^\el + \sign(y,b)\varDelta_\yb^\el}.\label{equ_cor30_4}
\end{eqnarray}
Since $b\in \cN_{0}(x,T\elm,\zeta)$, we have $y\notin B\elm\subset T\elm$ for all $y\in N(b)\setminus \{x\}$, and thus $|\de_\yb^\elm| \leq 0.1\delta$ and $|\sq_\yb^\elm| \leq 0.1\delta\pi\elm$. Letting
\begin{eqnarray}
\alpha_b = 1 - \prod_{y\in N(b)\setminus \{x\}} 1 - 2/\tau\el\br{E_\yb^\el + \sign(y,b)\varDelta_\yb^\el}
\end{eqnarray}
we find with (\ref{bound_tau_ell}) that
\begin{eqnarray}\label{equ_cor30_3}
-10\delta\eta k \overset{\text{\textbf{P1}}}{\leq} 1-(1 + 0.5\delta)^{\nb} \leq \alpha_b \leq 1-(1 - 0.5\delta)^{\nb} \overset{\text{\textbf{P1}}}{\leq} 10\delta\theta k.
\end{eqnarray}
Thus, by (\ref{equ_cor30_4}), (\ref{equ_cor30_3}) and \textbf{P1} we compute 
\begin{eqnarray}\label{equ_cor30_13}
1\geq \mu_\bx^\elm(0) \geq 1- \br{\zn}^{1-\nb}(1+10\delta\theta k) \geq 0.99.
\end{eqnarray}
Using the elementary inequality $-z-z^2\leq \ln (1-z)\leq -z$ for $0\leq z\leq 0.5$, we obtain from (\ref{equ_cor30_4}), (\ref{equ_cor30_3}) and (\ref{equ_cor30_13})
\begin{eqnarray*}
\ln \mu_\bx^\elm (0) &\leq& -\br{\zn}^{1-\nb}(1-\alpha_b) \leq -\br{\zn}^{1-\nb}(1-10\delta\theta k)\\
\ln \mu_\bx^\elm (0) &\geq& -\br{\zn}^{1-\nb}(1-\alpha_b) -\br{\zn}^{2(1-\nb)}(1-\alpha_b)^2\\
&\geq&  -\br{\zn}^{1-\nb}(1+10\delta\theta k).
\end{eqnarray*}
Summing these bounds up for $b\in \cN_{0}(x,T\elm,\zeta)$, we obtain
\begin{eqnarray*}
\ln\prod_{b\in\cN_{0}(x,T\elm,\zeta)}\mu_\bx^\elm(0)&\leq& -\sum_{b\in \cN_{0}(x,T\elm,\zeta)}\br{\zn}^{1-\nb} \\
&&\qquad\qquad\qquad+ 10k\delta\sum_{b\in \cN_{0}(x,T\elm,\zeta)}\br{\zn}^{1-\nb}\\
&\leq& -\sum_{b\in \cN_{0}(x,T\elm,\zeta)}\br{\zn}^{1-\nb} + 10(k\theta)^{-3}\qquad  \text{[by \textbf{H1}]}\\
&=& -\sum_{b\in \cN_{\leq 1}(x,T\elm,\zeta)}\br{\zn}^{1-\nb} \\
&&\qquad\qquad\qquad+ \sum_{b\in \cN_{1}(x,T\elm,\zeta)}\br{\zn}^{1-\nb} + 2(k\theta)^{-3} \\
&\leq& - \Pi\el + 10^{-3}\delta + \rho\br{\theta k}^5\delta + 10(\theta k)^{-3} \qquad \text{[by \textbf{H2}, \textbf{H4}]} \\
&\leq& - \Pi\el +10^{-6}  \text{ [because $\delta=\exp\br{-c\theta k}$ and $\theta k \geq \ln(\rho)/c^2$]}.
\end{eqnarray*}
Analogously, we obtain $\ln\prod_{b\in\cN_{0}(x,T\elm,\zeta)}\mu_\bx^\elm(0)\geq - \Pi\el -10^{-6}$ and thus
\begin{eqnarray}\label{equ_cor30_9}
\left|\Pi\el + \ln\prod_{b\in\cN_{0}(x,T\elm,\zeta)}\mu_\bx^\elm(0)  \right| \leq 10^{-6}.
\end{eqnarray}
Consequently, (\ref{equ_cor30_9}) and Lemma \ref{lem_mu_el_w_el} yield
\begin{eqnarray}\label{equ_cor30_7}
\left|\pi\el-\prod_{b\in\cN_{0}(x,T\elm,\zeta)}\mu_\bx^\elm(0)\right| \leq 10^{-5}\pi\el.
\end{eqnarray}
Plugging (\ref{equ_cor30_6}) and (\ref{equ_cor30_7}) into (\ref{equ_cor30_8}) we see that $\left|P_{\leq1}^\el(\xa,\zeta)-\pi\el\right|\leq 10^{-4}\pi\el$, while $\left|P_{>1}^\el(\xa,\zeta)-1\right| \leq 10^{-4}$ by (\ref{equ_cor30_5}). Therefore, (\ref{def_mu_el_xa_zeta}) yields
\begin{eqnarray}\label{equ_cor30_10}
\left|\pi^\el_\xa(\zeta) - \pi\el\right| \leq 10^{-3}\pi\el.
\end{eqnarray}
By (\ref{update_x_to_a}), (\ref{equ_cor30_10}) and Lemma \ref{lem_bound_psi} we have
\begin{eqnarray}
	\left|\mu_\xa^\el(1) - \psi_1(\pi\el)\right|&=& \left|\psi_1(\pi^\el_\xa(1),\pi^\el_\xa(-1)) - \psi_1(\pi\el)\right| \leq 2\cdot 10^{-3}\label{equ_cor30_11}\\
	\left|\mu_\xa^\el(0) - \psi_0(\pi\el)\right|&=& \left|\psi_0(\pi^\el_\xa(1),\pi^\el_\xa(-1)) - \psi_0(\pi\el)\right| \leq 2\cdot 10^{-3}\pi\el\label{equ_cor30_12}
\end{eqnarray}
and therefore, by (\ref{equ_def_delta}), (\ref{equ_cor30_11}) and (\ref{equ_cor30_12}) we find
\begin{eqnarray}
	|\de_\xa^\el| = \left|\mu_\xa^\el(1) - \frac12(1-\mu_\xa^\el(0))\right|&\leq&\left|\psi_1(\pi\el) -\frac12 (1-\psi_0(\pi\el))\right| \hspace{-0.1cm} + \hspace{-0.1cm}5\cdot10^{-3} \\
	&\leq& 0.01 \qquad\text{[by (\ref{pro_psi_3})]}.
\end{eqnarray}
as claimed.
\end{proof}

\begin{corollary} \label{cor_est_mu_0}
Let $\ell \geq 1$ and $b$ be a clause such that $N(b) \not\subset T[\ell]$. Let $x \in N(b)$. Assume that $B[\ell-1] \subset T[\ell-1]$. Then
\begin{eqnarray}
	1-\mu_{b\ra x}^\elm(0) \leq \exp\br{-k_1/2}.
\end{eqnarray}
\end{corollary}
\begin{proof}
Since $N(b) \not\subset T\el$, there exists a $y\notin T\el$ and because $b\in N(y)$ by \textbf{T3a} we have 
\begin{eqnarray}\label{equ_cor31_1}
	0.1\theta k \leq \nb \leq 10\theta k.
\end{eqnarray}
We consider two cases
\begin{description}{}
\item [\textbf{Case 1}]$|N(b) \setminus T\elm|> k_1$. By (\ref{equ_cor31_1}) and Lemma \ref{lem_28} we find
\begin{eqnarray}
\exp\br{-\exp\br{-0.6k_1}}\leq\exp\br{-2^{3-k_1}\exp\br{\delta\nb}} \leq  \mu_\bx^\elm(0)\leq 1,
\end{eqnarray}
whence the assertion follows.
\item [\textbf{Case 2}]$|N(b) \setminus T\elm|\leq k_1$. The assumption $N(b) \not\subset T\el$ implies that $b \not \subset T_3\el$. But since $|N(b) \setminus T\elm|\leq k_1$ and by (\ref{equ_cor31_1}), the only possible reason why $b\notin T_3\el$ is that $b\in T_3\elm$ (cf. the definition of $T_3\el$). As $N(b) \not \subset T_3\el$, \textbf{T3e} implies 
\begin{eqnarray}\label{equ_cor31_2}
|N(b) \cap H\elm| \geq 3\nb/4.
\end{eqnarray}
Let $J=N(b) \cap H\elm$. Since $b\in T_3\elm$, we have $\ell\geq 2$ and $|N(b) \setminus T[\ell-2]| \leq k_1$. Therefore, Corollary \ref{cor_30} implies that $\de_\yb^{\elm}\leq 0.01$ for all $y\in J$. Thus, for all $x\in N(b)$ we have
\begin{eqnarray*}
\mu_\bx^\elm(0) &=& 1-\prod_{y\in N(b)\setminus\{x\}} \mu_\yb^\elm (-\sign(y,b))\\
&\geq& 1-0.501^{|J| - 1} \overset{(\ref{equ_cor31_2})}{\geq} 1-0.501^{3\nb/4 - 1} \geq 1- 0.501^{0.07\theta k}.
\end{eqnarray*}
Consequently,  
\begin{eqnarray}
\left| \mu_\bx^\elm(0) - 1\right| \leq 0.501^{0.07\theta k} \leq \exp\br{-\theta k/100}\leq \exp\br{-k_1}.
\end{eqnarray}
\end{description} 
Thus, we have established the assertion in either case.
\end{proof}

\begin{proof}[Proof of \Prop~\ref{prop_mu_el}]
Let us fix an $\ell\geq 0$ and assume that $B[\ell] \subset T[\ell]$. Let $x \in V_t\setminus T[\ell+1]$.
Corollary \ref{cor_est_mu_0} implies that 
\begin{eqnarray}\label{equ_bound_a_to_x}
	1-\mu_{a\ra x}^{[\ell]}(0) \leq \exp\br{-k_1/2} \quad \text{ for all } x \notin T[\ell +1], a \in N(x).
\end{eqnarray}
We claim
\begin{eqnarray} \label{equ_p_small}
|P_{>1}^{[\ell + 1]}(x \ra a,\zeta) - 1| \leq \delta/500 \quad \text{ for all } x \notin T[\ell + 1], a \in N(x), \zeta \in \{1,-1\}.
\end{eqnarray}
To establish (\ref{equ_p_small}), we consider two cases.
\begin{description}{}
\item [\textbf{Case 1}] $x \notin N(T_4\el)$. Let $\cT =  N_{>1}^\elp(x\ra a,\zeta)$ be the set of all clauses $b$ that contribute to the product $P_{>1}^\elp(x\ra a,\zeta)$. Since $x\notin N(T\el\cup T\elp)$, none of the clauses $b \in \cT$ features more than $|N(b)|- k_1$ variables from $T\el$ (just from the definition of $T_4\el$). Furthermore, because $x\notin T_3\elp$, \textbf{T3c} is not satisfied and thus we obtain the bound
\begin{eqnarray} \label{equ_bound_T}
\sum_{b \in \cT} \br{\zn}^{|N(b)| \cap T\el \setminus \{x\}|-|N(b)|} &\leq& \sum_{b \in N_{>1}(x, T\el,\zeta)} 2^{|N(b)| \cap T\el \setminus \{x\}|-|N(b)|} \nonumber\\
 &\leq& \delta/(\theta k)\leq  \delta/10^4.
\end{eqnarray}
Since $x \notin T\elp$, \textbf{T3a} ensures that $|N(b)|\leq 10\theta k$ for all $b\in \cT$. Therefore, (\ref{equ_p_small}) follows from (\ref{equ_bound_T}) and Corollary \ref{cor_29}.
\item [\textbf{Case 2}] $x \in N(T_4\el)$. Let $\cT=N_{>1}^\elp(x\ra a,\zeta)\setminus T_4\el$ be the set of all clauses $b$ that occur in the product $P_{>1}^\elp(x\ra a,\zeta)$, apart from those in $T_3\el$. Since $x\notin T_3\elp \cup N(T_4\elp)$, this set $\cT$ also satisfies (\ref{equ_bound_T}). Thus Corollary \ref{cor_29} yields 
\begin{eqnarray} \label{equ_proof_26_1}
\left|\ln \prod_{b\in \cT} \mu_{b \ra a}^\el(0)\right| \leq \delta/10^3.
\end{eqnarray}
Let $\cT' = N_{>1}^\elp(x\ra a,\zeta)\cap T_4\el$. As condition \textbf{T3d} ensures that $|\cT'|\leq |N(x) \cap T_4\el| \leq 100$, (\ref{equ_bound_a_to_x}) implies
\begin{eqnarray} \label{equ_proof_26_2}
\left|\ln \prod_{b\in \cT'} \mu_{b \ra a}^\el(0)\right| \leq 2|\cT'|\exp(-k_1/2)\leq \delta/1000.
\end{eqnarray}
Since $N_{>1}^\elp(x\ra a,\zeta) = \cT \cup \cT'$, (\ref{equ_proof_26_1}) and (\ref{equ_proof_26_2}) yield $|1-P_{>1}^\elp(x\ra a,\zeta)|\leq \delta/500$.
\end{description}
Thus we have established (\ref{equ_p_small}) in either case.

Let $a\in N(x)$. If $x \notin T_1\elp$ by definition 
\begin{eqnarray}\label{equ_pp_small}
	|\pi\elp - P_{\leq1}^\elp(x\ra a,\zeta)| \leq \pi\elp\delta/100.
\end{eqnarray}
Thus by (\ref{equ_p_small}) and (\ref{equ_pp_small}) we obtain for all $x\notin T\elp$ 
\begin{eqnarray} \label{equ_proof_26_3}
\left|\pi\elp - \pi^\elp_\xa(\zeta)\right| &=& \left|\pi\elp - P_{\leq 1}^\elp(x\ra a,\zeta)\cdot P_{>1}^\elp(x\ra a,\zeta)\right|\nonumber\\
&\leq& \pi\elp\delta/80.
\end{eqnarray}

To show the second assertion let $x \notin T'\elp$ and $a\in N(x)$. In particular, $x\notin T_1\elp$ and thus by (\ref{equ_pp_small}) we find
\begin{eqnarray} \label{equ_proof_26_4}
\left|\pi\elp - \pi^\elp_\xa(\zeta)\right| &=& \left|\pi\elp - P_{\leq 1}^\elp(x\ra a,\zeta)\cdot P_{>1}^\elp(x\ra a,\zeta)\right| \nonumber\\
&\leq& 2\pi\elp \qquad\qquad\qquad \text{[since $0\leq P_{>1}^\elp(x\ra a,\zeta)\leq 1$]}
\end{eqnarray}
as claimed.
\end{proof}

\begin{proof}[Proof of \Prop~\ref{prop_B_subset_T}]
To prove that $B\el \subset T\el$ and $B'\el \subset T'\el$ we proceed by induction on $\ell$. Since $B[0] = B'[0]= \emptyset$ the assertion is trivial for $\ell =0$. We assume that $\ell\geq 0$ and that $B[\ell] \subset T[\ell]$. 

Let $x \in V_t\setminus T[\ell+1]$ and $a\in N(x,\zeta)$ and $\zeta \in \{-1,1\}$. We will prove that $x\notin B\elp$. By Proposition \ref{prop_mu_el} simultaneously for $\zeta \in \{-1,1\}$  we have
\begin{eqnarray}
	\left|\pi^\elp_\xa(\zeta) -\pi\elp\right| \leq \delta\pi\elp/80.\label{equ_prop_final_3}
\end{eqnarray}
By (\ref{update_x_to_a}), (\ref{equ_prop_final_3}) and Lemma \ref{lem_bound_psi} we have
\begin{eqnarray}
\left|\mu_\xa^\elp(\zeta)- \psi_\zeta(\pi\elp)\right| &\leq& \delta/20 \label{equ_prop_final_1}\\
\left|\mu_\xa^\elp(0)- \psi_0(\pi\elp)\right| &\leq& \pi\elp\delta/40.\label{equ_prop_final_2}
\end{eqnarray}
Thus,
\begin{eqnarray*}
\left|\de_\xa^\elp\right| &=& \left|\mu_\xa^\elp(\zeta)- \frac12\br{1-\mu_\xa^\elp(0)}\right| \\
&\leq& \left| \psi_\zeta(\pi\elp) - \frac12\br{1-\psi_0(\pi\elp)}\right| + \delta/20 + \pi\elp\delta/40 \\
&&\qquad\qquad\qquad\qquad\qquad\text{[by (\ref{equ_prop_final_1}) and (\ref{equ_prop_final_2})]}\\
&\leq& \delta/10. \qquad\text{[since $\pi\elp \leq 2k^{-(1+\varepsilon)}$ by (\ref{bound_pi_ell}) and by (\ref{pro_psi_3})]}
\end{eqnarray*}
and
\begin{eqnarray*}
\left|\sq_\xa^\elp\right| &=& \left|\frac12\br{\mu_\xa^\elp(0)- \psi_0(\pi\elp)}\right| \\
&\leq& \pi\elp\delta/80 \qquad\text{[by (\ref{equ_prop_final_2})]}.
\end{eqnarray*}
Consequently, $x\notin B\elp$.

Similarly, let $x \in V_t\setminus T'[\ell+1]$ and $a\in N(x,\zeta)$ for some $\zeta\in \{-1,1\}$. We will prove that $x\notin B'\elp$. By Proposition \ref{prop_mu_el} simultaneously for $\zeta \in \{-1,1\}$  we have
\begin{eqnarray}
	\left|\pi^\elp_\xa(\zeta) -\pi\elp\right| \leq 2\pi\elp.
\end{eqnarray}
Therefore, Lemma \ref{lem_bound_psi} yields $\left|\mu_\xa^\elp(0)- \psi_0(\pi\elp)\right| \leq 4 \pi\elp\delta$ and thus
\begin{eqnarray}
\left|\sq_\xa^\elp\right| = \left|\frac12\br{\mu_\xa^\elp(0)- \psi_0(\pi\elp)}\right| \leq 2 \pi\elp\delta.
\end{eqnarray} 
Consequently, $x\notin B'\elp$.
\end{proof}

\subsection{Proof of Proposition \ref{prop_bound_T}}
\label{sec_sp_proof_theo_25_2}
Conditioned on the quasirandomness properties we bound the sizes of $|T'\el| \leq \delta^2\theta n$ and $|T\el| \leq \delta \theta n$ by induction on $\ell$. Thus, we may assume that $|T\el| \leq \delta \theta n$ and $|T'\el|\leq \delta^2 \theta n$.

We begin by bounding the sizes of the sets $T_2\elp, T_3\elp$ and $T_4\elp$.
\begin{lemma}\label{lem_bound_T4}
Assume that $|T_1\el\cup T_2\el \cup T_3\el|\leq \delta\theta n/3$ and $|N(T_4\el)|\leq \delta\theta n/2$. Then $|N(T_4\elp)| \leq \delta\theta n/2$.
\end{lemma}
\begin{proof}
By construction we have $T_4\el \cap T_4\elp = \emptyset$ (cf. \ref{equ_T_4}). Furthermore, also by construction $N(T_4\el)\subset T\el$, and each clause in $T_4\elp$ has at least a $0.99$-fraction of its variables in $T\el$. Thus, $|N(b)\cap T\el|\geq 0.99\nb$ for all $b\in T_4\el \cup T_4\elp$. Hence, \textbf{Q4} yields
\begin{eqnarray*}
	|N(T_4\el)| + |N(T_4\elp)| &\leq& \sum_{b\in T_4\el\cup T_4\elp} \nb\\
	&\leq& \frac{1.01}{0.99}|T\el| \leq 1.03 (|T_1\el|+ |T_2\el| + |T_3\el| + |N(T_4\el)|).
\end{eqnarray*}
Hence, $|N(T_4\elp)|\leq 1.03(|T_1\el|+ |T_2\el| + |T_3\el|)+0.03|N(T_4\el)|\leq \delta \theta n /2$.
\end{proof}

\begin{lemma}\label{lem_bound_T2}
Assume that $|T\el|\leq \delta\theta n$ and $|T'\el|\leq \delta\theta n$. Then $|T_2\elp| \leq \delta^2\theta n/100$.
\end{lemma}
\begin{proof}
Applying \textbf{Q2} to the set $T\el\leq \delta\theta n$ yields that the number of variables that satisfy either \textbf{T2a}, the first part of \textbf{T2b} or \textbf{T2c} is $\leq 3\delta^2\theta n /1000$. Applying  \textbf{Q2} to the set $T'\el\leq \delta^2\theta n$ yields that the number of variables that satisfy the second part of \textbf{T2b} is $\leq \delta^2\theta n /1000$. The assertion follows. 
\end{proof}

\begin{lemma}\label{lem_bound_T3}
Assume that $|T_1\el\cup T_2\el \cup T_3\el|\leq \delta\theta n/3$ and $|N(T_4\el)|\leq \delta\theta n/2$. Moreover, suppose that $|T\elm|\leq \delta\theta n$. Then $|T_3\elp| \leq \delta\theta n/6$.
\end{lemma}
\begin{proof}
Conditions \textbf{Q0} and \textbf{Q1} readily imply that the number of variables that satisfy either \textbf{T3a} or \textbf{T3b} is $\leq \delta\theta n/1000$. Moreover, we apply \textbf{Q3} to the set $T\el$ of size 
\begin{eqnarray}\label{equ_lem_T3_1}
	|T\el| \leq |T_1\el\cup T_2\el \cup T_3\el| + |N(T_4)|\leq 0.9\delta \theta n
\end{eqnarray}
to conclude that the number of variables satisfying \textbf{T3c} is $\leq \delta\theta n/1000$.

To bound the number of variables that satisfy \textbf{T3d}, consider the subgraph of the factor graph induced on $T_4\el\cup N(T_3\el)$. For each $x\in N(T_4\el)$ let $D_x$ be the number of neighbors of $x$ in $T_4\el$. Let $\nu$ be the set of all $x\in V_t$ so that $D_x\geq 100$. Then \textbf{Q4} yields
\begin{eqnarray*}
	100\nu\leq \sum_{x\in N(T_4\el)}D_x &=& \sum_{a\in T_4\el} \na\leq 1.01 |T\el| + \delta\theta n/10000\leq \delta \theta n \\
	&&\qquad\qquad\qquad\qquad \qquad\text{[as $N(b)\subset T\el$ for all $b\in T_4\el$]}.
\end{eqnarray*}
Hence, there are at most $\nu \leq 0.01\delta\theta n$ variables that satisfy \textbf{T3d}. In summary, we have shown that 
\begin{eqnarray}
	|\left\{x\in V_t: x \text{ satisfies one of \textbf{T3a} - \textbf{T3d}} \right\}| \leq 15\delta\theta n/1000.
\end{eqnarray}

To deal with \textbf{T3e}, observe that if a clause $a$ has at least $\na/4$ variables that are \textit{not} harmless, then one of the following statements is true
\begin{itemize}
\item[i.] $a$ contains at least $\na/20$ variables $x$ that violate either \textbf{H1}, \textbf{H2} or \textbf{H4}.
\item[ii.]  $a$ contains at least $\na/5$ variables $x$ that violate condition \textbf{H3}.
\end{itemize}
Let $\cC_1$ be the set of clauses $a$ for which i. holds and let $\cC_2$ be the set of clauses satisfying ii., so that the number of variables satisfying \textbf{T3e} is bounded by $\sum_{a\in \cC_1\cup \cC_2}\na$.

To bound $\sum_{a\in \cC_1}\na$, let $\cQ$ be the set of all variables $x$ that violate either \textbf{H1}, \textbf{H2} or \textbf{H4} at time $\ell$. Then conditions \textbf{Q1}-\textbf{Q3} entail that $|\cQ| \leq 3\delta\theta n/1000$ (because we are assuming $|T\elm| \leq \delta\theta n$). Therefore, condition \textbf{Q4} implies that 
\begin{eqnarray}\label{equ_lem_T3_4}
	\sum_{a\in \cC_1} \na \leq 21|\cQ| + \delta\theta n/10000 \leq 64\delta\theta n/10000.
\end{eqnarray}

To deal with $\cC_2$ let $\cB'$ be the set of all clauses $b$ such that $\nb\geq 100k_1$ but $|N(b)\setminus T\el|\leq k_1$. Since we know from (\ref{equ_lem_T3_1}) that $|T\el|\leq \delta\theta n$, condition \textbf{Q4} appied to $T\el$ implies
\begin{eqnarray}\label{equ_lem_T3_2}
	|N(\cB')| \leq \sum_{b\in \cB'}\nb\leq 1.03 |T\el| +  \delta\theta n/10000\leq 1.0301\delta\theta n.
\end{eqnarray}
In addition, let $\cB''$ be the set of length less than $100k_1 = 100 \sqrt{c}\theta k\leq 0.1\theta k$ by our choice of $c$, \textbf{Q1} implies that $|N(\bar{\bar{\cB}})|\leq\delta\theta n/10000$. Hence, (\ref{equ_lem_T3_2}) shows that $\cB = \cB' \cup \cB''$ satisfies
\begin{eqnarray}\label{equ_lem_T3_3}
|N(\cB)|\leq 1.0302 \delta\theta n.
\end{eqnarray}
Furthermore, let $\cU$ be the set of all clauses $a$ such that $N(a)\subset N(\cB)$. Let $U$ be the set of variables $x\in N(\cB)$ that occur in at least two clauses from $\cU$. Then by \textbf{Q4}
\begin{eqnarray*}
	|U| + \nb \leq \sum_{a\in \cU} \na \leq 1.01 |N(\cB)| + \delta\theta n /10000,
\end{eqnarray*}
whence $|U| \leq 0.01|N(\cB)| + \delta\theta n /10000\leq 2\delta\theta n/100$ due to (\ref{equ_lem_T3_3}). Since $\cB \subset \cU$, the set $U$ contains all variable that occur in at least two clauses from $\cB$, i.e., all variables that violate condition \textbf{H3}. Therefore, any $a\in \cC_2$ contains at least $\na/5$ variables from $U$. Applying \textbf{Q4} once more, we obtain 
\begin{eqnarray*}
	\sum_{a\in\cC_2} \na \leq 5.05\cdot2\delta\theta n/100 + \delta\theta n/10000 = 0.1201 \delta\theta n.
\end{eqnarray*}
Combining this estimate with the bound (\ref{equ_lem_T3_4}) on $\cC_1$, we conclude that the number of variables satisfying \textbf{T3e} is bounded by $\sum_{a\in \cC_1\cup\cC_2} \na \leq 0.127\delta\theta n$. Together with (\ref{equ_lem_T3_4}) this yields the assertion. 
\end{proof}

In section \ref{sec_sp_proof_theo_25_3} we will derive the following bound on $|T_1\elp|$.
\begin{proposition}\label{prop_bound_T1}
If $|T\el|\leq \delta\theta n$ and $|T'\el|\leq \delta^2\theta n$, then $|T_1\elp \setminus T_2\elp|\leq \delta^2\theta n/6$.
\end{proposition}

\begin{proof}[Proof of \Prop~\ref{prop_bound_T}]
We are going to show that 
\begin{eqnarray}
|T_1\el \cup T_2\el| &\leq& \delta^2 \theta n/3 \label{equ_pro27_1}\\
|T_1\el \cup T_2\el \cup T_3\el| &\leq& \delta \theta n/3  \qquad \text{and}\qquad |N(T_4\el)|\leq \delta\theta n/2 \label{equ_pro27_2}
\end{eqnarray}
for all $\ell\geq 0$. This implies that $|T\el| \leq \delta \theta n$ and $|T'\el| \leq \delta^2 \theta n$ for all $\ell \geq 0$, as desired. 

In order to proof (\ref{equ_pro27_1}) and (\ref{equ_pro27_2}) we proceed by induction on $\ell$ showing additionally that 
\begin{eqnarray}
\pi\el &\leq& 2\exp\br{-\rho} \label{equ_pro27_3}
\end{eqnarray}
for all $\ell\geq 0$. 
The bounds on $\ell=0$ are immediate from definition. Now assume (\ref{equ_pro27_1}) to (\ref{equ_pro27_3}) hold for all $l\leq \ell$. Then Lemma \ref{lem_bound_mu_l} shows that $\pi\elp\leq 2\exp\br{-\rho}$. Additionally, Lemma \ref{lem_bound_T2} and Proposition \ref{prop_bound_T1} show that $|T_1\elp \cup T_2\elp| \leq \delta^2 \theta n/3$. Moreover, Lemma \ref{lem_bound_T3} applies (with the convention that $T[-1] = \emptyset$), giving $|T_1\elp \cup T_2\elp \cup T_3\elp| \leq \delta \theta n/3$. Finally, Lemma \ref{lem_bound_T4} shows that $|N(T_4\el)|\leq \delta\theta n/2$.
\end{proof}

\subsection{Proof of Proposition \ref{prop_bound_T1}}
\label{sec_sp_proof_theo_25_3}
Throughout this section we assume that $|T\el|\leq \delta \theta n, |T'\el| \leq \delta^2 \theta n$ and $\pi\el\leq2\exp\br{-\rho}$.
For a variable $x\in V_t, a \in N(x)$ and $\zeta\in \{1,-1\}$ we let 
\begin{eqnarray}
	\sigma_{x\ra a}^\elp(\zeta) &=& \sum_{b\in \cN_{\leq 1}^\elp(x\ra a,\zeta)} \br{2/\tau\el}^{1-|N(b)|}\\
	\alpha_{x\ra a}^\elp(\zeta) &=& \sum_{b\in \cN_{\leq 1}^\elp(x\ra a,\zeta)} \sum_{y\in N(b)\setminus\{x\}} \br{2/\tau\el}^{1-|N(b)|}\sign(y,b)\varDelta_{y\ra b}^\el\\
	\beta_{x\ra a}^\elp(\zeta) &=& \sum_{b\in \cN_{\leq 1}^\elp(x\ra a,\zeta)} \sum_{y\in N(b)\setminus\{x\}} \br{2/\tau\el}^{1-|N(b)|}E_{y\ra b}^\el \\
	L_{x\ra a}^\elp(\zeta) &=& \sigma_{x\ra a}^\elp(\zeta) + \alpha_{x\ra a}^\elp(\zeta) + \beta_{x\ra a}^\elp(\zeta).
\end{eqnarray}

\begin{proposition}\label{prop_lin}
For any variable $x \notin T'\elp$, any clause $a\in N(x)$ and $\zeta\in \{1,-1\}$ we have
\begin{eqnarray}
\left|L_{x\ra a}^\elp(\zeta) + \ln P_{\leq 1}^\elp(\xa,\zeta)\right| \leq 10^{-3}\delta 
\end{eqnarray}
\end{proposition}

We will prove \Prop~\ref{prop_lin} in Section~\ref{sec_sp_proof_theo_25_4}.

\begin{lemma}\label{lem_37}
Let $x$ be a variable and let $b_1,b_2 \in N(x)$ be such that $|N(b_i) \cap T\el| \leq 2$ and $|N(b_i)|\geq 0.1\theta k$ for $i = 1,2$. Then 
\begin{eqnarray}
\left|\varDelta_{x\ra b_1}^\el-\varDelta_{x\ra b_2}^\el\right| \leq \delta^3.
\end{eqnarray}
\end{lemma}
\begin{proof}
By Proposition \ref{prop_B_subset_T} we have $B\elm \subset T\elm$. Furthermore, our assumptions ensure that $N(b_i)\setminus T\el \neq \emptyset$. Hence, Corollary \ref{cor_est_mu_0} yields
\begin{eqnarray}\label{equ_37_1}
	\mu_{b_i\ra x}^\elm > 0 \text{ and }  1- \mu_{b_i\ra x}^\elm (0) \leq \exp\br{-k_1/2} \leq \delta^7
\end{eqnarray}
for $i = 1,2$. There are two cases. 
\begin{description}{}
\item [\textbf{Case 1}] \textbf{ There is $c\in N(x,\zeta)\setminus\{b_1,b_2\}$ such that $\mu_{c\ra x}^\elm(0)=0$ for one $\zeta \in \{-1,1\}$.} Then $\pi^\el_ {x\ra b_1}(\zeta) = \pi^\el_{x\ra b_2}(\zeta) = 0$ and by (\ref{update_b_to_x}) to (\ref{update_x_to_a}) we find $\mu_{x \ra b_1}^\el(-\zeta) = \mu_{x \ra b_2}^\el(-\zeta) = 0, \mu_{x \ra b_1}^\el(0) = \mu_{x \ra b_2}^\el(0) = 0$ and $\mu_{x \ra b_1}^\el(\zeta) = \mu_{x \ra b_2}^\el(\zeta) = 1$ and therefore $\de_{x\ra b_1}^\el = \de_{x\ra b_2}^\el$.
\item [\textbf{Case 2}] \textbf{ For all $c\in N(x)\setminus\{b_1,b_2\}$ we have $0<\mu_{c\ra x}^\elm(1)$.} Then (\ref{update_b_to_x}) to (\ref{update_x_to_a}) yield $0<\mu_{x\ra b_i}^\el(0)<1$ for $i=1,2$. Let 
\begin{eqnarray}
	\cP^\el_x(\zeta) = \prod_{b\in N(x,\zeta)\setminus\{b_1,b_2\}}\mu_\bx^\elm(0).
\end{eqnarray}
Then for $i=1,2$ we have
\begin{eqnarray}
	\pi^\el_{b_i\ra x}(\zeta) = \cP^\el_x(\zeta) \cdot \mu_{b_i\ra x}^\elm(0).
\end{eqnarray}
We bound
\begin{eqnarray}
\left|\ln\br{\frac{\pi^\el_{b_i\ra x}(\zeta)}{\cP^\el_x(\zeta)}}\right| = \left| \ln \mu_{b_i\ra x}^\elm(0)\right| \leq \delta^6 \qquad\text{[by (\ref{equ_37_1})]}.\label{equ_37_2}
\end{eqnarray}
and obtain
\begin{eqnarray}
\left|1-\frac{\pi^\el_{b_i\ra x}(\zeta)}{\cP^\el_x(\zeta)}\right| \leq \delta^5.
\end{eqnarray}
Therefore, $\left|\cP^\el_x(\zeta)-\pi^\el_{b_i\ra x}(\zeta)\right| \leq \delta^5\cP^\el_x(\zeta)$. Now, Lemma \ref{lem_bound_psi} applies for each $i=1,2$ such that
\begin{eqnarray}
\left|\psi_0(\pi^\el_{b_i\ra x}(1),\pi^\el_{b_i\ra x}(-1))-\psi_0(\cP^\el_x(1),\cP^\el_x(-1))\right| &\leq& 2 \delta^5 \leq\delta ^4 \label{equ_37_3}\\
\left|\psi_1(\pi^\el_{b_i\ra x}(1),\pi^\el_{b_i\ra x}(-1))-\psi_1(\cP^\el_x(1),\cP^\el_x(-1))\right| &\leq& 2 \delta^5 \leq\delta ^4.\label{equ_37_4}
\end{eqnarray}
Consequently, since
\begin{eqnarray*}
\mu_{x\ra b_i}^\el(\zeta) = \psi_\zeta(\pi^\el_{b_i\ra x}(1),\pi^\el_{b_i\ra x}(-1))
\end{eqnarray*}
and
\begin{eqnarray*}
\left|\de_{x\ra b_1}^\el - \de_{x\ra b_2}^\el\right| = \left|\mu_{x\ra b_1}^\el(1) - \mu_{x\ra b_2}^\el(1) - \frac12 \br{\mu_{x\ra b_2}^\el(0)- \mu_{x\ra b_1}^\el(0)}\right| 
\end{eqnarray*}
by (\ref{equ_37_3}) and (\ref{equ_37_4}) we obtain
\begin{eqnarray*}
\left|\de_{x\ra b_1}^\el - \de_{x\ra b_2}^\el\right| \leq 3\delta^4\leq \delta^3. 
\end{eqnarray*}
\end{description} 
Hence, we have established the desired bound in both cases. 
\end{proof}

\begin{lemma}\label{lem_sigma_small}
For all variables $x \notin T_2\elp$ and one $\zeta \in \{-1,1\}$ we have
\begin{eqnarray}
\max_{a\in N(x)}\left| \sigma_\xa^\elp(\zeta) - \Pi\elp\right| \leq 3\delta/1000 \quad \text{ for } \zeta \in \{1,-1\}.
\end{eqnarray}
\end{lemma}
\begin{proof}
Let $x \notin T_2\elp$ and $a\in N(x)$. Since $\cN_{\leq 1}^\elp(\xa,\zeta)= \cN_{\leq1}(x,T\el,\zeta)\setminus \{a\}$, we obtain 
\begin{eqnarray*}\label{equ_proof_lem_sigma_small_2}
	\left| \Pi\elp -  \sigma_\xa^\elp(\zeta)\right|&\leq&\left| \Pi\elp-  \sum_{b\in \cN_{\leq 1}(x,T\el,\zeta)}\br{2/\tau\el}^{1-|N(b)|} \right|\\
	&&\qquad\qquad\qquad\qquad+\vecone_{a\in\cN_\les^\elp(\xa,\zeta)}\cdot2^{1-|N(a)|} \\
	&\leq& 2\delta/1000+ \vecone_{a\in\cN_\les^\elp(\xa,\zeta)}\cdot2^{1-|N(a)|} \quad \quad\quad\quad \text{[by \textbf{T2a}]} \\
	&\leq& 2\delta/1000 + \exp\br{-0.05\theta k} \\
	&&\qquad\quad \quad\quad\quad \text{[as $|N(a)| \geq 0.1\theta k$ if $a\in \cN_\les^\elp(\xa,\zeta)$]} \\
	&\leq& 3\delta/1000
\end{eqnarray*}
as desired. 
\end{proof}

\begin{lemma}\label{lem_alpha_samll}
For all but at most $0.1\delta^2 \theta n$ variables $x\notin T_2\elp$ and any $\zeta \in \{-1,1\}$ we have
\begin{eqnarray}
\max_{a\in N(x)}\left|\alpha_\xa^\elp(\zeta)\right| \leq 10^{-3}\delta.
\end{eqnarray}
\end{lemma}
\begin{proof}
For a variable $y$ let  $\cN(y)$ be the set of all clauses $b \in N(y)$ such that $b\in \cN_\les(x,T\el,\zeta)$ for some variable $x\in V_t$. If $\cN(y) = \emptyset $ we define $\varDelta_y = 0$; otherwise select $a_y \in \cN(y)$ arbitrarily and set $\varDelta_y=\varDelta_{y\ra a_y}^\el$. Thus, we obtain a vector $\varDelta = (\varDelta_y)_{y\in V}$ with norm $||\varDelta||_{\infty} \leq \frac12$. Let $A^\elp(\zeta) = (\alpha_x^\elp(\zeta))_{x\in V_t} =  \Lambda(T\el,\pi\el,\zeta)\varDelta$, where $\Lambda(T\el,\pi\el,\zeta)$ is one of the linear operators from condition \textbf{Q5} in Definition \ref{def_quasi}. That is, for any $x\in V_t$ we have 
\begin{eqnarray}
	\alpha_x^\elp(\zeta) = \sum_{b\in \cN_\les(x,T\el,\zeta)} \sum_{y\in N(b)\setminus \{x\}} \br{2/\tau\el}^{-\nb}\sign(y,b) \varDelta_y.
\end{eqnarray}
Because $|T\el| \leq \delta \theta n$, condition \textbf{Q5} ensures that $||\Lambda(T\el,\pi\el,\zeta)||_\square \leq \delta^4\theta n$. Consequently, 
\begin{eqnarray}\label{equ_38_1}
	||A^\elp(\zeta)||_1 = ||\Lambda(T\el,\pi\el,\zeta)\varDelta||_1 \leq || \Lambda(T\el,\pi\el,\zeta)||_\square ||\varDelta||_\infty \leq \delta^4\theta n.
\end{eqnarray}
Since $||A^\elp(\zeta)||_1 = \sum_{x\in V_t} |\alpha_x^\elp(\zeta)|$, (\ref{equ_38_1}) implies that
\begin{eqnarray}\label{equ_38_2}
	|\{x\in V_t:|\alpha_x^\elp(\zeta)|>\delta^{1.5}\}| \leq \delta^{2.5} \theta n.
\end{eqnarray}
To infer the Lemma from (\ref{equ_38_2}), we need to establish a relation between $\alpha_x^\elp(\zeta)$ and $\alpha_{x\ra a}^\elp(\zeta)$ for $x \notin T_2\el$ and $a \in N(x)$. Since for each $b \in \cN(y)$ there is a $x \in V_t$ such that $b \in \cN_\les(x,T\el,\zeta)$, we see that $|N(b)\cap T\el|\leq 2$ and $|N(b)|\leq 0.1 \theta k$ for all $b \in \cN(y)$.  
Consequently, Lemma \ref{lem_37} applies to $b \in \cN(y)$, whence $\left|\varDelta_{\yb}^\el - \varDelta_{y\ra b'}^\el\right| \leq \delta^3$ for all $y\in V_t, b,b' \in \cN(y)$. Hence, 
\begin{eqnarray}\label{equ_38_3}
	\left| \varDelta_{y\ra b}^\el - \varDelta_y\right| \leq \delta^3 \qquad\text{for all $y\in V_t, b\in \cN(y)$}.
\end{eqnarray} 
Consequently, we obtain for $x\notin T_2\elp$
\begin{align}
&\max_{a\in N(x)}\left|2\alpha_x^\elp(\zeta) - \alpha_{\xa}^\elp(\zeta)\right|\nonumber\\
&\qquad= \max_{a\in N(x)}\left| \vecone_{a \in \cN_\les(x,T\el,\zeta)} \cdot \sum_{y\in N(a)\setminus \{x\}} \br{2/\tau\el}^{1-\na} \sign(y,a) \varDelta_y  \right.\nonumber\\
& \qquad\quad +  \left. \sum_{b\in \cN_\les^\elp(\xa,\zeta)}\sum_{y\in N(b)\setminus \{x\}} \br{2/\tau\el}^{1-\nb} \sign(y,b) \br{\varDelta_y - \varDelta_\yb^\el} \right|\nonumber\\
&\leq \vecone_{a \in \cN_\les(x,T\el,\zeta)} \cdot \sum_{y\in N(a)\setminus \{x\}} \br{2/\tau\el}^{1-\na}\left|\varDelta_y  \right| \nonumber\\
&\qquad\qquad\qquad+\sum_{b\in \cN_\les^\elp(\xa,\zeta)}\sum_{y\in N(b)\setminus \{x\}}  \br{2/\tau\el}^{1-\nb}\left|\varDelta_y - \varDelta_\yb^\el\right| \nonumber\\
&\leq \vecone_{a \in \cN_\les(x,T\el,\zeta)} \cdot \na \br{2/\tau\el}^{-\na} \nonumber\\
&\qquad\qquad+ \delta^3\sum_{b\in \cN_\les(x,T\el,\zeta)}\nb  \br{2/\tau\el}^{1-\nb} \qquad \text{[by (\ref{equ_38_3})]} \nonumber\\
&\leq 10\theta k 2^{-0.1\theta k} + 10\delta^3\theta k\sum_{b\in \cN_\les(x,T\el,\zeta)}  \br{2/\tau\el}^{1-\nb} \nonumber\\
&\qquad\qquad\qquad\qquad  \qquad \text{[as $0.1\theta k\leq |N(a)| \leq 10\theta k$ if $a\in \cN_\les(x,T\el,\zeta)$]} \nonumber\\
&\leq \delta^2 + 10^5\rho\delta^3\theta k \qquad\qquad\qquad \text{[by \textbf{T2c}]}\nonumber \\
&\leq \delta/10000 \qquad\qquad\qquad \text{[as $\delta=\exp\br{-c\theta k}$ and $\theta k\geq \ln(\rho)/c^2$]}. \label{equ_38_4}
\end{align}
If $x\notin T_2\elp$ is such that $|\alpha_x^\elp(\zeta)|\leq \delta^{1.5}$, then (\ref{equ_38_4}) implies that $|\alpha_{\xa}^\elp(\zeta)|\leq \delta/5000$ for any $a \in N(x)$. Therefore, the assertion follows from (\ref{equ_38_2}).
\end{proof}

\begin{lemma}\label{lem_beta_small}
For any variable $x\notin T_2\elp$ and any $\zeta \in \{-1,1\}$ we have
\begin{eqnarray}
\max_{a\in N(x)}\left|\beta_\xa^\elp(\zeta)\right| \leq \delta/1000.
\end{eqnarray}
\end{lemma}
\begin{proof}
Let us recall that $\cN_{\leq1}^\elp(\xa,\zeta)=\cN_{0}^\elp(\xa,\zeta) \cup \cN_{1}^\elp(\xa,\zeta)$ where we have  
\begin{eqnarray}
  \cN_0^\elp(\xa,\zeta)&=&\cN_0(x,T\el,\zeta)\setminus \{a\} \quad\quad \text{and} \\
  \cN_1^\elp(\xa,\zeta)&=&\cN_1(x,T\el,\zeta)\setminus \{a\} \\
  &=& (\cN_1(x,T\el\setminus T'\el,\zeta) \cup \cN_1(x,T'\el,\zeta))\setminus \{a\}
\end{eqnarray}
since $T'\el \subset T\el$. Therefore, let 
\begin{eqnarray}
	\Gamma_1 = \cN_0(x,T\el,\zeta) \quad \text{and} \quad
	\Gamma_2 = \cN_1(x,T\el\setminus T'\el,\zeta)  \quad \text{and} \quad
	\Gamma_3 = \cN_1(x,T'\el,\zeta).
\end{eqnarray}
Since for all $b\in \Gamma_1$ we have $\left|\sq_\yb^\el\right| \leq 0.1\delta\pi\el$ for all $y\in N(b)$ we obtain
\begin{eqnarray}\label{equ_beta_1}
\sum_{b\in\Gamma_1} \sum_{y\in N(b)\setminus\{x\}} \br{2/\tau\el}^{1-|N(b)|} |\sq_\yb^\el| \leq \sum_{b\in\Gamma_1} \br{2/\tau\el}^{1-|N(b)|} |N(b)| \delta\pi\el.
\end{eqnarray}
For all $b\in \Gamma_2$ there exists one $y_1 \in N(b)$ such that $\left|\sq_{y_1\ra b}^\el\right| \leq 10\pi\el$ and $\left|\sq_\yb^\el\right| \leq 0.1\delta\pi\el$ for all $y\in N(b)\setminus\{y_1\}$. We obtain
\begin{eqnarray}\label{equ_beta_2}
\sum_{b\in\Gamma_2}\hspace{-0.1cm} \sum_{y\in N(b)\setminus\{x\}}\hspace{-0.6cm} \br{2/\tau\el}^{1-|N(b)|} |\sq_\yb^\el|\hspace{-0.07cm} \leq \hspace{-0.07cm}\sum_{b\in\Gamma_2}\hspace{-0.18cm} \br{2/\tau\el}^{1-|N(b)|} \br{\br{|N(b)|-1} \delta\pi\el + 10\pi\el}.
\end{eqnarray}
For all $b\in \Gamma_3$ there exists one $y_1 \in N(b)$ such that $\left|\sq_{y_1\ra b}^\el\right| \leq 1$ and $\left|\sq_\yb^\el\right| \leq 0.1\delta\pi\el$ for all $y\in N(b)\setminus\{y_1\}$. We obtain
\begin{eqnarray}\label{equ_beta_3}
\sum_{b\in\Gamma_3} \sum_{y\in N(b)\setminus\{x\}}\hspace{-0.3cm} \br{2/\tau\el}^{1-|N(b)|} |\sq_\yb^\el| \leq \sum_{b\in\Gamma_3} \br{2/\tau\el}^{1-|N(b)|} \br{\br{|N(b)|-1} \delta\pi\el + 1}
\end{eqnarray}

Let $x\notin T_2\elp$. Since $\cN_\les^\elp(\xa,\zeta) \subset \Gamma_1\cup\Gamma_2\cup\Gamma_3$ we get by (\ref{equ_beta_1}) to (\ref{equ_beta_3}) that 
\begin{align*}
\left|\beta_\xa^\elp(\zeta)\right| &= \left|\sum_{b \in\cN_\les^\elp(\xa,\zeta)} \sum_{y\in N(b)\setminus\{x\}} \br{2/\tau\el}^{1-|N(b)|} \sq_\yb^\el \right|\\
&\leq\sum_{b\in \cN_{\leq1}(x,T\el,\zeta)}   \br{2/\tau\el}^{1-|N(b)|} |N(b)|\delta\pi\el \\
&\qquad+\sum_{b\in \Gamma_2}  \sum_{y\in N(b)\setminus\{x\}} \br{2/\tau\el}^{1-|N(b)|} 10\pi\el\\
&\qquad+ \sum_{b\in \Gamma_3}  \sum_{y\in N(b)\setminus\{x\}} \br{2/\tau\el}^{1-|N(b)|} \\
&\leq 10^6\rho\theta k \delta\pi\el + 10^5\rho\theta k \delta \pi\el + 10^5\rho\theta k\delta^2 \qquad\text{[by \textbf{T2b} and as $\nb \leq10\theta k$]}\\
&\leq \delta/1000\qquad\text{[as $\pi\el\leq k^{-(1+\varepsilon_k)}, \theta k\geq \log(\rho)/c^2$ and $c\ll 1$]},
\end{align*}
as claimed.
\end{proof}

\begin{proof}[Proof of \Prop~\ref{prop_bound_T1}]
Let $S$ be the set of all variables $x\notin T_2\elp$ such that simultaneously for $\zeta \in \{-1,1\}$ we have
\begin{eqnarray}
	\max_{a\in N(x)}|\alpha_{\xa}^\elp(\zeta)|&\leq& \delta/1000.
\end{eqnarray}
For any $x\notin T_2\elp$ and $\zeta \in \{-1,1\}$, Lemma \ref{lem_sigma_small} and \ref{lem_beta_small} imply that for both $\zeta\in \{-1,1\}$
\begin{eqnarray}
	\max_{a\in N(x)}|\sigma_{\xa}^\elp(\zeta)-\Pi\elp|&\leq& 3\delta/1000\\
	\max_{a\in N(x)}|\beta_{\xa}^\elp(\zeta)| &\leq& \delta/1000 
\end{eqnarray}
and Proposition \ref{prop_lin} entails that for any $x\in S$ and $a\in N(x)$ we have
\begin{eqnarray}
	\left|\Pi\elp - \ln P_{\leq1}^\elp(\xa,\zeta) \right|&\leq& \left| L_\xa^\elp(\zeta)\right| + 10^{-3}\delta \nonumber \\
	&\leq& \left|\sigma_\xa^\elp(\zeta) -\Pi\elp\right| + \left|\alpha_\xa^\elp(\zeta)\right| + \left|\beta_\xa^\elp(\zeta)\right| +10^{-3}\delta \nonumber \\
	&\leq& \delta/100 .
\end{eqnarray}
Therefore, $\left|P_{\leq 1}^\elp(\xa,\zeta) / \exp\br{-\Pi\elp}-1\right| \leq \delta/50$ and thus
\begin{eqnarray}
	\left|P_{\leq 1}^\elp(\xa,\zeta)- \exp\br{-\Pi\elp}\right|	&\leq& \delta \exp\br{-\Pi\elp}/50  
\end{eqnarray}
and by Lemma \ref{lem_mu_el_w_el}
\begin{eqnarray}
	\left|P_{\leq 1}^\elp(\xa,\zeta)- \pi\elp\right|	&\leq& \delta \pi\elp/40 .
\end{eqnarray}
Consequently,  
\begin{eqnarray}
	T_1\elp\setminus T_2\elp\subset V_t\setminus (S\cup T_2\elp)
\end{eqnarray} 
and thus Lemma \ref{lem_alpha_samll} implies $|T_1\elp\setminus T_2\elp| \leq |V_t\setminus (S\cup T_2\elp)|\leq \delta^2\theta n/1000$.
\end{proof}

\subsection{Proof of Proposition~\ref{prop_lin}}
\label{sec_sp_proof_theo_25_4}
Let $O_t(\cdot)$ denote an asymptotic bound that holds in the limit for large $t$. That is, $f(t) = O(g(t))$ if there exist $C >0,t_*>0$ such that $|f(t)|\leq C|g(t)|$ for $t>t_*$.

\begin{lemma}\label{lem_lin}
Let $x\in V_t, a\in N(x), \zeta \in \{1,-1\}$ and $b\in \cN_{\leq 1}^\elp(\xa,\zeta)$. Then
\begin{eqnarray}
\ln \mu_\bx^\el(0)& =& \br{\zn}^{1-|N(b)|} \brk{1 + 2/\tau\el\sum_{y\in N(b)\setminus \{x\}}E_\yb^\el + \sign(y,b)\varDelta_\yb^\el}  \\
&&\quad+ \br{\zn}^{1-|N(b)|}\br{\theta k \delta + |N(b) \cap T\el\setminus \{x\}|}\cdot O_k(k\theta \delta)
\end{eqnarray}
\end{lemma}
\begin{proof}
The definition of the set $\cN_\les^\elp(\xa,\zeta)$ ensures that for all $b \in \cN_\les^\elp(\xa,\zeta)$ we have 
\begin{eqnarray}\label{equ_39_1}
|N(b)\cap T\el| \leq 2 \qquad\text{ and } \qquad 0.1\theta k \leq \nb \leq 10\theta k.
\end{eqnarray} 
Therefore, Lemma \ref{lem_28} shows that $|1-\mu_{b\ra x}^\el(0)| \leq \delta^2$ (recall from Proposition \ref{prop_B_subset_T} that $B\el \subset T\el$). Furthermore, $b$ is not redundant, and thus not a tautology, because otherwise $N(b) \subset T_3\el\subset T\el$ due to \textbf{T3a} in contradiction to (\ref{equ_39_1}).

Recall (\ref{equ_def_mu_ax}) the representation of
\begin{eqnarray}\label{equ_39_5}
	\mu_\bx^\el(0) = 1- \br{\zn}^{1-\nb} \prod_{y\in N(b)\setminus \{x\}}1-2/\tau\el\br{\sq_\yb^\el + \sign(y,b)\de_\yb^\el}.
\end{eqnarray}
Let $\Gamma =  N(b)\setminus (T\el\cup \{x\})$. As Proposition \ref{prop_B_subset_T} shows $B\el\subset T\el$ contains all biased variables, we have $\left|\de_\yb^\el\right| \leq 0.1\delta$ and $\left|\sq_\yb^\el\right| \leq 0.1\pi\el\delta$ for all $y\in \Gamma$. By (\ref{bound_tau_ell}) we have $\tau\el\geq \frac12$, thus we can use the approximation $|\ln(1-z) + z|\leq z^2$ for $|z|\leq \frac12$ to obtain 
\begin{eqnarray}
&&\left|\br{\ln \prod_{y\in \Gamma} 1-2/\tau\el\br{\sq_\yb^\el + \sign(y,b)\de_\yb^\el} }+  \sum_{y\in \Gamma} 2/\tau\el\br{\sq_\yb^\el + \sign(y,b)\de_\yb^\el} \right|\nonumber\\ 
&&\quad\leq\sum_{y\in \Gamma} \left|\ln \br{ 1-2/\tau\el\br{\sq_\yb^\el + \sign(y,b)\de_\yb^\el} }+  2/\tau\el\br{\sq_\yb^\el + \sign(y,b)\de_\yb^\el} \right|\nonumber\\ 
&&\quad \leq4\sum_{y\in \Gamma}\br{\sq_\yb^\el + \sign(y,b)\de_\yb^\el}^2 \leq 40\theta k \delta^2\qquad \label{equ_39_2}
\end{eqnarray}
since $|\Gamma| \leq |N(b)|\leq 10\theta k, |\de_\ya^\el|\leq 0.1\delta$ and $|\sq_\ya^\el|\leq 0.1\pi\el\delta$ for all $y\in\Gamma$. Hence
\begin{eqnarray}
	\left|\sum_{y\in \Gamma}2\sign(y,b) \de_\yb^\el\right| \leq 2\theta k\delta. \label{equ_39_4}
\end{eqnarray}
Therefore, taking exponentials in (\ref{equ_39_2}), we obtain
\begin{eqnarray}
	&&\prod_{y\in \Gamma} 1-2/\tau\el\br{\sq_\yb^\el+ \sign(y,b)\de_\yb^\el} \nonumber\\
	&&\qquad= \exp\br{O_k(\theta k\delta)^2 - \sum_{y\in \Gamma} 2/\tau\el\br{\sq_\yb^\el + \sign(y,b)\de_\yb^\el}} \nonumber\\
	&&\qquad= 1- \sum_{y\in \Gamma} 2/\tau\el\br{\sq_\yb^\el + \sign(y,b)\de_\yb^\el} + O_k(\theta k\delta)^2.\label{equ_39_3}
\end{eqnarray}
Furthermore, the definition of $\cN_\les^\elp(\xa,\zeta)$ ensures that 
\begin{eqnarray*}
|N(b) \setminus (\Gamma\cup\{x\})| = |N(b) \cap T\el \setminus \{x\}| \leq 1.
\end{eqnarray*}
If there is $y_0 \in N(b) \cap T\el\setminus \{x\}$, then (\ref{equ_39_4}) and (\ref{equ_39_3}) yield
\begin{eqnarray*}
	&&\prod_{y\in N(b)\setminus\{x\}} 1-2/\tau\el\br{\sq_\yb^\el+ \sign(y,b)\de_\yb^\el} \nonumber\\
	&&\qquad\qquad = \br{1-2/\tau\el\br{\sq_{y_0 \ra b}^\el+ \sign(y_0,b)\de_{y_0\ra b}^\el}} \\
	&&\qquad\qquad\qquad\cdot\prod_{y\in \Gamma} 1-2/\tau\el\br{\sq_\yb^\el+ \sign(y,b)\de_\yb^\el} \\
	&&\qquad\qquad = 1- \sum_{y\in N(b)\setminus \{x\}} 2/\tau\el\br{\sq_\yb^\el + \sign(y,b)\de_\yb^\el} + O_k(\theta k\delta).
\end{eqnarray*}
Hence, in any case we have 
\begin{eqnarray*}
	&&\prod_{y\in n(b)\setminus \{x\}}1-2/\tau\el\br{\sq_\yb^\el + \sign(y,b)\de_\yb^\el}\\
	&&\qquad= 1- \sum_{y\in N(b)\setminus \{x\}} 2/\tau\el\br{\sq_\yb^\el + \sign(y,b)\de_\yb^\el} \\
	&&\qquad\qquad+ \br{\theta k \delta+ |N(b)\cap T\el\setminus \{x\}} \cdot O_k(\theta k\delta)
\end{eqnarray*}
which is a small constant. Thus, combining this with (\ref{equ_39_2}) and using the approximation $|\ln(1-z) + z| \leq z^2$ for $|z| \leq \frac12$ we see that
\begin{eqnarray}\label{equ_39_5}
	\mu_\bx^\el(0) &=& - \br{\zn}^{1-\nb} \br{ 1-\hspace{-0.2cm} \sum_{y\in N(b)\setminus \{x\}} 2/\tau\el\br{\sq_\yb^\el + \sign(y,b)\de_\yb^\el} } \\
		&&\qquad\qquad + \br{\zn}^{1-\nb}\br{\theta k \delta+ |N(b)\cap T\el\setminus \{x\}} \cdot O_k(\theta k\delta),
\end{eqnarray}
whence the assertion follows.
\end{proof}

\begin{proof}[Proof of \Prop~\ref{prop_lin}]
By the definition of $P_{\leq 1}^\elp(\xa,\zeta)$ we have 
\begin{eqnarray}
\ln P_{\leq 1}^\elp(\xa,\zeta) = \sum_{b\in \cN_{\leq 1}^\elp(\xa,\zeta)}\ln \mu_\bx^\el(0).
\end{eqnarray}
Hence, Lemma \ref{lem_lin} yields
\begin{eqnarray}
&&\hspace{-1cm}\ln P_{\leq 1}^\elp(\xa,\zeta) \nonumber\\
&&\hspace{-1cm}\quad= L_{\xa}^\elp + \sum_{b\in \cN_{\leq 1}^\elp(\xa,\zeta)}\br{2/\tau\el}^{1-|N(b)|}\br{\theta k \delta + |N(b) \cap T\el\setminus \{x\}|}\cdot O_k(k\theta \delta).\label{prop_lin_0}
\end{eqnarray}
Let $x \notin T'\elp$. Condition \textbf{T2c} implies 
\begin{eqnarray}\label{prop_lin_1}
O_k(\delta\theta k)^2 \sum_{b\in \cN_{\leq 1}^\elp(\xa,\zeta)}\br{2/\tau\el}^{1-|N(b)|} &\leq& O_k(\delta\theta k)^2 \sum_{b\in \cN_{\leq 1}(x,T\el,\zeta)}2^{-|N(b)|} \\
&\leq& O_k(\delta\theta k)^2\cdot\rho \leq \delta/1000
\end{eqnarray} 
Furthermore, \textbf{T2b} yields 
\begin{eqnarray}
O_k(\delta\theta k) \sum_{b\in \cN_{\leq 1}^\elp(\xa,\zeta)}\br{2/ \tau\el}^{1-|N(b)|}|N(b) \cap T\el\setminus \{x\}| &\leq& O_k(\theta\delta k) \sum_{b\in \cN_1(x,T\el,\zeta)}2^{-|N(b)|}\nonumber\\
&\leq& O_k(\theta k \delta ) \cdot \rho\theta k \delta \nonumber\\
&\leq& \delta/1000.\label{prop_lin_2}
\end{eqnarray}
Finally, the assertion follows by plugging (\ref{prop_lin_1}) and (\ref{prop_lin_2}) into (\ref{prop_lin_0}).
\end{proof}

\subsection{Completing the proof of Theorem \ref{theo_25}}
\label{sec_sp_proof_theo_25_5}
We are going to show that for $\zeta \in \{1,-1\}$ simultaneously
\begin{eqnarray} \label{equ_theo25_1}
	\left|\mu_x^{[\omega]}(\varPhi_t,\zeta) -\frac12 \br{1-\mu_x^{[\omega]}(\varPhi_t,0)}\right| \leq \delta = \delta_t
\end{eqnarray}
for all $x\in V_t\setminus T[\omega+1]$. This will imply Theorem \ref{theo_25} because $|T[\omega+1]|\leq \delta_t(n-t)$ by Proposition \ref{prop_bound_T}.

Thus, let $x\in V_t\setminus T[\omega+1]$ and recall from (\ref{def_pi_omegap}) that 
\begin{eqnarray}
	\pi_x^{[\omega+1]}(\varPhi_t,\zeta) =\prod_{b\in N(x,\zeta)}\mu_\xb^{[\omega]}(0)
\end{eqnarray}
and from (\ref{def_marg_psi}) that
\begin{eqnarray}\label{equ_theo25_4}
	\mu_x^{[\omega]}(\varPhi_t,\zeta) =\psi_\zeta\br{\pi^{[\omega+1]}_x(\varPhi_t,1),\pi^{[\omega+1]}_x(\varPhi_t,-1)}.
\end{eqnarray} 

If $N(x) = \emptyset $, then trivially $\pi^{[\omega+1]}_x(\varPhi_t,1)=\pi^{[\omega+1]}_x(\varPhi_t,-1) = 1$ and $\mu_x^{[\omega]}(\varPhi_t,\zeta) = 0$ for $\zeta\in \{1,-1\}$ and $\mu_x^{[\omega]}(\varPhi_t,0) = 1$. Consequently, (\ref{equ_theo25_1}) holds true. 

Therefore, assume that $N(x)\neq \emptyset$ and pick an arbitrary $a \in N(x)$. Since $x\notin T[\omega + 1]$ Proposition \ref{prop_mu_el} yields
\begin{eqnarray}\label{equ_theo25_2}
 \left|\pi^{[\omega+1]}_\xa(\zeta)-\pi{[\omega+1]}\right|\leq \delta\pi{[\omega+1]}/50.
\end{eqnarray}
Furthermore, since $x\notin T[\omega+1]$ Corollary \ref{cor_est_mu_0} yields  
\begin{eqnarray}\label{equ_theo25_3}
	1- \mu_\ax^{[\omega]}(0)\leq \exp\br{-k_1/2}\leq \delta^2.
\end{eqnarray}
Thus we compute 
\begin{eqnarray}
	\left|\pi^{[\omega+1]}_x(\varPhi_t,\zeta)-\pi{[\omega+1]}\right|&\leq& \left|\pi^{[\omega+1]}_x(\varPhi_t,\zeta)-\pi_\xa^{[\omega+1]}(\zeta)\right| + \delta\pi[\omega+1]/50 \qquad\text{[by (\ref{equ_theo25_2})]}\nonumber\\
	&\leq& \left|\pi_\xa^{[\omega+1]}(\zeta)\cdot(1-\mu_\ax^{[\omega]}(0))\right| + \delta\pi[\omega+1]/50 \nonumber\\
	&\leq& \delta^2(\pi[\omega+1] + \delta\pi[\omega + 1]/50) + \delta\pi[\omega+1]/50\nonumber\\
	&&\qquad\qquad\qquad\qquad\text{[by (\ref{equ_theo25_2}) and (\ref{equ_theo25_3})]} \nonumber\\
	&\leq& \delta\pi[\omega+1]/20. \label{equ_theo25_5}
\end{eqnarray}
Finally, (\ref{equ_theo25_5}) and (\ref{equ_theo25_4}) with Lemma \ref{lem_bound_psi} yield 
\begin{align}
\left|\mu_x^{[\omega]}(\varPhi_t,\zeta)- \psi_\zeta(\pi{[\omega+1]})\right| &=  \left|\psi_\zeta(\pi^{[\omega+1]}_x(\varPhi_t,1),\pi^{[\omega+1]}_x(\varPhi_t,-1))-\psi_\zeta(\pi{[\omega+1]})\right| \nonumber\\
&\leq \delta/5 \label{equ_theo_final_1}\\
\left|\mu_x^{[\omega]}(\varPhi_t,0)- \psi_0(\pi{[\omega+1]})\right|& =  \left|\psi_0(\pi^{[\omega+1]}_x(\varPhi_t,1),\pi^{[\omega+1]}_x(\varPhi_t,-1))-\psi_0(\pi{[\omega+1]})\right| \nonumber\\
&\leq \delta\pi[\omega+1]/10 .\label{equ_theo_final_2}
\end{align}
Thus,
\begin{eqnarray*}
\left|\mu_x^{[\omega]}(\varPhi_t,\zeta)- \frac12\br{1-\mu_x^{[\omega]}(\varPhi_t,0)}\right|
&\leq& \left| \psi_\zeta(\pi{[\omega+1]}) - \frac12\br{1-\psi_0(\pi{[\omega+1]})}\right| \\
&& \qquad\qquad\qquad\qquad + \delta/5 + \delta\pi{[\omega+1]}/10 \\
&& \qquad\qquad\qquad\qquad\qquad\text{[by (\ref{equ_theo_final_1}) and (\ref{equ_theo_final_2})]}\\
&\leq& \delta \qquad\text{[by (\ref{pro_psi_3})]},
\end{eqnarray*}
as desired.

\subsection{Proof of Proposition \ref{prop_quasirandom}}
\label{sec_proof_quasi}
Recall from (\ref{equ_def_delta}) that $\delta_t= \exp\br{-c(1-t/n)k}$ and that $\hat t = \br{1-\frac{\ln \rho}{c^2 k}}n$. Suppose that $1\leq t\leq \hat t$. Then $\theta =1-t/n$. Set $\delta=\delta_t=\exp\br{-c\theta k}$ for brevity. Lemma \ref{lem_12} yields
\begin{eqnarray} \label{equ_bound_delta_1015}
	\delta\theta n > 10^{15}\Delta_t.
\end{eqnarray}
To prove Proposition \ref{prop_quasirandom}, we will study two slightly different models of random $k$-CNFs. In the first ``binomial'' model $\vec{\varPhi}_{bin}$, we obtain a $k$-CNF by including each of the $(2n)^k$ possible clauses over $V= \{x_1,\ldots,x_n\}$ with probability $p=m/(2n)^k$ independently, where each clause is an ordered $k$-tuple of not necessarily distinct literals. Thus, $\vec{\varPhi}_{bin}$ is a random set of clauses, and $\Erw\brk{\vec{\varPhi}_{bin}}=m$.

In the second model, we choose a {\it{sequence}} $\vec{\varPhi}_{seq}$ of $m$ independent $k$-clauses \sh{break} $\vec{\varPhi}_{seq}(1),\ldots,\vec{\varPhi}_{seq}(m)$, each of which consists of $k$ independently chosen literals. Thus, the probability of each individual sequence is $(2n)^{-km}$. The sequence $\vec{\varPhi}_{seq}'$ corresponds to the $k$-CNF $\vec{\varPhi}_{seq}(1),\ldots,\vec{\varPhi}_{seq}(m)$ with {\it{at most}} $m$ clauses. The following well-known fact relates $\vec{\varPhi}$ to $\vec{\varPhi}_{bin},\vec{\varPhi}_{seq}$
\begin{fact}\label{fac_40}
For any event $\cE$ we have
\begin{eqnarray}
	\Pr\brk{\vec{\varPhi} \in \cE} &\leq& =O(\sqrt{m})\cdot \Pr\brk{\vec{\varPhi}_{bin} \in \cE}, \\
	\Pr\brk{\vec{\varPhi} \in \cE} &\leq& =O(\sqrt{m})\cdot \Pr\brk{\vec{\varPhi}_{seq} \in \cE}.	
\end{eqnarray}
\end{fact}

Due to Fact \ref{fac_40} and (\ref{equ_bound_delta_1015}), it suffices to prove that the statements \textbf{Q1}-\textbf{Q5} hold for either of $\vec{\varPhi}, \vec{\varPhi}_{bin},\vec{\varPhi}_{seq}$ with probability at least $1-\exp\br{-10^{-13}\delta\theta n}$.

\subsubsection{Establishing Q1}
\label{sec_est_q1}
We are going to deal with the number of variables that appear in ``short'' clauses first.
\begin{lemma}\label{lem_41}
With probability at least $1-\exp(-10^{-6}\delta \theta n)$ in $\vphi^t$ there are no more than $\theta n\cdot 10^{-5}\delta/(\theta k)$  clauses of length less than $0.1\theta k$.
\end{lemma}
\begin{proof}
We are going to work with $\vphi_{bin}$. Let $L_j$ be the number of clauses of length $j$ in $\vphi^t_{bin}$. Then for any $j \in [k]$ we have 
\begin{eqnarray}
	\lambda_j = \Erw\brk{L_j} = m \cdot 2^{j-k} \binom{k}{j}\theta^j(1-\theta)^{k-j} = \frac{2^j\rho \theta n}{j}\binom{k-1}{j-1}\theta^{j-1}1-\theta)^{k-j}.
\end{eqnarray} 
Indeed, a clause has length $j$ in $\vphi_{bin}^t$ iff it contains $j$ variables from the set $V_t$ of size $\theta n$ and $k-j$ variables form $V\setminus V_t$ and none of the $k-j$ variables from $V\setminus V_t$ occurs positively. The total number of possible clauses with these properties is $2^j\binom{k}{j}(\theta n)^j((1-\theta)n)^{k-j}\rho$, and each of them is present in $\vphi_{bin}^t$ with probability $p=m/(2n)^k$ independently.  

Let's start by bounding the total number $L_*=\sum_{j<\theta k/10}L_j$ of ``short'' clauses. It's expectation is bounded by 
\begin{eqnarray}
	\Erw\brk{L_*} &=& \sum_{j<\theta k/10}\lambda_j \leq 2^{0.1\theta k} \rho \theta n\cdot \Pr \brk{\Bin(k-1,\theta) < \theta k/10} \\
	&\leq& 2^{0.1\theta k}\rho \theta n\cdot \exp\br{-\theta k/3} \qquad \text{[by Lemma \ref{Lemma_Chernoff}]} \\
	&\leq& \theta \exp \br{-\theta k/4} n \qquad\text{[as $\theta k\geq \ln(\rho)/c^2$]}.
\end{eqnarray}  
Furthermore, $L_*$ is binomially distributed, because clauses appear independently in $\vphi_{bin}$. Hence again by Lemma \ref{Lemma_Chernoff} we have
\begin{eqnarray}
	\Pr\brk{L_*>\theta n \cdot/(\theta k)} &\leq& \exp\br{-\frac{10^{-5}\delta}{\theta k}\cdot \ln\br{\frac{10^{-5}\delta/(\theta k)}{\exp\br{1-\theta k/4}}}\cdot \theta n} \nonumber\\
	&\leq& \exp\br{-\frac{\delta}{5\cdot 10^5 \theta k}\cdot \theta k\cdot \theta n} \leq \exp\br{-10^{-6}\delta \theta n} \label{equ_41_1}.
\end{eqnarray}
Hence, the assertion follows from (\ref{equ_41_1}) and Fact \ref{fac_40}.
\end{proof}

\begin{corollary}
With probability at least $1-\exp(-10^{-6}\delta\theta n)$ in $\vphi^t$ no more than $10^{-6}\delta\theta n$ variables appear in clauses of length less than $0.1\theta k$.
\end{corollary}
\begin{proof}
This is immediate from Lemma \ref{lem_41}. 
\end{proof}

As a next step, we are going to bound the number of variables that appear in clauses of length $\geq 10\theta k$.

\begin{lemma}\label{lem_43}
With probability at least $1-\exp(-10^{-11}\delta\theta n)$ we have 
\begin{eqnarray}
\sum_{b\in \vphi^t: |N(b)|>10 \theta k} |N(b)| \leq 10^{-6}\delta \theta n.
\end{eqnarray} 
\end{lemma}
\begin{proof}
For a given $\mu > 0$ let $\cL_\mu$ be the event that $\vphi_{seq}^t$ has $\mu$ clauses so that the sum of the lengths of these clauses is at least $\lambda =10\theta k\mu$. Then
\begin{eqnarray}
	\Pr\brk{\cL_\mu} \leq \binom{m}{\mu}\binom{k\mu}{\lambda}\theta^\lambda \br{\frac12 + \theta}^{k\mu - \lambda}.
\end{eqnarray}
Indeed there are $\binom{m}{\mu}$ ways to choose $\mu$ places for these $\mu$ clauses in $\vphi_{seq}$. Once these have been specified, there are $k\mu$ literals that constitute the $\mu$ clauses, and we choose $\lambda$ whose underlying variables are supposed to be in $V_t$; the probability that this is indeed the case for all of these $\lambda$ literals is $\theta^\lambda$. Moreover, in order for each of the clauses to remain in $\vphi_{seq}^t$, the remaining $k\mu-\lambda$ literals must either be negatives of have underlying variables from $V_t$, leading to the $(\theta + 1/2)^{k\mu-\lambda}$ factor. Thus
\begin{eqnarray}
\Pr\brk{\cL_\mu} &\leq& \binom {m}{\mu} \br{\br{1/2+\theta}\br{\frac{e}{5}}^{10\theta}}^{k\mu} \qquad \text{[as $\lambda = 10\theta k\mu$]} \\
&\leq& \br{\frac{en\rho}{k\mu}}^\mu \br{\br{1+2\theta}\br{\frac{e}{5}}^{10\theta}}^{k\mu}\qquad \text{[as $m=n\cdot 2k\rho/k$]}\\
&\leq& \br{\frac{en\rho\theta}{\lambda}\br{\frac{e}{4}}^{10\theta k}}^\mu \\
&=& \br{\br{\frac{10en\rho}{k\mu}}^{1/(10\theta k)}\br{\frac{e}{4}}}^\lambda \qquad\text{[as $\lambda = 10\theta k \mu$]}.
\end{eqnarray} 
Hence, if $\lambda \geq 10^{-6}\delta \theta n$ we get 
\begin{eqnarray}
	\Pr\brk{\cL_\mu}&\leq & \br{\br{\frac{10^7e\rho}{\delta}}^{1/(10\theta k)}\br{\frac{e}{4}}}^\lambda \leq \br{\frac{e}{3}}^\lambda \qquad\text{[as $\theta k \geq \ln(\rho)/c^2$ and $\delta = \exp \br{-c\theta k}$]}\nonumber\\
	&\leq & \exp\br{-10^{-10}\delta \theta n}.
\end{eqnarray}
Thus, we see that $\vphi_{seq}^t$ with probability at least $1-\exp\br{-10^{-10}\delta\theta n}$ we have 
\begin{eqnarray}\label{equ_43_1}
	\sum_{b_\nb > 10\theta k} \nb \leq 10^{-6}\delta\theta n.
\end{eqnarray}
Hence, Fact \ref{fac_40} implies that (\ref{equ_43_1}) holds $\vphi^t$ with probability at least $1-\exp\br{-10^{-11}\delta\theta n}$. 
\end{proof}

\begin{corollary} \label{cor_44}
With probability at least $1-\exp(-10^{-11}\delta \theta n)$ no more than $10^{-6}\delta \theta n$ variables appear in clauses of length greater than $10\theta k$.
\end{corollary}
\begin{proof}
The number of such variables is bounded by $\sum_{b:|N(b)|>10\theta k}|N(b)|$. Therefore, the assertion follows from Lemma \ref{lem_43}
\end{proof}

\begin{lemma}\label{lem_45}
Let $x\in V_t$. The expected number of clauses of length $j $ in $\vphi_{bin}^t$ where $x$ is the underlying variable of the $l$th literal is 
\begin{eqnarray}\label{equ_45_1}
\mu_j=\frac{2^j\rho}{j}\cdot\Pr\brk{\Bin(k-1,\theta)=j-1}.
\end{eqnarray}
\end{lemma}
\begin{proof}
There are $2^j\binom{k}{j}\br{\theta n}^{j-1}\br{(1-\theta)n}^{k-j}$ possible clauses that have exactly $j$ literals whose underlying variable is in $V_t$ such that the underlying variable of the $j$th such literal is $x$. Each such clause is present in $\vphi_{bin}$ with probability $p = m/(2n)^k =  \frac{\rho}{k}n^{1-k}$ independently. 
\end{proof}

\begin{lemma}
With probability at least $1-\exp(-10^{-12}\delta\theta n)$ no more than $10^{-4}\delta \theta n$ variables $x\in V_t$ are such that $\delta(\theta k)^3\sum_{b\in N(x)}2^{-|N(b)|}>1$.
\end{lemma}
\begin{proof}
For $x\in V_t$ let $X_j(x)$ be the number of clauses of length $j$ in $\vphi_{bin}^t$ that contain $x$, and let $X_{jl}(x)$ be te number of such clauses where $x$ is the underlying variable of the $l$th literal of that clause $(1\leq l\leq j)$. Then $\Erw\brk{X_{jl}(x)} = \mu_{j}$, with $\mu_j$ as in (\ref{equ_45_1}). Since $1/\delta = \exp\br{c\theta k}$ and $\theta k \geq \ln(\rho)/c^2$, we see that $2j\delta^{-1}(\theta k)^{-5}/j>100\mu_j$. Hence, Lemma \ref{Lemma_Chernoff} (the Chernoff bound) yields
\begin{eqnarray}
	\Pr\brk{X_{jl}(x)> 10(\mu_j + 2^j\delta^{-1}(\theta k)^{-5}/j)}\leq \zeta, \qquad \text{with } \zeta=\exp\br{-10/(\delta(\theta k)^5)}.
\end{eqnarray}
Let $V_{jl}$ be the set of all variables $x\in V_t$ such that $X_{jl}(x) > 10(\mu_j + 2^j\delta{-1}(\theta k)^{-5}/j)$. Since the random variables $(X_{jl}(x))_{x\in V_t}$ are mutually independent, Lemma \ref{Lemma_Chernoff} yields
\begin{eqnarray}
	\Pr\brk{|V_{jl}|> \frac{\delta}{(\theta k)^9}\cdot \theta n} \leq \exp\br{-\frac{\delta \theta n}{(\theta k)^9}\cdot\ln\br{\frac{\delta}{e(\theta k)^9\zeta}}}.
\end{eqnarray}
Since $\zeta^{-1} = \exp\br{10/(\delta(\theta k)^5)}= \exp\br{10\exp(c\theta k)/(\theta k)^5}$ and $\theta k \geq \ln(\rho)/c^2 \gg1$, we have
\begin{eqnarray}
	\ln\br{\frac{\delta}{e(\theta k)^9\zeta}}\geq -\ln(\zeta)/2,
\end{eqnarray}
whence
\begin{eqnarray}\label{equ_46_1}
		\Pr\brk{|V_{jl}|> \frac{\delta}{(\theta k)^9}\cdot \theta n} \leq \exp\br{\frac{\delta \theta n}{2(\theta k)^9}\cdot\ln\zeta}\leq \exp\br{-\frac{\theta n}{(\theta k)^{15}}}\leq \exp\br{-\delta\theta n}.
\end{eqnarray}
Furthermore, if $x\notin V_{jl}$ for all $1\leq l\leq 10\theta k$ and all $1\leq l\leq j$, then 
\begin{eqnarray}
	\sum_{b\in N(x):\nb\leq 10 \theta k} 2{-\nb}&\leq& 10\sum_{j\leq 10\theta k} 2^{-j}(j\mu_j + 2^j\delta^{-1}(\theta k)^{-5}) \\
	&\leq & 100\delta^{-1}(\theta k)^{-4} + 10 \sum_{j\leq 10\theta k} j2^{-j}\mu_j \\
	&\leq& 100\delta^{-1}(\theta k)^{-4} + 10 \rho < \delta^{-1}(\theta k)^{-3},
\end{eqnarray} 
where we used that $\theta k \geq \ln(\rho)/c^2$, so that $1/\delta\geq (\theta k)^5\rho$. Hence, the assertion follows from (\ref{equ_46_1}), Fact \ref{fac_40} and the bound on the number of variables in clauses of length $>10\theta k$ provided by Corollary \ref{cor_44}.
\end{proof}

\subsubsection{Establishing Q2}
\label{sec_est_q2}

Let $T\subset V_t$ be a set of size $|T|\leq s\theta n$ for some $\delta^5\leq s\leq 10\delta$. For a variable $x$ we let $\cQ(x,i,j,l,T)$ be the number of clauses $b$ of $\vphi_{bin}^t$ such that the $i$th literal is either $x$ or $\neg x$, $|N(b)| = j $, and $|N(b)\cap T\setminus \{x\}|= l$. Let $\mu_{j,l}(T) = \sum_{i = 1}^j\Erw\brk{\cQ(x,i,j,l,T)} = j\cdot\Erw\brk{\cQ(x,i,j,l,T)}$.

\begin{lemma} \label{lem_erw_Q}
For all $x\in V_t$ we have 
\begin{eqnarray}
\Erw\brk{\cQ(x,i,j,l,T)} &=& \frac{2^j\rho}{j}\cdot \Pr\brk{\Bin(k-1,\theta)=j-1} \cdot \Pr\brk{\Bin(j-1,|T|/(\theta n))=l}\nonumber\\
&=& \mu_j\cdot \Pr\brk{\Bin(j-1,|T|/(\theta n))=l} .
\end{eqnarray}
\end{lemma}
\begin{proof}
Let $\nu=\frac{|T|}{\theta n}$. There are 
\begin{align*}
&2^j\binom{k}{j}\binom{j-1}{l}\br{(1-\nu)\theta n}^{j-1-l}(\nu\theta n)^l\br{(1-\theta)n}^{k-j}\\
&\qquad=2^j\binom{k}{j}\binom{j-1}{l}\br{1-\nu}^{j-1-l}(\nu)^l(\theta n)^{j-1}\br{(1-\theta)n}^{k-j} 
\end{align*}
possible clauses that have exactly $j-l$ literals whose underlying variable is in $V_t\setminus T$ and $l$ literals whose underlying variable is in $T$ such that the underlying variable of the $j$th such literal is $x$. Each such clause is present in $\vphi_{bin}$ with probability $p = m/(2n)^k =  \frac{\rho}{k}n^{1-k}$ independently.
\end{proof}

\begin{lemma}\label{lem_bound_clauses_l}
Suppose that $l\geq0, j-l>k_1$ and $0.1\theta k\leq j\leq 10\theta k$. Let 
\begin{eqnarray}
m(\theta, j) &=& \max \{(\theta k)^{-1}, \Pr\brk{\Bin(k-1,\theta)=j-1}\} \qquad\text{and}\\
\gamma_{i,l}(s) &=& \begin{cases}
		10\cdot2^j\rho m(\theta, j)/j & \text{if } l=0 \\
		10\cdot2^js\rho m(\theta, j) & \text{if } l=1 \\
		10\cdot2^{j-l}s^{1.9}&\text{if } l\geq2.
\end{cases}
\end{eqnarray}
Then for any $i,x,T$ we have $\Pr\brk{\cQ(x,i,j,l,T)>\gamma_{i,l}(s)} \leq \exp\br{-\exp\br{c^{2/3}\theta k}}$.
\end{lemma}
\begin{proof}
The random variable $\cQ(x,i,j,l,T)$ has a binomial distribution, because clauses appear independently in $\vphi_ {bin}$. By Lemma \ref{lem_erw_Q} we have for $l > 1$
\begin{eqnarray}
\Erw \brk{\cQ(x,i,j,l,T)} \leq \binom{j}{l}\delta^l \mu_j \leq \rho \binom{j}{l}s^l2^j \leq 2^{j-l}s^{1.9};
\end{eqnarray}
in the last step we used that $s^{0.05} \leq \delta^{0.05}\leq 1/\rho$, which follows from our assumption that $\theta k \leq \ln(\rho)/c^2$, and that $2^j\binom{j}{l} \leq (2j)^l \leq (20\theta k)^l \leq s^{0.02l}$. Hence by Lemma \ref{Lemma_Chernoff} in the case $j - l > k_1 = \sqrt c \theta k$, we get
\begin{eqnarray*}
\Pr\brk{\cQ(x,i,j,l,T) > 10\cdot 2^{j-l}s^{1.9}} &\leq& \exp\br{-2^{j-l}s^{1.9}} \leq \exp\br{-2^{k_1}s^{1.9}} \\
&\leq& \exp\br{-\exp\br{c^{2/3}\theta k}}, 
\end{eqnarray*}
as $\delta = \exp\br{-c\theta k}$ and $s \geq \delta^5$.

By a similar token, in the case $l = 1$ we have 
\begin{eqnarray}
\Erw \brk{\cQ(x,i,j,l,T)} \leq js \mu_j = \rho s2^j \Pr\brk{\Bin(k-1,\theta) = j-1}.
\end{eqnarray}
Hence, once more by the Chernoff bound
\begin{eqnarray*}
\Pr\brk{\cQ(x,i,j,l,T) > 10 \cdot 2^js \rho m(\theta, j)} &\leq& \exp\br{-2^js\rho m(\theta,j)}\leq \exp\br{-2^{k_1}s/(\theta k)}\\
&\leq& \exp\br{-\exp\br{c^{2/3}\theta k} },
\end{eqnarray*}
as claimed.

Finally, analogously in the case $l = 0$ we have
\begin{eqnarray}
 \Erw \brk{\cQ(x,i,j,l,T)} \leq \mu_j =  \frac{2^j\rho}{j}\Pr\brk{\Bin(k-1,\theta) = j-1}.
\end{eqnarray}
Thus, applying the Chernoff bound yields
 \begin{eqnarray*}
 \Pr\brk{\cQ(x,i,j,l,T) > 10 \cdot 2^j \rho m(\theta, j)/j } &\leq& \exp\br{-2^j\rho m(\theta, j)/j}\leq \exp\br{-0.1\cdot2^{0.1\theta k}/(\theta k)^2}\\
 &\leq& \exp\br{-\exp\br{c^{2/3}\theta k} }
 \end{eqnarray*}
 as claimed.
\end{proof}

Let $\cZ(i,j,l.T)$ be the number of variables $x\in V_t$ for which $\cQ(x,i,j,l,T) > \gamma_{j,l}(s)$.

\begin{lemma}\label{lem_prob_Z}
Suppose that $l\geq 1, j-l > k_1$ and $0.1\theta k \leq j \leq 10\theta k$. Then for any $i,T$ we have 
\begin{eqnarray}
\Pr\brk{\cZ(i,j,l,T) > \delta^2/(\theta k)^4} \leq \exp\br{-\frac{\delta^2\theta n}{2(\theta k)^4}\cdot\exp\br{c^{2/3}\theta k}}
\end{eqnarray}
\end{lemma} 
\begin{proof}
Whether a variable $x\in V_t$ contributes to $\cZ(i,j,l,T)$ depends only on those clauses of $\vphi^t_{bin}$ whose $i$th literal reads either $x$ or $\neg x$. Since these sets of clauses are disjoint for distinct variables and as clauses appear independently in $\vphi^t_{bin}$, $\cZ(i,j,l,T)$ is a binomial random variable. By Lemma~\ref{lem_bound_clauses_l},
\begin{eqnarray}
\Erw\brk{\cZ(i,j,l,T)} \leq \theta n\exp\br{-\exp\br{c^{2/3}\theta k}}.
\end{eqnarray}
Hence, Lemma \ref{Lemma_Chernoff} yields 
\begin{eqnarray*}
	\Pr\brk{\cZ(i,j,l,T) >\delta \theta n/(\theta k)^4} &\leq& \exp\br{-\frac{\delta^2 \theta n}{2(\theta k)^4}\ln\br{\frac{\delta}{(\theta k)^4\exp\br{1-\exp(c^{2/3}\theta k)}}}}\\
	&\leq& \exp\br{-\frac{\delta^2 \theta n}{2(\theta k)^4}\exp(c^{2/3}\theta k)} ,
\end{eqnarray*}
as desired.
\end{proof}

\begin{corollary}\label{cor_49}
With probability $1-\exp(-\delta\theta n)$ the random formula $\vphi^t_{bin}$ has the following property.
\begin{equation}\label{equ_cor49_a}
\parbox{12cm}{
For all $i,j,l,T$ such that $l \geq 1, j- l > k_1, 0.1\theta k\leq j\leq 10 \theta k$ and $|T| \leq \delta \theta n$ we have $\cZ(i,j,l,T)\leq \delta^2\theta n/(\theta k)^4$.}
\end{equation}
\end{corollary}
\begin{proof}
We apply the union bound. There are at most $n\binom{n}{\delta \theta n}$ ways to choose the set $T$, and no more than $n$ ways to choose $i,j,l$. Hence, by Lemma \ref{lem_prob_Z} the probability that there exist $i,j,T$ such that $\cZ(i,j,l,T)>\theta n \exp\br{-\exp\br{c^{2/3}\theta k}}$ is bounded by
\begin{eqnarray*}
n^2\binom{n}{\delta \theta n}\exp\br{-\frac{\delta^2 \theta n}{2(\theta k)^4}\exp(c^{2/3}\theta k)} &\leq& \exp\br{O(n)-\delta \theta n\ln(\delta \theta)-\delta \theta n\exp\br{c^{3/4}\theta k}}\\
&\leq& \exp\br{\delta \theta n \br{O(1)-\ln(\delta \theta)-\exp\br{c^{3/4}\theta k}}} \\
&\leq& \exp\br{-\delta\theta n},\\
&&\quad\quad\quad\text{[as $\theta k \geq \ln(\rho)/c^2$ and $\delta=\exp\br{-c\theta k}$]}
\end{eqnarray*} 
as claimed.
\end{proof}

\begin{corollary}
With probability $1-\exp(-10^{-12}\delta \theta n)$ the random formula $\vphi^t$ has the following property. 
\begin{equation}\label{prop_s_less_Lower}
\parbox{12cm}{
If $T \subset V_t$ has size $|T|\leq s\theta n$ for some $\delta^5\leq s\leq 10 \delta$, then for all but $10^{-4}\delta^2\theta n$ variables $x\in V_t$ we have}
\end{equation}
\begin{eqnarray*}
\sum_{\cN_\les(x,T,\zeta)}  2^{-|N(b)|} < 10^{4}\rho  \quad\text{and}\quad \sum_{\cN_1(x,T,\zeta)}  2^{-|N(b)|} < s \rho \theta k
\end{eqnarray*}
\end{corollary}
\begin{proof}
Given $T\subset V_t$ of size $|T|\leq s\theta n$ for some $\delta^5\leq s\leq 10\delta$, let $\cV_T$ be the set of all variables $x$ with the following property.
\begin{equation}\label{prop_s_less_Lower_1}
\parbox{12cm}{
For all $1\leq i\leq j, 1\leq l\leq j-k_1$, and $0.1\theta k\leq j\leq 10\theta k$ we have $\cQ(x,i,j,l,T)\leq \gamma_{j,l}(s)$.}
\end{equation}
Let 
\begin{eqnarray*}
J_> &=& \{j \in \mathbb{N}: 0.1\theta k\leq j \leq 10\theta k \text{ and } m(\theta,j)=\Pr\brk{\Bin(k-1,\theta)=j-1}\}\} \\
J_\leq &=& \{j \in \mathbb{N}: 0.1\theta k\leq j \leq 10\theta k \text{ and } m(\theta,j)=(\theta k )^{-1}\}.
\end{eqnarray*}
Then for all $x\in \cV_t$ we have
\begin{eqnarray*}
\sum_{\cN_\les(x,T,\zeta)}  2^{-|N(b)|} &=& \sum_{0.1\theta k\leq j\leq 10\theta k} \sum_{i=1}^j (\cQ(x,i,j,0,T)+\cQ(x,i,j,1,T))2^{-j} \\
&\leq& \sum_{0.1\theta k\leq j\leq 10\theta k} \sum_{i=1}^j 10\cdot(j^{-1} + s)\rho m(\theta,j) \qquad\text{[due to i.]}\\
&\leq& \sum_{0.1\theta k\leq j\leq 10\theta k}10 \cdot 10\theta k \cdot 2\cdot(0.1\theta k)^{-1} \rho m(\theta,j) \\
&&\qquad\qquad\text{[as $0.1\theta k\leq j\leq 10\theta k$ and $s\leq \exp\br{-c\theta k}$]}\\
&\leq& \sum_{j\in J_>} 200\rho \Pr\brk{\Bin(k-1,\theta)=j-1}\} +  \sum_{j\in J_\leq} 200\rho (\theta k)^{-1} \\
&\leq& 200\rho + 2000\rho \qquad\qquad\text{[as $|J_\leq| \leq 10\theta k$]} \\
&\leq& 10^4\rho.
\end{eqnarray*}
Similarly,
\begin{eqnarray*}
\sum_{\cN_1(x,T,\zeta)}  2^{-|N(b)|} &=& \sum_{0.1\theta k\leq j\leq 10\theta k} \sum_{i=1}^j \cQ(x,i,j,1,T))2^{-j} \\
&\leq& 10\theta k\sum_{0.1\theta k\leq j\leq 10\theta k} 10\rho sm(\theta,j) \qquad\text{[due to i.]}\\
&\leq& 10\theta k\sum_{j\in J_>} 20\rho s \Pr\brk{\Bin(k-1,\theta)=j-1}\} +  10\theta k\sum_{j\in J_\leq} 20\rho s(\theta k)^{-1} \\
&\leq& 200\theta k s\rho + 2000\theta k \rho s \qquad\qquad\text{[as $|J_\leq| \leq 10\theta k$]} \\
&\leq& 10^4\theta k \rho s.
\end{eqnarray*}

Thus to complete the proof we need to show that with sufficiently high probability $\cV_t$ is sufficiently big for all $T$. By Corollary \ref{cor_49} and Fact \ref{fac_40} with probability $\geq 1- \exp\br{-\delta \theta n/2}$ the random formula $\vphi^t$ satisfies (\ref{equ_cor49_a}). In this case, for all $T$ the number of variables that fail to satisfy (\ref{prop_s_less_Lower_1}) is bounded by $\delta\theta n/(\theta k)^4 <10^{-5}\delta\theta n$. Thus, with probability $\geq 1-\exp\br{-10^{-12}\delta\theta n}$ we have $|\cV_t| > \theta n(1-10^{-4}\delta)$ for all $T$, as desired. 
\end{proof}

For a set $T\subset V_t$ and numbers $i\leq j$ we let $\cN_{\leq1}(x,i,j,T,\zeta)$ be the number of clauses $b \in N(x,\zeta)$ in $\vphi^t_{bin}$ such that $\nb = j$, the underlying variable of the $i$th literal of $b$ is $x$ such that $\sign(x)=\zeta$ and $|N(b) \cap T\setminus \{x\}| \leq 1$. Let $\mu_{j,\les}(T) = \mu_{j,0}(T) + \mu_{j,1}(T)$ and  $\cB(i,j,T)$ be the set of variables such that for at least one $\zeta \in \{-1,1\}$ we have 
\begin{eqnarray}
|\cN_{\leq1}(x,i,j,T,\zeta)-\mu_{j,\les}(T)/(2j)| >2^j\delta(\theta k)^{-3}
\end{eqnarray}

\begin{lemma}\label{lem_51}
Let $T\subset V_t$ be a set of size $|T|\leq \delta \theta n$. Let $i,j$ be such that $i\leq j$ and $0.1\theta k\leq j\leq 10\theta k$. Then in $\vphi^t_{bin}$ we have $\Pr\brk{\cB(i,j,T) > \delta^2 \theta n /(\theta k)^3}\leq \exp\br{-\delta^2 \theta n\exp\br{\theta k/22}}$. 
\end{lemma}
\begin{proof}
Let $x\in V_t$. In the random formula $\vphi^t_{bin}$ we have $\cN_\les(x,i,j,T,1) + \cN_\les(x,i,j,T,-1) = \cQ(x,i,j,T,0) + \cQ(x,i,j,T,1)$. Furthermore, $\cN_\les(x,i,j,T,1)$ and $\cN_\les(x,i,j,T,-1)$ are binomially distributed with identical means, because in $\vphi^t_{bin}$ each literal is positive/negative with probability $\frac12$. By Lemma \ref{lem_erw_Q} we have 
\begin{eqnarray*}
\Erw\brk{\cN_\les(x,i,j,T,\zeta)} &=&\frac12 \cdot \Erw\brk{\cQ(x,i,j,T,0) + \cQ(x,i,j,T,1)} \\
&=&  \frac{1}{2j}(\mu_{j,0}(T) + \mu_{j,1}(T)) \\
&=& \mu_{j,\les}(T)/(2j)\leq \mu_j \leq 2^j\rho =\bar{\mu}_j \qquad \text{[by (\ref{equ_45_1})]}.
\end{eqnarray*}
Let $\eta_j = 2^j\delta/(\theta k)^3$. Hence, Lemma \ref{Lemma_Chernoff} (the Chernoff bound) yields
\begin{eqnarray*}
&&\pr\brk{\cN_\les(x,i,j,T,\zeta)>\mu_{j,\les}(T)/(2j) + \eta_j} \\
&&\qquad\leq \exp\br{-\frac{\eta_j}{2j}\br{\br{1+\frac{2j\eta_j}{\mu_{j,\les}(T)}}\ln\br{1+\frac{\mu_{j,\les}(T)}{2j\eta_j}}-1}}\\
&&\qquad\leq \exp\br{-\frac{\eta_j^2}{6j\bar\mu_j}} \qquad \text{[as $\eta_j/\bar\mu_{j} = o_k(1)$]} \\
&&\pr\brk{\cN_\les(x,i,j,T,\zeta)<\mu_{j,\les}(T)/(2j) - \eta_j}\\
&&\qquad\leq \exp\br{-\frac{\eta_j}{2j}\br{\br{1-\frac{2j\eta_j}{\mu_{j,\les}(T)}}\ln\br{1-\frac{\mu_{j,\les}(T)}{2j\eta_j}}+1}}\\
&&\qquad\leq \exp\br{-\frac{\eta_j^2}{6j\bar\mu_j}} \qquad \text{[as $\eta_j/\bar\mu_{j} = o_k(1)$]}
\end{eqnarray*}
and thus, for $j\geq 0.1\theta k$
\begin{align}
\pr\brk{|\cN_\les(x,i,j,T,\zeta) - \mu_{j,\les}(T)/(2j)|> \eta_j}&\leq \exp\br{-\frac{\eta_j^2}{7j\bar\mu_j}} \nonumber\\
&\leq \exp\br{-\frac{2^j\delta^2}{70(\theta k)^7\rho}} \qquad \text{[as $ j\leq 10\theta k$]} \nonumber \\
&\leq \exp\br{-\exp\br{\theta k/20}}\label{equ_lem51_1} \\
&\qquad \quad \text{[as $\delta = \exp\br{-c\theta k}, j\geq 0.1\theta k$]}.\nonumber 
\end{align}
For different $x\in V_t$ the random variables $\cN_\les(x,i,j,T,\zeta)$ are independent (because we fix the position $i$ where $x$ occurs). Hence, $\cB(i,j,T)$ is a binomial random variable, and (\ref{equ_lem51_1}) yields
\begin{eqnarray}
\Erw\brk{\cB(i,j,T)} \leq \theta n \exp\br{-\exp\br{\theta k/20)}}.
\end{eqnarray}
Consequently, Lemma \ref{Lemma_Chernoff} (the Chernoff bound) gives 
\begin{eqnarray*}
\pr\brk{\cB(i,j,T) > \delta^2 \theta n/(\theta k)^3} &\leq& \exp\br{-\frac{\delta^2\theta n}{(\theta k)^3}\ln\br{\frac{\delta^2\theta n/(\theta k)^3}{\exp\br{1-\exp\br{\theta k/20}}\theta n}}} \\
&\leq& \exp \br{-\frac{\delta^2\theta n}{(\theta k)^3}\cdot\exp\br{\theta k/21}} \\
&\leq& \exp\br{-\delta^2\theta n \exp\br{\theta k/22}}
\end{eqnarray*}
as claimed.
\end{proof}

\begin{corollary}\label{cor_52}
With probability $\geq 1-\exp\br{-\delta \theta n}$ the random formula $\vphi_{bin}^t$ has the following property.
\begin{equation}\label{stat_cor_52}
\parbox{12cm}{
For all $T \subset V_t$ of size $|T|\leq \delta \theta n$ and all $i, j$ such that $i\leq j,0.1\theta k\leq j\leq  10\theta k$ we have $\cB(i,j,T)\leq \delta \theta n/(\theta k)^3$}
\end{equation}
\end{corollary}
\begin{proof}
Let $i,j$ be such that $i\leq j, 0.1\theta k\leq j\leq 10\theta k$. By Lemma \ref{lem_51} and the union bound, the probability that there is a set $T$ such that $\cB(i,j,T)>\delta\theta n/(\theta k)^3$ is bounded by
\begin{eqnarray*}
n\binom{\theta n}{\delta\theta n} \exp\br{-\delta \theta n\exp\br{\theta k/22}} &\leq& \exp\br{o(n)+\delta \theta n(1-\ln(\theta \delta)-\exp\br{\theta k/22})} \\
&\leq& \exp\br{-2\delta \theta  n} \qquad\text{[as $\delta = \exp\br{-c\theta k}$]}.
\end{eqnarray*}
Since there are no more than $(10\theta k)^2$ ways to choose $i,j$, the assertion follows. 
\end{proof}

\begin{corollary}\label{cor_53}
With probability $\leq 1-\exp\br{-10^{-12}\delta \theta n}$ the random formula $\vphi^t$ has the following property.
\begin{equation}\label{stat_prop_52}
\parbox{12cm}{
If $T \subset V_t$ has size $|T|\leq \delta \theta n$ and $p \in (0,1]$, then there are no more than $10^{-5}\delta^2 \theta n$ variables $x\in V_t$ such that}
\end{equation}
\begin{eqnarray}
\left|\Pi(T,p) - \sum_{b\in \cN_\les(x,T,\zeta)} \br{2/\tau(p)}^{-\nb} \right|> \delta/1000
\end{eqnarray}
\end{corollary}
\begin{proof}
Given $T\subset V_t$, let $\cV(T,\zeta)$ be the set of all $x\in V_t$ such that for all $1\leq i\leq j \leq 10\theta k$ and we have
\begin{align}\label{equ_cor53_5}
|\cN_\les(x,i,j,T,\zeta)-\mu_{j,\les}(T)/(2j)|\leq 2^j\delta/(\theta k)^3.
\end{align}
Then for all $x\in \cV(T,\zeta)$ we have 
\begin{align*}
&\hspace{-0.5cm}\left|\Pi(T,p) - \sum_{b\in \cN_\les(x,T,\zeta)} \br{2/\tau(p)}^{-\nb}\right|\\
&=\left|\sum_{0.1\theta k\leq j\leq 10 \theta k}(2/\tau(p))^{-j} \brk{ \mu_{j,\les}(T)/2 - \sum_{i=1}^{j} \cN_\les(x,i,j,T,\zeta)}\right| \\
&\leq\sum_{0.1\theta k\leq j\leq 10 \theta k}(2/\tau(p))^{-j} \sum_{i=1}^{j}\left|\mu_{j,\les}(T)/(2j) - \cN_\les(x,i,j,T,\zeta)\right| \\
&\leq\sum_{0.1\theta k\leq j\leq 10 \theta k}(2/\tau(p))^{-j} \cdot 2^j\delta/(\theta k)^3 \qquad\qquad\text{[by (\ref{equ_cor53_5})]}\\
&\leq 100\delta/(\theta k) \qquad\qquad\text{[as $\tau(p) \in (0,1]$]}\\
&\leq \delta/1000. 
\end{align*}
By Corollary \ref{cor_52} and Fact \ref{fac_40} with probability $\geq 1- \exp\br{-\delta\theta n/2}$ the number of variables not in $\cV(T,\zeta)$ for at least one $\zeta \in \{-1,1\} $ is bounded by $10^{-5} \delta \theta n$ for all $T$, as claimed.
\end{proof}

\subsubsection{Establishing Q3}
\label{sec_est_q3}

\begin{corollary}\label{cor_50}
With probability $1-\exp(-10^{-12}\delta \theta n)$ the random formula $\Phi^t$ has the following property. 
\begin{equation}\label{equ_cor50}
\parbox{12cm}{
If $T \subset V_t$ has size $|T|=\delta\theta n$, then for all but $10^{-4}\delta\theta n$ variables $x$ we have}
\end{equation}
\begin{align*}
\sum_{N_{>1}(x,T,\zeta)}  2^{|N(b)\cap T\setminus \{x\}|-|N(b)|} < \delta/(\theta k) 
\end{align*}
\end{corollary}
\begin{proof}
Given $T\subset V_t$ of size $|T|\leq\delta \theta n$, let $\cV_T$ be the set of all variables $x$ with the following two properties.
\begin{enumerate}
\item[i.] For all $b \in N(x,\zeta)$ we have $0.1\theta k\leq |N(b)|\leq10\theta k$.
\item[ii.] For all $1\leq i\leq j, 1\leq l\leq j-k_1$, and $0.1\theta k\leq j\leq 10\theta k$ we have $\cQ(x,i,j,l,T)\leq \gamma_{j,l}(\delta)$.
\end{enumerate}
Then for all $x\in \cV_t$ we have
\begin{eqnarray*}
\sum_{N_{>1}(x,T,\zeta)}  2^{|N(b)\cap T\setminus \{x\}|-|N(b)|} &=& \sum_{0.1\theta k\leq j\leq 10\theta k} \sum_{i=1}^j\sum_{l=2}^{j-k_1} \cQ(x,i,j,l,T)2^{l-j} \qquad\text{[due to i.]}\\
&\leq& 10\theta k\sum_{0.1\theta k\leq j\leq 10\theta k} \sum_{l=2}^{j-k_1} \gamma_{j,l}(\delta)2^{l-j} \qquad\text{[due to ii.]}\\
&\leq& 1000(\theta k)^2\delta^{1.9} < \delta/(\theta k) \qquad\qquad\text{[as $ \delta = \exp\br{-c\theta k}$]}
\end{eqnarray*}
Thus to complete the proof we need to show that with sufficiently high probability $\cV_t$ is sufficiently big for all $T$. By Lemma \ref{lem_41} and \ref{lem_43} with probability $1-2\exp\br{-10^{-11}\delta\theta n}$ the number of variables $x$ that fail to satisfy i. is less than $2\cdot 10^{-6}\delta \theta n$. Furthermore, by Corollary \ref{cor_49} and Fact \ref{fac_40} with probability $\geq 1- \exp\br{-\delta \theta n/2}$ the random formula $\vphi^t$ satisfies (\ref{equ_cor49}). In this case, for all $T$ the number of variables that fail to satisfy ii. is bounded by $\delta\theta n/(\theta k)^4 <10^{-5}\delta\theta n$. Thus, with probability $\geq 1-\exp\br{-10^{-12}\delta\theta n}$ we have $|\cV_t| > \theta n(1-10^{-4}\delta)$ for all $T$, as desired. 
\end{proof}

\subsubsection{Establishing Q4}
\label{sec_est_q4}

We carry the proof out in the model $\vphi_{seq}$. Let $0.01\leq z \leq 1$ and let $T$ be a set of size $|T| = q\theta n$ with $q \leq 100\delta$. 
\begin{lemma}\label{lem_54}
Let $S,Z >0$ be integers and let $\cE_z(T,S,Z)$ be the event that $\vphi_{seq}^t$ contains a set $\cZ$ of $Z$ clauses with the following properties. 
\begin{enumerate}
\item[i.] $S=\sum_{b\in \cZ} \nb>1.009|T|/z$,
\item[ii.] For all $b\in \cZ$ we have $0.1\theta k\leq \nb\leq 10\theta k$,
\item[iii.] All $b\in \cZ$ satisfy $|N(b) \cap T| \geq z\nb$. 
\end{enumerate}
Then $\Pr\brk{\cE_z(T,S,Z)}\leq q^{0.99999zS}$.
\end{lemma}
\begin{proof}
We claim that in $\vphi_{seq}^t$, 
\begin{eqnarray}
	\Pr\brk{\cE_z(T,S,Z)} \leq \binom {m}{Z}\binom{kZ}{S}\binom{S}{zS} 2^{S-kZ}\theta^S(1-\theta)^{kZ-S}q^{zS}.
\end{eqnarray}
Indeed, $\vphi_{seq}^t$ is based on the random sequence $\vphi_{seq}$ of $m$ independent clauses. Out of these $m$ clauses we choose a subset $\cZ$ of size $Z$, inducing a $\binom {m}{Z}$ factor. Then out of the $kZ$ literal occurrences of the clauses in $\cZ$ we choose $S$ (leading to the $\binom{kZ}{S}$ factor) whose underlying variables lie in $V_t$, which occurs with probability $\theta = |V_t|/n$ independently for each literal (inducing a $\theta^S$ factor).  Furthermore, all $kZ-S$ literals whose variables are in $V\setminus V_t$ must be negative, because otherwise the corresponding clauses woud have been eliminated form $\vphi_{seq}^t$; this explains the $2^{S-kZ}(1-\theta)^{kZ-S}$ factor. Finally, out of the $S$ literal occurrences in $V_t$ a total of at least $zS$ has an underlying variable from $T$ (a factor of $\binom{S}{zS}$), which occurs with probability $q=|T|/(\theta n)$ independently (hence the $q^{zS}$ factor). 

Hence we obtain
\begin{eqnarray}
	\Pr\brk{\cE_z(T,S,Z)} &\leq& \binom {m}{Z} 2^{-kZ} \br{2^{1/z}\cdot \frac{e}{z}\cdot q}^{zS} \cdot\binom{kZ}{S} \theta ^S(1-\theta)^{kZ-S} \nonumber\\ 
	&\leq& \binom {m}{Z} 2^{-kZ} \br{2^{1/z}\cdot \frac{e}{z}\cdot q}^{zS}\leq	\binom {m}{Z} 2^{-kZ}\br{Cq}^{zS}\label{equ_54_3}
\end{eqnarray}
for a certain absolute constant $C>0$, because $z \geq 0.01$. Since all clauses lengths are required to be between $0.1\theta k$ and $10 \theta k$, we obtain $0.1S/(\theta k)\leq Z\leq 10S/(\theta k)$. Therefore,
\begin{eqnarray}
	\binom {m}{Z} 2^{-kZ} &\leq& \br{\frac{em}{2^kZ}}^Z\leq  \br{\frac{e\rho n}{kZ}}^Z \qquad \text{[as $m=2^k\rho n/k$]} \nonumber\\
	&\leq& \br{\frac{10e\rho \theta n}{S}}^Z \nonumber \\
	&\leq& \br{\frac{10e\rho}{1.009q}}^Z \qquad \text{[as $S\geq 1.009q\theta n/z\geq 1.009q\theta n$ by i.]}. \label{equ_54_1}
\end{eqnarray}
Since $q\leq 100\delta = 100 \exp\br{-c\theta k}$ and $\theta k\geq \ln(\rho)/c^2$, we have $1/q\geq 100\rho$.
Hence, (\ref{equ_54_1}) yields
\begin{eqnarray}
	\binom {m}{Z} 2^{-kZ} &\leq& q^{-2Z}\leq q^{-20S/(\theta k)} \label{equ_54_2}.
\end{eqnarray}
Plugging (\ref{equ_54_2}) into (\ref{equ_54_3}), we obtain for $\theta k \leq \ln(\rho)/c^2$ and $S\geq 1.009|T|/z$
\begin{eqnarray}
	\Pr\brk{\cE_z(T,S,Z)} \leq q^{-20S/(\theta k)}\cdot \br{Cq}^{zS} \leq q^{0.99999zS},
\end{eqnarray}
as claimed.
\end{proof}

\begin{corollary} \label{cor_55}
Let $\cE$ be the event that there exist a number $0.01\leq z\leq 1$, a set $T\subset V_t$ of size $|T| \leq 100\delta \theta n$ and $S\geq \frac{1.01}{z}|T| + 10^{-6}\delta \theta n, Z>0$ such that $\cE_z(T,S,Z)$ occurs. Then $\cE$ occurs in $\vphi^t$ with probability $\leq \exp\br{-10^{-7}\delta\theta n}$. 
\end{corollary}
\begin{proof}
Let $0.01\leq z\leq 1$ and $0<q\leq 100\delta$. Let $s,Z >0$ be integers such that $S\geq \frac{1.01}{z}q\theta n + 10^{-6}\delta\theta n$. Let $\cE_z(q,S,Z)$ denote the event that tere is a set $T\subset V_t$ of size $|T| = q\theta n$ such that $\cE_z(T,S,Z)$ occurs. Then by Lemma \ref{lem_54} and the union bound, in $\vphi_{seq}^t$ we have 
\begin{eqnarray}
	\Pr\brk{\cE(q,S,Z)}&\leq& \binom{\theta n}{q\theta n} q^{0.99999zS} \leq \exp\br{q\theta n(1-\ln q + 1.008\ln q) + 0.9\cdot 10^{-6}\delta \theta n\ln q}\nonumber\\
	&\leq& \exp\br{-0.9 \cdot10^{-6}\delta \theta n} \qquad\qquad\text{[as $q\leq 100\delta<1/e$]}. \label{equ_55_1}
\end{eqnarray}
Since there are only $O(n^4)$ possible choices of $S,Z,z$ and $q$, (\ref{equ_55_1}) and Fact \ref{fac_40} imply the assertion. 
\end{proof}

\begin{corollary}
With probability at least $1-\exp\br{-10^{-12}\delta\theta n}$, $\vphi^t$ has the following property.
\begin{equation}\label{equ_cor56}
\parbox{12cm}{
Let $0.01 \leq z\leq 1$ and let $T\subset V_t$ have size $0.01\delta\theta n\leq |T|\leq 100\delta \theta n$. Then 
\begin{eqnarray*}
	\sum_{b:|N(b)\cap T|\geq z\nb}\nb \leq \frac{1.01}{z}|T| + 2\cdot10^{-5}\delta\theta n.
\end{eqnarray*}
}
\end{equation}
\end{corollary}
\begin{proof}
Lemmas \ref{lem_41} and \ref{lem_43} and Corollary \ref{cor_55} imply that with probability at least $1-\exp\br{-10^{-11}\delta \theta n}, \vphi^t$ has the following properties.
\begin{enumerate}
\item[i.] $\cE$ does not occur.
\item[ii.] $\sum_{b:\nb\notin [0.1\theta k,10\theta k]}\nb \leq 10^{-5}\delta\theta n$.
\end{enumerate}
Assume that i. and ii. hold and let $T\subset V_t$ be a set of size $|T| \leq 100\delta \theta n$. Let $0.01\leq z\leq 1$. Let $\cN_T$ be the set of all clauses $b$ of $\vphi^t$ such that $|N(b) \cap T|\geq z\nb$ and $0.1\theta k\leq\nb\leq 10\theta k$. Then i. implies that 
\begin{eqnarray}
	\sum_{b\in\cN_T} \nb \leq \frac{1.009}{z} |T| + 10^{-6}\delta \theta n.
\end{eqnarray}
Furthermore, ii. yields 
\begin{eqnarray*}
	\sum_{b:|N(b)\cap T| \geq z\nb} \nb &\leq&  \sum_{b:\nb\notin [0.1 \theta k,10\theta k]}\nb + \sum_{b\in \cN_T}\nb\\
	&\leq & 1.009 |T| /z + 2 \cdot 10^{-5}\delta \theta n,
\end{eqnarray*}
as desired.
\end{proof}

\subsubsection{Establishing Q5}
\label{sec_est_q5}

We are going to work with the probability distribution $\vphi_{seq}$ (sequence of $m$ independent clauses). Let $\cM$ be the set of all indices $l \in [m]$ such that the $l$th clause $\vphi_{seq}(l)$ does not contain any ot the variables $x_1,\ldots,x_l$ positively. In this case, $\vphi_{seq}(l)$ is still present in the decimated formula $\vphi_{seq}^t$ (with all occurrences of $\neg x_1,\ldots,\neg x_t$ eliminated, of course). For each $l\in \cM$ let $\cL(l)$ be the number of literals in $\vphi_{seq}(l)$ whose underlying variable is in $V_t$. We may assume without loss of generality that for any $l \in \cM$ the $\cL(l)$ ``leftmost'' literals $\vphi_{seq}(l,i), l\leq i\leq \cL(l)$, are the ones with an underlying variable from $V_t$. 

Let $T\subset V_t$. Analysing the operator $\Lambda_T$ directly is a little awkward. Therefore, we will decompose $\Lambda_T$ into a sum of several operators that are easier to investigate. For any $0.1\theta k \leq L \leq 10\theta k, l\leq i <j\leq L, l\in \cM$, and any distinct $x,y \in V_t$ we define
\begin{eqnarray}
	m_{xy}(i,j,l,L,1) = \begin{cases}
	1 &\text{ if } \cL(l) = L, \vphi_{seq}(l,i) = x \text{ and } \vphi_{seq}(l,j) = y \\
	-1 &\text{ if } \cL(l) = L, \vphi_{seq}(l,i) = x \text{ and } \vphi_{seq}(l,j) = \neg y \\
	0 & \text{ otherwise}
	\end{cases}
\end{eqnarray}
and
\begin{eqnarray}
	m_{xy}(i,j,l,L,-1) = \begin{cases}
	1 &\text{ if } \cL(l) = L, \vphi_{seq}(l,i) = \neg x \text{ and } \vphi_{seq}(l,j) = \neg y \\
	-1 &\text{ if } \cL(l) = L, \vphi_{seq}(l,i) = \neg x \text{ and } \vphi_{seq}(l,j) = y \\
	0 & \text{ otherwise},
	\end{cases}
\end{eqnarray}
while we let $m_{xx}(i,j,l,L,\zeta)=0$. 
For a variable $x \in V_t$ we let $\cN(x,T,\zeta)$ be the set of all $l \in \cM$ such that $0.1\theta k \leq \cL(l) \leq 10\theta k$ and the clause $\vphi_{seq}(l)$ contains at most one literal whose underlying variable is in $T\setminus \{x\}$ and $\sign(x)=\zeta$. Moreover, for $l \in \cM$ let $\cN(x,l)$ be the set of all variables $y\in V_t\setminus \{x\}$ that occur in clauses $\vphi_{seq}(l)$ (either positively or negatively). We are going to analyse the operators
\begin{eqnarray*}
	&&\Lambda^{ijL}(T,\mu,\zeta): \mathbb{R}^{V_t} \ra \mathbb{R}^{V_t},\\
	&&\Gamma=(\Gamma_y)_{y\in V_t} \mapsto \left\{\sum_{l\in \cN(x,T,\zeta)} \sum_{y\in \cN(x,l)}\br{2/\nu(\mu)}^{-L}m_{xy}m(i,j,l,L,\zeta)\Gamma_y\right\}_{x\in V_t}
\end{eqnarray*}

\begin{lemma}\label{lem_57}
For any $0.1\theta k\leq L\leq 10\theta k, 1\leq i , j \leq L, i\neq j$ and for any set $T\subset V_t$ we have
\begin{eqnarray*}
\Pr \brk{\left\|\Lambda^{ijL}(T,\mu,\zeta)\right\|_{\square}\leq \delta^5\theta n} \geq 1-\exp\br{-\theta n}
\end{eqnarray*}
\end{lemma}
\begin{proof}
This proof is based on Fact \ref{fac_10}. Fix two sets $A,B \subset V_t$. For each $l\in \cM$ and any $x,y\in V_t$ the two $0/1$ random variables 
\begin{eqnarray}
	\sum_{(x,y)\in A\times B} \max\{m_{x,y}(i,j,l,L,\zeta),0\}, \qquad \sum_{(x,y)\in A\times B} \max\{-m_{x,y}(i,j,l,L,\zeta),0\}
\end{eqnarray}
are identically distributed, because the clause $\vphi_{seq}(l)$ is chosen uniformly at random. In effect, the two random variables
\begin{eqnarray}
	\mu_L^\zeta(A,B) &=& \sum_{l\in \cM}\sum_{(x,y)\in A\times B} \vecone_{l\in\cN(x,T)} \max\{m_{xy}(i,j,l,L,\zeta),0\},\\
	\nu_L^\zeta(A,B) &=& \sum_{l\in \cM}\sum_{(x,y)\in A\times B} \vecone_{l\in\cN(x,T)} \max\{-m_{xy}(i,j,l,L,\zeta),0\}
\end{eqnarray}
are identically distributed. Furthermore, both $\mu_L^\zeta(A,B)$ and $\nu_L^\zeta(A,B)$ are sums of independent Bernoulli variables, because the clauses $(\vphi_{seq}(l))_{l\in [m]}$ are mutually independent.

We need to estimate the mean $\Erw\brk{\mu_L^\pm(A,B)} = \Erw\brk{\nu_L^\pm(A,B)}$. As each of the clauses $\vphi_{seq}(l)$ is chosen uniformly, for each $l\in [m]$ we have
\begin{eqnarray*}
	\Pr\brk{l\in \cM \text{ and } \cL(l) = L} = \binom{k}{L} \theta^L(1-\theta)^{k-L}2^{L-k}.
\end{eqnarray*}
Therefore, 
\begin{eqnarray}
 	\Erw\brk{\mu_L^\zeta(A,B) + \nu_L^\pm(A,B)} &\leq&  m  \binom{k}{L} \theta^L(1-\theta)^{k-L}2^{L-k}\nonumber\\
 	&=& \frac{2^L\rho\theta n}{L}\binom{k-1}{L-1}\theta^{L-1}(1-\theta)^{k-L} \qquad \text{[as $m=2^k\rho/k$]}\nonumber\\
 	&\leq& \frac{2^L\rho\theta n}{L}. \label{equ_lem57_1} 
\end{eqnarray}
Hence, Lemma \ref{Lemma_Chernoff} (the Chernoff bound) yields
\begin{eqnarray*}
 	&&\Pr\brk{|\mu_L^\zeta(A,B)-\Erw\brk{\mu_L^\zeta(A,B)}|>10\sqrt{2^L\rho/L}\cdot\theta n} \\
 	&&\qquad= \Pr\brk{|\nu_L^\zeta(A,B)-\Erw\brk{\nu_L^\zeta(A,B)}|>10\sqrt{2^L\rho/L}\cdot\theta n} \\
 	&&\qquad\leq 16^{-\theta n}.
\end{eqnarray*}
Let $\cA$ be the event that $\exists A,B\subset V_t: \max\{|\mu_L^\zeta(A,B)-\Erw\brk{\mu_L^\zeta(A,B)}|,|\nu_L^\zeta(A,B)-\Erw\brk{\nu_L^\zeta(A,B)}|\}>10\sqrt{2^L\rho/L}\cdot\theta n$. Hence, by the union bound 
\begin{eqnarray*}
 	\Pr\brk{\cA} \leq 2\cdot 4^{\theta n}\cdot 16^{-\theta n}\leq \exp\br{-\theta n}.
\end{eqnarray*}
Thus, with probability $\geq 1- \exp\br{-\theta n}$ we have
\begin{eqnarray*}
	\langle \Lambda^{ijL}(T,\mu,\zeta) \vecone_B,\vecone_A\rangle &=& 2^{-L}(\mu_L^\zeta(A,B)-\nu_L^\zeta(A,B))\\
	&\leq& 2^{-L}(|\mu_L^\zeta(A,B)-\Erw\brk{\mu_L^\zeta(A,B)}|+|\nu_L^\zeta(A,B)- \Erw\brk{\nu_L^\zeta(A,B)}|) \\
	&\leq& \theta n \cdot 20 \sqrt{\frac{\rho}{L2^L}} \leq 0.01\delta^5\theta n \\
	&&\qquad \text{[as $L\geq 0.1\theta k, \theta k \geq \ln(\rho)/c^2$, and $\delta = \exp\br{-c\theta k}$]}.
\end{eqnarray*}
Finally, the assertion follows from Fact \ref{fac_10}.
\end{proof}

\begin{corollary}\label{cor_58}
	With probability at least $1-\exp\br{-0.1\theta n}$ the random formula $\vphi_{seq}^t$ has the following property.
	\begin{equation}\label{equ_cor49}
	\parbox{14cm}{
	Let $T\subset V_T$ and let 
	\begin{eqnarray}
	\bar\Lambda(T,\mu,\zeta) = \sum_{0.1\theta k\leq L \leq 10\theta k} \sum_{j=1}^L \sum_{i=1,i\neq j}^L\Lambda^{ijL}(T,\mu,\zeta).
	\end{eqnarray}
	Then $\left\|\bar\Lambda(T,\mu,\zeta)\right\|_\square\leq\delta^{4.9}\theta n$.
	}
	\end{equation}
\end{corollary}
\begin{proof}
By Lemma \ref{lem_57} and the union bound, we have
\begin{eqnarray*}
	\Pr\brk{\exists T,i,j, L : \left\|\Lambda^{ijL}(T,\mu,\zeta)\right\|_\square > \delta^5 \theta n} \leq (10\theta k)^3 2^{\theta n}\cdot \exp\br{-\theta n} \leq \exp\br{-0.2\theta n}.
\end{eqnarray*}
Furthermore, if $\left\|\Lambda^{ijL}(T,\mu,\zeta)\right\|_\square \leq \delta^5\theta n$ for all $i,j,L$ then by the triangle inequality 
\begin{eqnarray*}
	\left\|\bar\Lambda^{ijL}(T,\mu,\zeta)\right\|_\square  \leq (10\theta n)^3\delta^5 \theta n \leq \delta^{4.9} \theta n \qquad\text{[as $\delta=\exp\br{-c\theta k}$]},
\end{eqnarray*}
as claimed. 
\end{proof}

To complete the proof of \textbf{Q5}, we observe that for $(x,y)\in V_t \times V_t$ the $(x,y)$ entries of the matrices $\Lambda(T,\mu,\zeta)$ and $\bar\Lambda(T,\mu,\zeta)$ differ only if either $x$ or $y$ occurs in a redundant clause. Consequently, \textbf{Q0} ensures that $\left\|\Lambda(T,\mu,\zeta) - \bar\Lambda(T,\mu,\zeta)\right\|_\square = o(n)$. Therefore, Fact \ref{fac_40} and Corollary \ref{cor_58} imply $\vphi^t$ satisfies \textbf{Q5} with probability at least $1-\exp\br{-11\de_t}$.




\begin{thebibliography}{50}
 \bibitem{Barriers}
 D.~Achlioptas, A.~Coja-Oghlan:
 Algorithmic barriers from phase transitions.
 Proc.~49th FOCS (2008) 793--802. 
 
 \bibitem{AchPer}
 D.~Achlioptas, Y.~Peres:
 The threshold for random $k$-SAT is $2^k\ln 2 - O(k)$.
 Journal of the AMS {\bf 17} (2004) 947--973. 
   
\bibitem{AchSor}
D.~Achlioptas, G.~Sorkin:
Optimal myopic algorithms for random $3$-SAT.
Proc.41st FOCS (2000) 590--600. 

\bibitem{BMZ}
A.~Braunstein, M.~M\'ezard, R.~Zecchina:
Survey propagation: an algorithm for satisfiability.
Random Structures and Algorithms {\bf 27} (2005)  201--226.

 \bibitem{BMPWZ} 
 A.\ Braunstein, R.\ Mulet, A.\ Pagnani, M.\ Weigt, R.\ Zecchina:
 Polynomial iterative algorithms for coloring and analysing random graphs.
 Phys.\ Rev.\ E {\bf 68} (2003) 036702.
 
 \bibitem{BraZec}
 A.~Braunstein, R.~Zecchina:
 Survey and Belief Propagation on Random $k$-SAT.
 Lecture Notes in Computer Science, Springer Berlin (2003).
 
 \bibitem{Cheeseman}
 P.~Cheeseman, B.~Kanefsky, W.~Taylor: Where the {\em really} hard problems are.
 Proc.\ IJCAI (1991) 331--337.
 
 \bibitem{BP}
 A.~Coja-Oghlan: On belief propagation guided decimation for random $k$-SAT.
 Proc.\ 22nd SODA (2011) 957--966.

\bibitem{BetterAlg}
A.~Coja-Oghlan:
A better algorithm for random $k$-SAT.
SIAM J.\ Computing {\bf 39} (2010) 2823--2864.

\bibitem{Covers}
A.~Coja-Oghlan:
The asymptotic $k$-SAT threshold.
Proc.\  46th STOC (2014) 804--813.
 
 
 
 \bibitem{Angelica}
 A.~Coja-Oghlan, A.~Y.\ Pachon-Pinzon:
 The decimation process in random $k$-SAT.
 SIAM Journal on Discrete Mathematics {\bf 26} (2012) 1471--1509.
   
\bibitem{DSS}
J.~Ding, A.~Sly, N.~Sun:
Proof of the satisfiability conjecture for large $k$.
Proc.\ 47th STOC (2015) 59--68.


 \bibitem{FMV}
 U.~Feige, E.~Mossel, D.~Vilenchik:
 Complete convergence of message passing algorithms for some satisfiability problems.
 Theory of Computing {\bf 9} (2013) 617--651.
 
 \bibitem{Ehud}
 E.~Friedgut:
 Sharp thresholds of graph properties, and the {$k$-SAT} problem.
 J.\ AMS {\bf12} (1999) 1017--1054.
 
\bibitem{FrSu}
A.~Frieze, S.~Suen:
Analysis of two simple heuristics on a random instance of $k$-SAT.
Journal of Algorithms \textbf{20} (1996) 312--355.

\bibitem{Gamarnik1}
D.\ Gamarnik, M.\ Sudan:
Limits of local algorithms over sparse random graphs.
Proc.\ 5th ITCS (2014) 369--376.

\bibitem{Gamarnik2}
D.\ Gamarnik, M.\ Sudan:
Performance of the Survey Propagation-guided decimation algorithm for the random NAE-K-SAT problem.
arXiv 1402.0052 (2014).

 \bibitem{JLR}
 S.~Janson, T.~{\L}uczak, A.~Ruci\'nski: Random Graphs, Wiley  2000.
  
\bibitem{Kroc}
L.~Kroc, A.~Sabharwal, B.~Selman:
Message-passing and local heuristics as decimation strategies for satisfiability.
Proc\ 24th SAC (2009) 1408--1414.

 \bibitem{pnas}
 F.~Krzakala, A.~Montanari, F.~Ricci-Tersenghi, G.~Semerjian, L.~Zdeborova:
 Gibbs states and the set of solutions of random constraint satisfaction problems.
 Proc.~National Academy of Sciences {\bf104} (2007) 10318--10323.
  
\bibitem{Marino}
R.\ Marino, G.\ Parisi, F.\ Ricci-Tersenghi:
The backtracking Survey Propagation algorithm for solving random $K$-SAT problems.
arXiv 1508.05117 (2015).
  
 \bibitem{Mertens}
S.~Mertens, M.~M\'ezard, R.~Zecchina:
Threshold values of random $K$-SAT from the cavity method.
Random Struct.\ Alg.\ {\bf28} (2006) 340--373.

 \bibitem{MM}
 M.~M\'ezard, A.~Montanari:
 Information, physics and computation.
 Oxford University Press~2009.
 
 \bibitem{MPZ}
 M.~M\'ezard, G.~Parisi, R.~Zecchina:
 Analytic and algorithmic solution of random satisfiability problems.
 Science {\bf 297} (2002) 812--815.
 
 \bibitem{MitchellSelmanLevesque}
 D.~Mitchell, B.~Selman, H.~Levesque:
 Hard and easy distribution of SAT problems.
 Proc.\ 10th AAAI (1992) 459--465. 


\bibitem{Molloy}
M.~Molloy: The freezing threshold for $k$-colourings of a random graph.
Proc.\ 43rd STOC (2012) 921--930.


\bibitem{Allerton}
A.~Montanari, F.~Ricci-Tersenghi, G.~Semerjian:
Solving constraint satisfaction problems through Belief Propa\-gation-guided decimation.
Proc.\ 45th Allerton (2007).

\bibitem{Pearl}
J.~Pearl:
Probabilistic reasoning in intelligent systems: networks of  plausible inference.
Morgan Kaufmann Publishers Inc., San Francisco, CA, USA, 1988.
  
 \bibitem{RTS}
 F.~Ricci-Tersenghi,  G.~Semerjian:
 On the cavity method for decimated random constraint satisfaction problems and the analysis of belief propagation guided decimation algorithms.
 J.\ Stat.\ Mech.\ (2009) P09001.
\end{thebibliography}


\end{document}